\documentclass[final]{tJOT2e}
\usepackage{rotating}
\newcounter{subfigure}

\begin{document}
\title{Dynamical Eigenfunction Decomposition of \\ Turbulent Pipe Flow}
\markboth{Dynamical Eigenfunction Decomposition of  Turbulent Pipe
  Flow}{A. Duggleby, K.S. Ball, M.R. Paul, and P.F. Fischer}
\author{Andrew Duggleby,$^{\dagger \ddagger}$ Kenneth S. Ball,$^\dagger$ Mark R. Paul,$^\dagger$ and
  Paul F. Fischer$\S$
\affil{$\dagger$ Department of Mechanical Engineering, Virginia
  Polytechnic Institute and State University, Blacksburg, VA 24061\\
 $\S$ Mathematics and Computer Science Division, Argonne
  National Laboratory, Argonne, IL 60439}}
\thanks{$\ddagger$ Corresponding author.  E-mail: duggleby@vt.edu}
\received{in review}
\doi{}
 \issn{}
\issnp{} \jvol{} \jnum{} \jyear{2007} \jmonth{Jan}

\maketitle

\begin{abstract}

The results of an analysis of turbulent pipe flow based on a
Karhunen-Lo\`{e}ve decomposition are presented.  The turbulent flow is
generated by a direct numerical simulation of the Navier-Stokes
equations using a spectral element algorithm at a Reynolds number 
Re$_\tau=150$.  This simulation yields a set of basis functions that captures
90\% of the energy after 2,453 modes.  The eigenfunctions are
categorised into two classes and six subclasses based on their
wavenumber and coherent vorticity structure.  Of the total energy, 81\% 
is in the propagating class, characterised by constant phase speeds;
 the remaining energy is found in the non propagating subclasses, the shear and
roll modes.  The four subclasses of the propagating modes are the
wall, lift, asymmetric, and ring modes.  The wall modes display
coherent vorticity structures near the wall, the lift modes display coherent
vorticity structures that lift away from the wall, the asymmetric
modes break the symmetry about the axis, and the ring modes display
rings of coherent vorticity.  Together, the propagating modes form a
wave packet, as found from a circular normal speed locus.  The energy
transfer mechanism in the flow is a four step process.  The process begins with energy being transferred from mean flow
to the shear modes, then to the roll modes.  Energy is then
transfer ed from the roll modes to the wall modes, and then eventually
to the lift modes.  The ring and asymmetric modes act as catalysts that
aid in this four step energy transfer.  Physically, this mechanism
shows how the energy in the flow starts at the wall and then
propagates into the outer layer.

\end{abstract}

\begin{keywords} Direct numerical simulation, Karhunen-Lo\`{e}ve decomposition,
  turbulence, pipe flow, mechanism
\end{keywords}

\section{Introduction}
Turbulence, hailed as one of the last major unsolved problems of
classical physics, has been the subject of numerous publications as
researchers seek to understand the underlying physics, structures, and
mechanisms inherent to the flow \cite{panton_overview}.  The standard test problem for
wall-bounded studies historically has been  
turbulent channel flow because of its simple geometry and
computational efficiency.  Even though much insight has been achieved
through the study of turbulent channel flow, it remains an academic
problem because of its infinite (computationally periodic) spanwise
direction.  

The next simplest geometry is turbulent pipe flow, which is of interest
because of its real-world applications and its slightly
different dynamics.  The three major differences between turbulent pipe
and channel flow are that pipe flow  displays a log layer, but overshoots the theoretical profile until a much 
higher Reynolds numbers ($\mathrm{Re}=3000$ for a channel versus
$\mathrm{Re}=7442$ for a pipe \cite{patel_head,durst}), has  an observed
higher critical Reynolds number, and is linearly stable to an 
infinitesimal disturbance \cite{orszag_patera, oSullivan_breuer}.
Unfortunately, few direct numerical simulations of turbulent pipe flow
have been carried out because of the complexity in handling the  
numerical radial singularity at the origin.  Although the singularity itself is
avoidable, its presence causes standard high-order spectral methods to fail to converge
exponentially.  As a result, only a handful of  algorithms found in the literature for turbulent pipe flow; the first reported use   for each algorithm is listed in
Table \ref{radialDiscr}.  These algorithms
typically use a low-order expansion in the radial direction.  Only
Shan et al. \cite{ma} using concentric Chebyshev domains
(``piecemeal'') and Loulou et al. \cite{loulou} using basis spline (B-spline) polynomials provide
a higher-order examination of turbulent pipe flow.  Using a spectral
element method, this study provides not only a high-order examination
but the first exponentially convergent investigation of turbulent
pipe flow through direct numerical simulation (DNS).

\begin{table}
\tbl{Summary of existing  algorithms for the DNS of turbulent
  pipe flow,  citing the first reported use for each algorithm.  As seen, most
  approaches use a 2nd-order radial discretization.  The reason is that the standard
  spectral methods, in the presence of the coordinate
  singularity, achieve only 2nd-order convergence instead of geometric convergence.}
{\begin{tabular}{ll}
\toprule
Authors & Radial Discretization \\
\colrule
Eggels et al. (1994) \cite{eggels} & 2nd-order FD - first pipe DNS \\
Verzicco and Orlandi (1996) \cite{verzicco} & 2nd-order FD flux based \\
Fukagata and Kasagi (2000) \cite{fukagata} & 2nd order FD - energy
conservative \\
Shan et al. (1999) \cite{ma} & ``Piecemeal'' Chebyshev \\
Loulou et al.(1997) \cite{loulou} & B-spline \\
\botrule
\end{tabular}}
\label{radialDiscr}
\end{table}

With this DNS result, one of the studies that can be performed with the full flow field and time
history it provides is an analysis based
on an orthogonal
decomposition method.  In such methods, the flow is expanded in terms of a natural or preferred
turbulent basis.  One method used frequently in the field of turbulence is Karhunen-Lo\`{e}ve (KL) decomposition, which uses a two-point spatial correlation
tensor to generate the eigenfunctions of the flow.  This is sometimes referred to as
proper orthogonal decomposition, empirical orthogonal function, or
empirical eigenfunction analysis.

Work in this area was pioneered by Lumley, who was the first to use the KL
method in homogeneous turbulence \cite{lumley1,lumley2}.  This method was later
applied to turbulent channel flow in a series of papers by Ball, Sirovich,
and Keefe \cite{ball,sirovich1} and Sirovich, Ball, and Handler \cite{sirovich2}, who discovered plane waves
and propagating structures 
that play an essential role in the production of turbulence through
bursting or sweeping events.  To study the interactions 
of the propagating structures, researchers have examined minimal
expansions of a turbulent flow
\cite{aubry, zhou_sirovich,sirovich_xhou, smith2005}.  These efforts have led to recent work by
Webber et al. \cite{webber1, webber2}, who examined the energy
dynamics between KL modes and discovered the energy transfer path from
the applied pressure gradient to the flow through triad interactions
of KL modes.

This present study uses a spectral element Navier-Stokes
solver to generate a globally high-order turbulent pipe flow data set.
 The Karhunen-Lo\`{e}ve method is used to examine the turbulent flow
structures and dynamics of turbulent pipe flow. 

\section{Methodology}
The direct numerical simulation of the three-dimensional
time dependent Navier-Stokes equations is a computationally intensive task.  By
fully resolving the 
necessary time and spatial scales of turbulent flow, however, no subgrid
dissipation model is needed, and thus a turbulent flow is calculated directly
from the equations.  DNS has one main advantage over experiments, in
that the whole flow field and time history are known, enabling analyses
such as the Karhunen-Lo\`{e}ve decomposition.

Because of the long time integration and the grid resolution necessary for
DNS, a high-order (typically spectral) method is often used to keep
numerical round-off and dissipation error  small.  Spectral methods
and spectral elements use trial
functions that are infinitely and analytically differentiable to span the
element.  This approach decreases the global error exponentially  with
respect to resolution, in contrast to an algebraic decrease with
standard methods such as finite difference or finite element methods \cite{boyd}.

\subsection{Direct Numerical Simulation}
\subsubsection{Numerical Methods}
This study uses a spectral element Navier-Stokes solver that has been developed
over the past 20 years \cite{fischer_patera, tufo}  to solve the
  Navier-Stokes equations:
\begin{eqnarray}
& \partial_t \mathbf{U}+\mathbf{U}\cdot \nabla
\mathbf{U}=-\nabla P+\mathrm{Re}_\tau^{-1} \nabla^2 \mathbf {U} \\
\label{Navier_Stokes}
& \nabla \cdot \mathbf{U} = 0.
\label {continuity}
\end{eqnarray}
In the above equations, $\mathbf{U}=(U_r, U_\theta, U_z)$ is the velocity vector
  corresponding to the radial, azimuthal, and streamwise direction
  respectively; Re$_\tau= U_\tau R
  / \nu$ is the Reynolds number; $R$ is the radius of the pipe; $\nu$ is
  the kinematic viscosity; and $U_\tau = \sqrt{\tau_w / \rho}$ is the shear
  velocity based upon the wall shear stress $\tau_w$ and density $\rho$.
  This solver employs a
geometrically 
flexible yet exponentially convergent spectral element discretization in
space, in this case allowing the geometry to be fitted to a cylinder.
The domain is subdivided into elements, each containing high-order
(usually 11--13) Legendre Lagrangian interpolants.  The resulting data 
localisation allows for minimal communication between elements, and therefore
efficient parallelization.  Time discretization is done with third-order
operator splitting methods, and  the remaining tensor-product polynomial bases are solved
by using conjugate gradient iteration with scalable Jacobi and hybrid
Schwarz/multigrid preconditioning \cite{lottes}. 

\subsubsection{Geometry}

\begin{figure}
\epsfxsize=2 in
\centerline{ \epsfbox{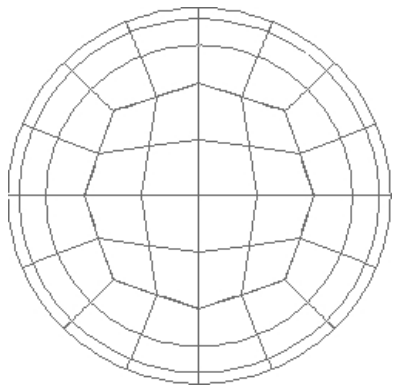}}
\caption{Cross section of the spectral element grid for pipe flow.  The
  spectral element algorithm avoids the singularity found in polar
  cylindrical coordinates.}
\label{grid}
\end{figure}

Spectral elements are effective in cylindrical geometries \cite{paul}
and their use elegantly avoids the radial singularity
at the origin.  The
mesh is structured as a box near the origin and transitions
to a circle near the pipe walls (Figure \ref{grid}), maintaining a
globally high-order method at both the wall and the origin.  In addition to
avoiding the numerical error associated with the singularity, 
the method also avoids the time-step restriction due to the smaller element width at the origin of a
polar-cylindrical coordinate system, which could lead to potential
violations of the Courant-Friedrichs-Levy stability criteria.

Each slice, as shown in Figure \ref{grid}, has 64 elements, and there are 40
slices stacked in the streamwise direction, adding up to a length of 20R.  Each element has 12th-order
Legendre polynomials in each direction for Re$_\tau=150$.  This
discretization results in
4.4 million degrees of freedom.  Near the wall, the grid spacing 
  normalised by 
wall units ($\nu / U_\tau$) denoted by the superscript $^+$
is $\Delta r^+\approx 0.78$ and  $(R \Delta \theta)^+ \approx 4.9$, where
$r$ is the radius and $\theta$ is the azimuthal angle.  Near the centre of the
pipe, the spacing in Cartesian coordinates is  $\Delta^+ \approx 3.1$.  The streamwise grid
spacing is a constant $\Delta z^+=6.25$ throughout the domain
 where $z$ is the streamwise coordinate.

\subsubsection{Benchmarking at $\mathrm{Re}_\tau=180$}

Benchmarking was performed at Re$_\tau=180$ with the
 experiments and DNS of Eggels et al. \cite{eggels} and the DNS of Fukagata and Kasagi \cite{fukagata}.  For
this higher Reynolds number flow, 14th-order polynomials were used, giving
grid spacings near the wall of $\Delta r^+\approx 0.80$ and {\bf $(R \Delta
\theta)^+ \approx 5.0$, $\Delta^+ \approx 3.2$} in the centre, and $\Delta
z^+=6.42$.  Eggels et al. and Fukagata and Kasagi used a 
spectral Fourier discretization in the azimuthal and axial directions and
then a 2nd-order finite difference discretization in the radial direction.
Also, both groups used a domain length of 10R and grid sizes in their DNS
studies of $96 \times 128 \times 256$ for $r$, $\theta$, and $z$
 directions, respectively.  

\begin{figure}
\epsfxsize=3 in
\centerline{ \epsfbox{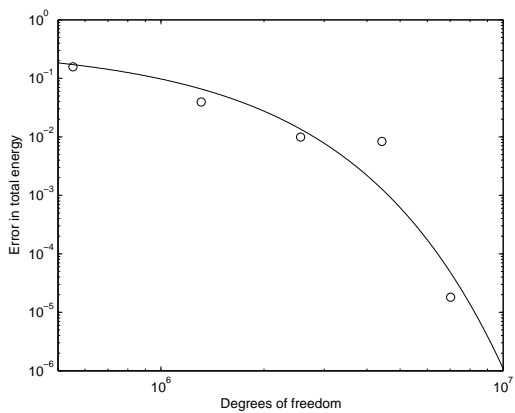}}
\caption{Convergence plot showing the error in fluctuating total energy of the flow versus degrees
of freedom after 1000 time steps for a fixed grid and polynomial orders of 8, 10, 12,
and 14 (circles), using 16th order polynomials to approximate the exact
solution.  The exponentially decay of the curve fit (solid) shows that our spectral
element algorithm is geometrically or exponential convergent.}
\label{convergence}
\end{figure}

 The turbulent flow was tripped with a solenoidal disturbance, and
  the simulation was run until transition was complete.  This was
  estimated by examining the mean flow and root-mean-squared
  statistics over sequential periods of 100 $t^+$ until a
  statistically steady  state was achieved, marked by a change of
  less than 1  percent between the current and previous period.  In
  this case, it took $7000 t^+$ to arrive at a  statistically steady
  state.  Convergence was checked by examining the total fluctuating
  energy after 1000 time steps for increasing polynomial orders 8, 10,
  12, and 14 with a fixed
  number of elements (2560).  The absolute value of the difference of
  the total fluctuating energy for each simulation and the value
  obtained using 16th order polynomials was
  plotted versus the total number of degrees of freedom in Figure
  \ref{convergence}.  The exponential decay of the error with
  increasing degrees of freedom shows that our spectral element
  algorithm achieves geometric or exponential convergence.

\begin{figure}[h]
\centering
\includegraphics[width=4 in]{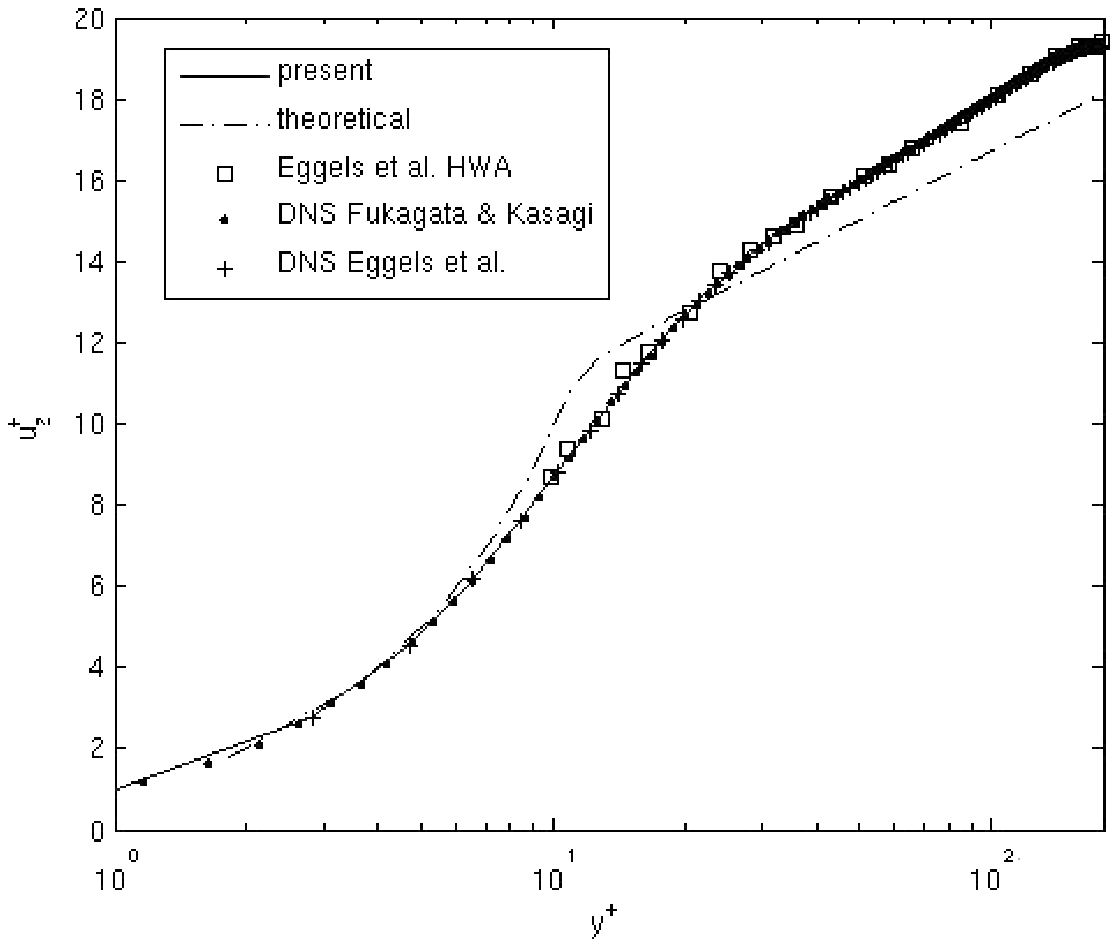}
\caption{Benchmark of mean velocity profile for the spectral element algorithm vs low-order methods of Eggels et
  al.\cite{eggels} and Fukagata and Kasagi \cite{fukagata} for Re$_\tau=180$.  The theoretical line is the
  law of the wall $U^+=y^+$ and the log layer $U^+=1/0.41 \log y^+ +
  5.5$.  Deviations from the log layer are expected in turbulent
  pipe flow until much higher Reynolds numbers.}
\label{mean_180}
 \end{figure}

\begin{figure}[h]
\centering
{\includegraphics[width=4 in]{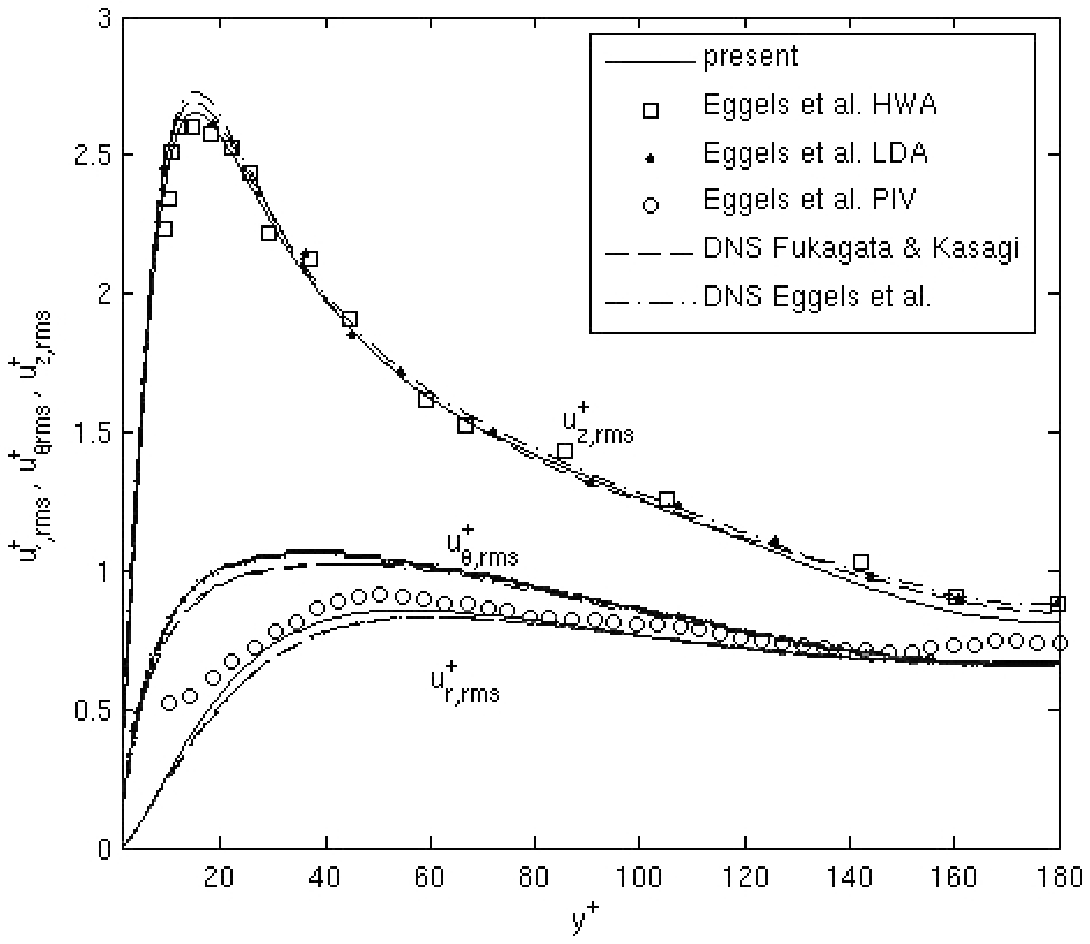}}
\caption{Benchmark of the rms profiles for Re$_\tau$=180 comparing the
  spectral element algorithm vs the low-order methods of Eggels et
  al.\cite{eggels} and Fukagata and Kasagi \cite{fukagata}. As seen
  near $y^+=20$ in 
  the $u^+_{z,rms}$ and near $y^+=50$ in the $u_{r,rms}^+$, the
  higher-order method is closer to the experimental results than the
  2nd-order methods.}
\label{rms_180}
 \end{figure}

The mean flow profiles in Figure \ref{mean_180} correspond well with
 the  hot wire anemometry (HWA) results of Eggels et al. \cite{eggels}, but as seen in the root-mean-squared (rms) statistics shown in Figure \ref{rms_180}, the spectral element calculation shows a lower peak
 $U_{z,rms}$  and higher peak $U_{\theta, rms}$ and $U_{r,rms}$
 compared to the HWA, particle image velocimetry (PIV), and
 laser Doppler anemometry (LDA) results.
 These results are in
 contrast to those with channel flow, as reported by Gullbrand \cite{gullbrand},
 where the
 2nd order finite difference methods undershoot the spectral method wall-normal
 velocity rms.

When compared to the experimental results of Eggels et al. \cite{eggels}, the
 spectral element results are in better agreement than the 2nd-order finite
 difference results.  We note that Eggels et al. report that 
 their (PIV) system had difficulties capturing $U_{r}$ near the wall and
 near the centerline of 
 the pipe due to reflection, which could explain the deviation of all of the DNS
 results in that area.

\begin{figure}
\epsfxsize=4 in
\centerline{\epsfbox{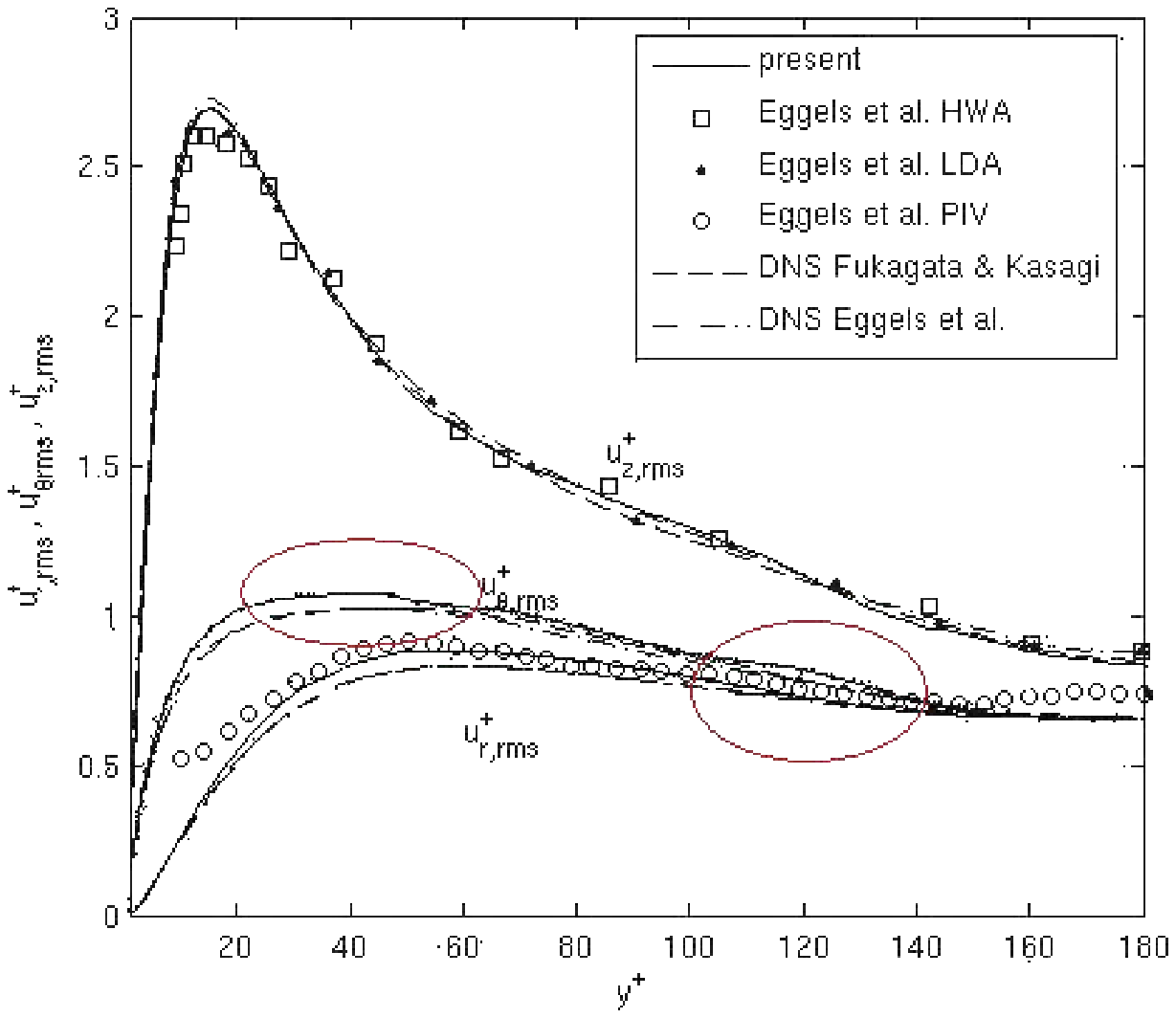}}
\caption{Effect of a short domain (10R) with a high-order method,
  showing unphysical deviations in the rms profiles near $y^+=50$ and
  $y^+=120$ for the spectral element algorithm for Re$_\tau=180$.  A
  bulge, highlighted by the red ovals, is present
  in the azimuthal rms that is not found in the low-order methods of Eggels et
  al.\cite{eggels} and Fukagata and Kasagi \cite{fukagata}.   By
  extending the domain to 20R, the unphysical bulge is removed.
  bf{It is therefore surmised that the unphysical result is possibly}
  due to the more dissipative nature of the low-order schemes that
  washes away a long-scale structure that the present high-order scheme
  captures and re-enters the inlet from the periodic outflow
  conditions.}
\label{L10}
\end{figure}

A second major difference our spectral element method and the previous
pipe DNS using 2nd-order finite difference is
the domain size.  With the spectral element method, which results in a global
high-order convergence, a domain size of 10R yielded  unphysical
  results in the
flow, as seen in the bulge in the azimuthal $U_{\theta, rms}$ in Figure \ref{L10}
at $y^+=55$ and $y^+=120$.  This
 unphysical result arose even with a more refined grid,  and is
therefore not a function of under-resolution.  However, when the domain of
the spectral element method was extended to 20R, this problem disappeared.
We surmise that the 2nd-order finite difference method dissipated 
  some
large-scale structures after 10R that the higher-order  spectral
element case appropriately resolves.  This  undissipated structure, 
because of the periodic boundary conditions,  then re-enters the inlet and
causes the unphysical bulges in the azimuthal rms profile.
This result is also 
supported by the work of Jim\'{e}nez \cite{jimenez} in turbulent channel flow
(Re$_\tau$=180) that shows large-scale 
structures that extend well past the domain size of $z^+ = 1800=10R$.

This benchmark confirms that the spectral element algorithm, at the given grid resolution and
domain size, will
generate the appropriate turbulent flow field and time history to perform the
KL decomposition.

\subsection{Karhunen-Lo\`{e}ve Decomposition}
For completeness, the Karhunen-Lo\`{e}ve Decomposition method is briefly described
here, but for more detail, see \cite{ball, sirovichKL1, sirovichKL2,
  sirovichKL3, sirovich_chaos, aubry, sirovich2, sirovich1, lumley1,
  lumley2 }.

\subsubsection{Numerical Method}
The Karhunen-Lo\`{e}ve method is the solution of the two-point
velocity correlation kernel equation defined by
\begin{eqnarray}
\label{KL}
\int_0^L\int_0^{2 \pi} \int_0^R K(\mathbf{x}, \mathbf{x'})\mathbf{\Phi} (\mathbf{x}') r' dr'
d\theta' dz' = \lambda \mathbf{\Phi} \\
\label{two_point}
K(\mathbf{x}, \mathbf{x'})=\langle \mathbf{u} (\mathbf{x})
\otimes \mathbf{u} 
(\mathbf{x'}) \rangle,
\end{eqnarray}
where $\mathbf{u}(\mathbf{x})=\mathbf{U}(\mathbf{x})- \overline{\mathbf{U}}(\mathbf{x})$ is the fluctuating component of the velocity
and where the mean $\overline{\mathbf{U}}(\mathbf{x})$ is determined by averaging over both homogeneous
planes and time.  The angle brackets represent an average over many
time steps, on the order of $t^+ \approx 40,000 \sim 50,000$, to
sample the the entire attractor.  The $\otimes$ denotes the outer
product, establishing the kernel as the velocity two-point correlation between every spatial point
$\mathbf{x}=(r,\theta,z)$ and $\mathbf{x'}=(r', \theta', z')$.

For turbulent pipe flow, with two homogeneous directions providing
translational invariance in the $\theta$ (azimuthal) and $z$
(streamwise) direction, Eq. (\ref{two_point}) becomes
\begin{eqnarray}
\label{two_point_pipe}
K (\mathbf{x}, \mathbf{x'})&=&K (r, r', \theta-\theta',z-z')
\\
&=& \mathcal{K}(n,m; r, r')e^{i n\theta}e^{i 2 \pi m z/L}.
\end{eqnarray}
Thus, given the kernel in Eq. (\ref{two_point_pipe}), the eigenfunctions
have the form
\begin{equation}
\label{eigenfunctions}
\mathbf{\Phi}_{(m,n)} (r,\theta,z)=\mathbf{\Psi} (n,m; r) e^{i n \theta}e^{i 2 \pi m z / L},
\end{equation}
where $n$ is the azimuthal wavenumber and $m$ the streamwise wavenumber.
The determination of $\mathbf{\Psi}$ is then given by

\begin{eqnarray}
\label{eigen_solve}
\int_0^R  \mathcal{K}(m,n; r,r')\mathbf{\Psi}^\star(m,n;
r') r' dr'=\lambda_{mn}\mathbf{\Psi}(m,n;r), \\
\label{reduced2pt}
\mathcal{K}(m,n; r,r')=\langle \hat{\mathbf{u}}(n,m; r) \otimes \hat{\mathbf{u}}(n,m; r') \rangle
\end{eqnarray}
where the $\star$ denotes the complex conjugate,
$\lambda$ is the eigenvalue, and $\hat{\mathbf{u}} (n,m; r)$ is the Fourier
transform of the fluctuating velocity in the azimuthal and axial direction.

In the present problem, 2,100 snapshots of the flow field were taken,
corresponding to one snapshot 
every eight viscous time steps ( $t^+= U_\tau^2 t / \nu$).  The results of each snapshot were
projected to an evenly spaced grid with $101 \times 64 \times 400$
points in $r,\theta,$ and $z$,
respectively.  The Fourier transform of the data was then taken and
the kernel assembled.  This kernel was averaged over every snapshot
to generate the final kernel to be decomposed.  Since the dimension of
K is 303 (given by three velocity components on 101 radial grid points) there
are 303 eigenfunctions and eigenvalues  for each Fourier wavenumber
pair $(m,n)$.  The eigenfunctions are ranked
in descending order with the quantum
number $q$ to specify the particular eigenfunction associated with the
corresponding eigenvalue.  Thus, it requires a triplet
$(m,n,q)$ to specify a given eigenfunction. 

The eigenfunctions $\mathbf{\Psi}_q(m,n; r)$
are complex vector functions and are normalised so that the inner product
of the full eigenfunction $\mathbf{\Phi}_{(m,n,q)}$ is of  unit
length,  namely
$(\mathbf{\Phi}_{(m,n,q)},\mathbf{\Phi}_{(n',m',q')})=\delta_{mm'}\delta_{nn'}\delta_{qq'}$,
where $\delta$ is the Kronecker delta.  The eigenvalues physically represent the average energy of the flow
  captured by the eigenfunction $\mathbf{\Phi}_{(m,n,q)}$,

\begin{equation}
\label{energy}
\lambda_{(m,n,q)}=\langle |(\mathbf{u}, \mathbf{\Phi}_{(m,n,q)})|^2 \rangle.
\end{equation}

We note that the eigenfunctions, as an orthogonal expansion
of the flow field, retain the properties of the flow field, such as
incompressibility and boundary conditions of no slip at  the wall. 

\subsubsection{Symmetry Considerations}
The pipe flow is  statistically invariant under azimuthal reflection,

\begin{equation}
\label{reflection}
R_\theta:  (r, \theta, z, u_r, u_\theta, u_z) \rightarrow (r, -\theta,
z, u_r, -u_\theta, u_z),
\end{equation}
and taking advantage of
it reduces the total number of calculations as well as the memory and storage
requirements.  We note that, because of its geometry, turbulent channel flow has two
more symmetries -- a vertical reflection, and a
x-axis rotation -- that are not present in the pipe, since a negative radius
is equivalent to a 180 degree rotation in the azimuthal direction.

A major consequence of this symmetry is that the resulting
eigenfunctions are also symmetric, and the modes with azimuthal wave
number $n$ will be the azimuthal reflection of the modes with wave
number $-n$, 

\begin{equation}
\label{reflect_wave}
R_\theta: (\Phi^r_{(m,n,q)}, \Phi^\theta_{(m,n,q)},
\Phi^z_{(m,n,q)}) \rightarrow (\Phi^r_{(m,-n,q)},
-\Phi^\theta_{(m,-n,q)}, \Phi^z_{(m,-n,q)}), 
\end{equation}
thus reducing the total computational memory needed for this
calculation.

\subsubsection{Time-Dependent Eigenfunction Flow Field Expansion}
The KL method provides an orthogonal set of eigenfunctions that span
the flow field. As such, the method allows the flow field to be represented as
an expansion in that basis,

\begin{equation}
\label{flowfield_expand}
\mathbf{u}(r, \theta, z, t)=\sum_{(m,n,q)} a_{(m,n,q)} (t) \mathbf{\Phi}_{(m,n,q)}(r,\theta,z),
\end{equation}
with
\begin{equation}
\label{coeff}
 a_{(m,n,q)} (t)=(\mathbf{\Phi}_{(m,n,q)}(r,\theta,z),\mathbf{u}(r, \theta, z,t)).
\end{equation}
Since the Fourier modes are orthogonal to each other, equation \ref{coeff} becomes

\begin{equation}
\label{coeff_FFT}
 a_{(m,n,q)} (t)=2 \pi L_z \int_0^R \mathbf{\Psi}_q(m,n;r) \mathbf{\hat{u}^\star}(m,n; r,t) r dr,
\end{equation}
with $\mathbf{\hat{u}}$ being the Fourier transform of $\mathbf{u}$ in
the azimuthal and streamwise direction with wavenumbers $n$ and $m$, respectively.

The time history of the eigenfunctions can be used to examine their
interactions, such as the energy interaction
examined by Webber et al. \cite{webber2} and bursting events by
Sirovich et al. \cite{sirovich2}.

\section{Results}
This section presents the results of our analysis of turbulent pipe
flow based on KL decomposition.

\subsection{Mean Properties of Flow and Flow Statistics}
\begin{figure}
\epsfxsize=4 in
\centerline{\epsfbox{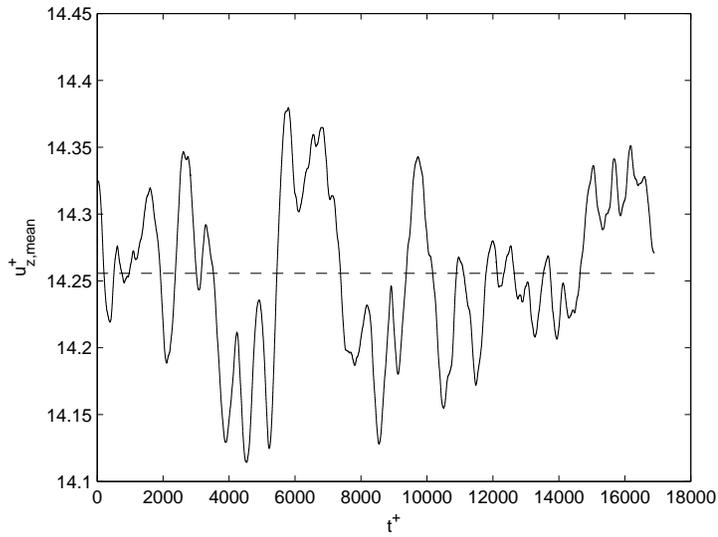}}
\caption{Variations in bulk mean velocity $u_m (t^+)$ consistent with
  turbulent pipe flow.  The dashed line is the time averaged mean velocity.}
\label{mean_time}
\end{figure}

DNS was performed for $\mathrm{Re}_\tau=150$.  The Reynolds number
based on the centreline velocity is $\mathrm{Re}_c\approx 5700$ and
based on the mean velocity is $\mathrm{Re}_m\approx 4300$.  This
range is above the observed
critical Reynolds number for pipe flow and exhibits self-sustaining
turbulence, as seen from the fluctuating time history of the mean velocity
shown in Figure \ref{mean_time}. The velocity profile, seen in Figure
 \ref{mean_profile}, shows the mean velocity with respect to wall
 units  away from the wall
($y^+=(r-R)U_\tau / \nu$).  The profile fits the law of
the wall but fails to conform to the log law, and as mentioned in
Section 1, the log law is not expected for turbulent pipe flow until much higher Reynolds number.

\begin{figure}
\epsfxsize=4 in
\centerline{\epsfbox{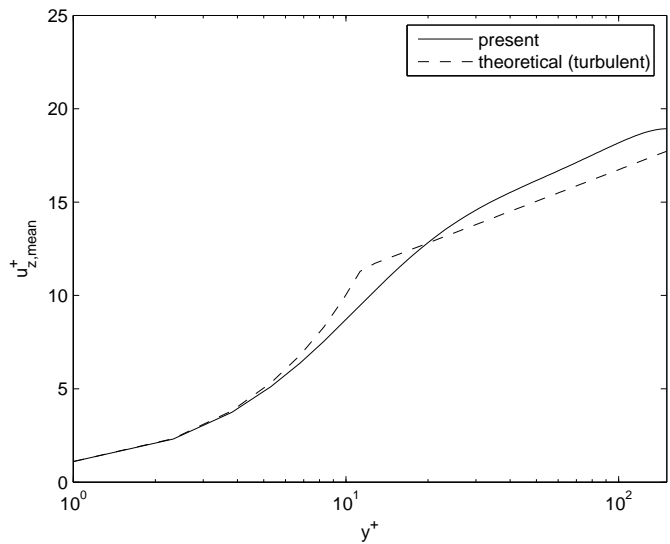}}
\caption{Mean velocity profile (solid) compared to the
  theoretical law of the wall $u^+=y^+$ and the log layer $u^+=1/0.41 \log y^+ +
  5.5$ (dashed).  As in the Re$_\tau=180$ case, the overshoot of the
  log layer above the theoretical profile is expected in turbulent
  pipe flow until much higher Reynolds number. }
\label{mean_profile}
\end{figure}

The rms velocity fluctuation profiles and the Reynolds stress profile is shown in Figures
\ref{rms} and \ref{reynolds}.  The streamwise fluctuations peak near $y^+=16$.  The
azimuthal and axial velocities show a weaker peak near $y^+=39$ and
$y^+=55$, respectively, and then remain fairly flat throughout the pipe.  The Reynolds stress $\overline{v_z v_r}$ has a maximum of 0.68 at $y^+=31$.

\renewcommand{\thefigure}{\arabic{figure}\alph{subfigure}}
\setcounter{subfigure}{1}
\begin{figure}[h]
\centering
\begin{tabular}{cc}
\begin{minipage}{2.75 in}
\includegraphics[width=2.75 in]{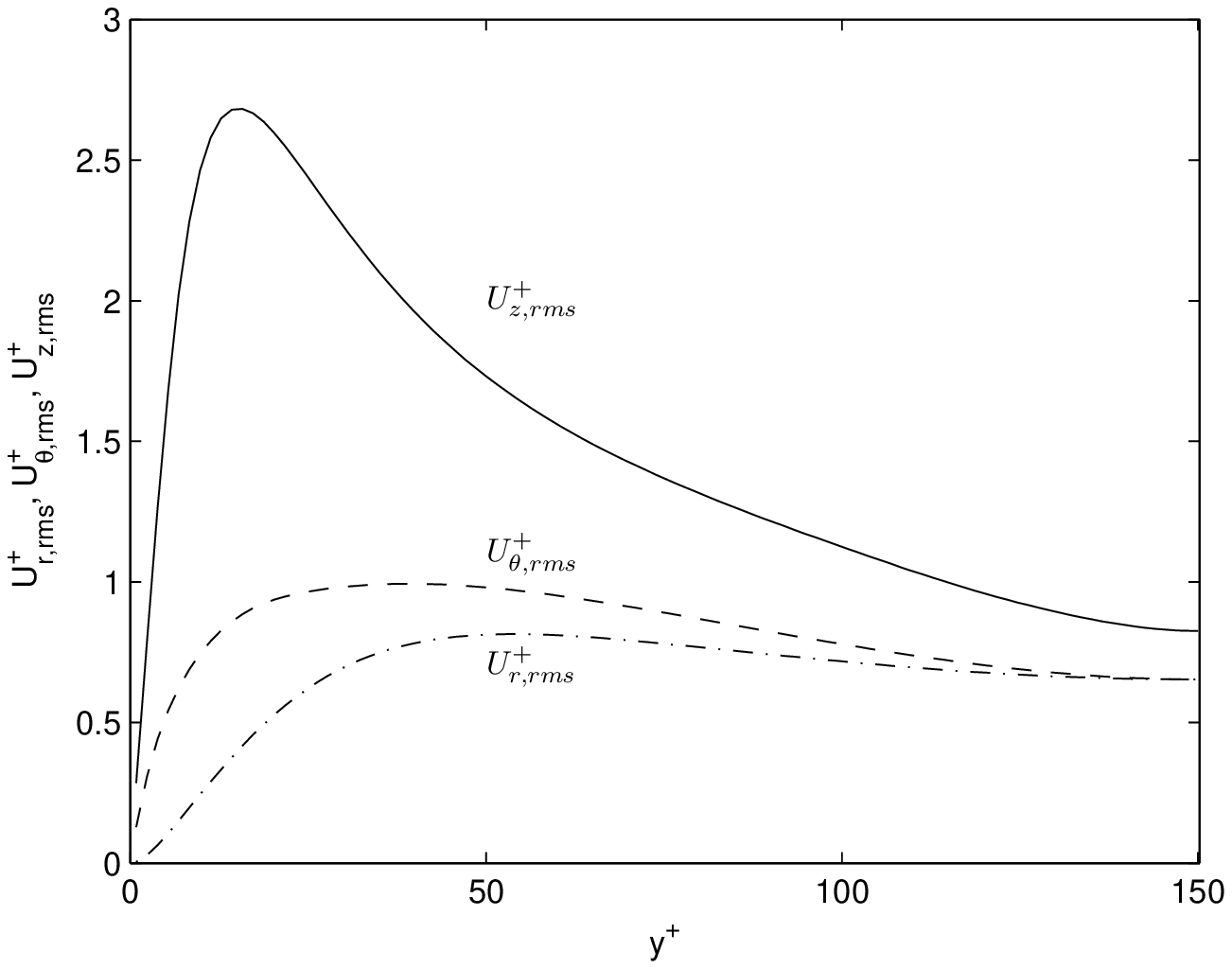}
\caption{Root-mean-square statistics across the pipe for the radial (solid),
  azimuthal (dashed), and streamwise (dot-dashed) velocities.  The results show the standard profiles,
  similar to the Re$_\tau=180$ rms profiles.}
\label{rms}
\end{minipage}
& 
\begin{minipage}{2.75 in}
\addtocounter{figure}{-1}
\addtocounter{subfigure}{1}
 \resizebox{2.75 in}{!}{\includegraphics{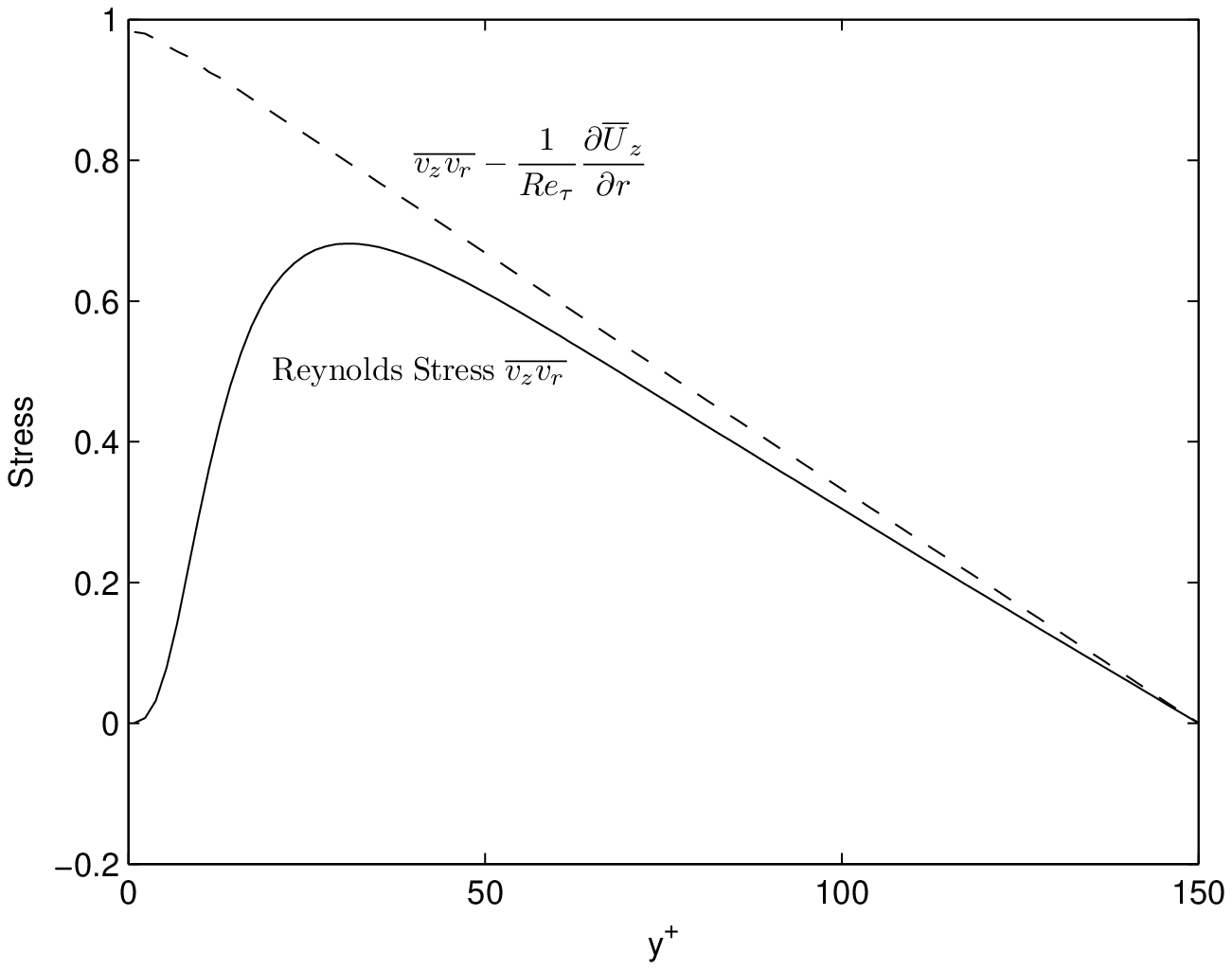}}
\caption{Reynolds stress $\overline{v_z v_r}$ across the pipe (solid), peaking
at 0.68 at $y^+=31$, and the total shear stress distribution (dashed).}
\label{reynolds}
\end{minipage} \\
\end{tabular}

 \end{figure}

\renewcommand{\thefigure}{\arabic{figure}}

\subsection{Eigenvalue Spectrum}
\begin{figure}
\epsfxsize=4 in
\centerline{\epsfbox{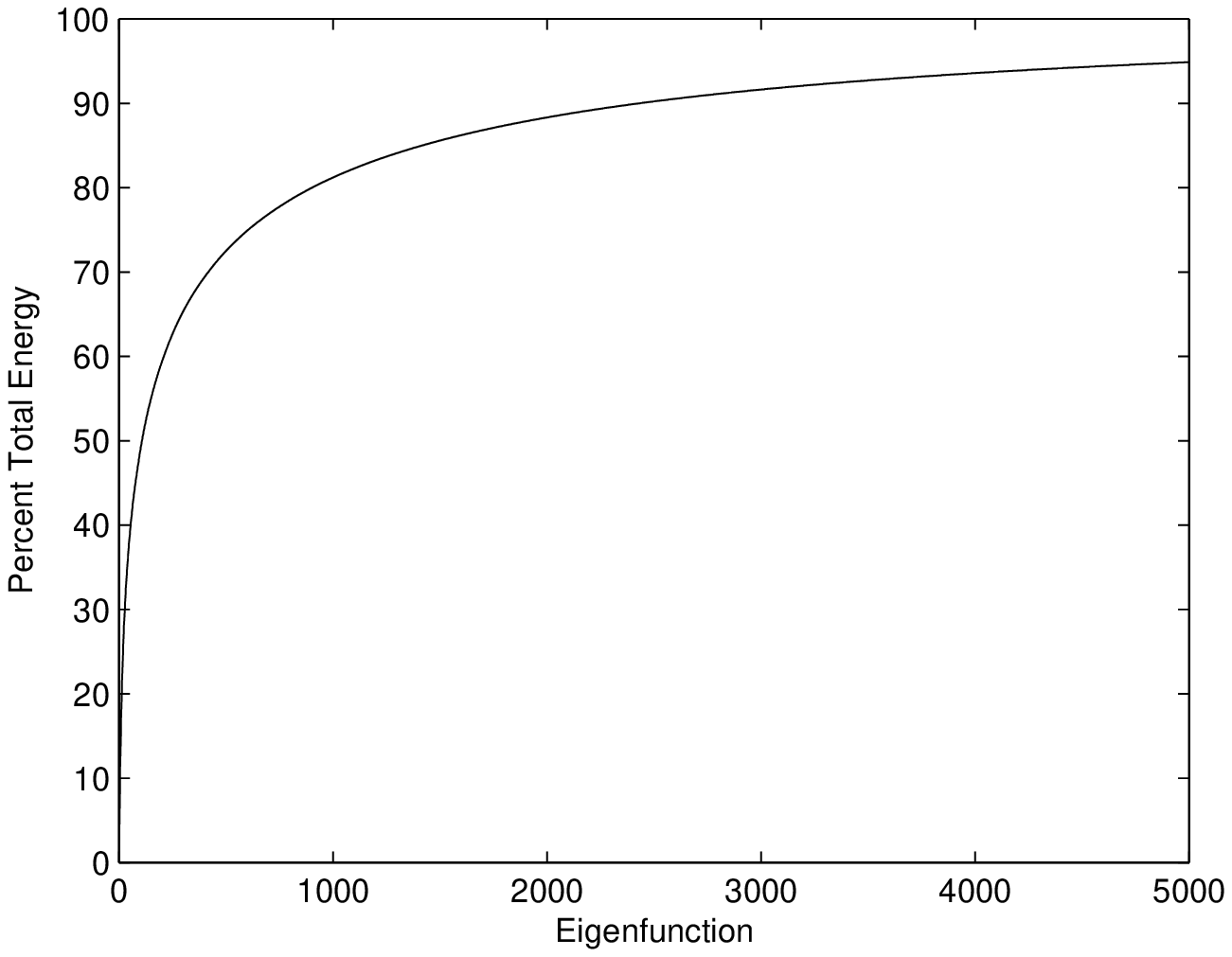}}
\caption{Percentage of energy retained in the KL expansion with the
  90\% mark achieved after 2453 modes (Karhunen-Lo\`{e}ve dimension).}
\label{energyKL}
\end{figure}

As discussed in Section 2.2, the eigenvalues represent the energy of
each eigenfunction.  By ordering the eigenvalues from largest to
smallest, one can minimise the number of eigenfunctions, $N$, needed to capture a given
percentage of the energy of the flow.  Table \ref{top25} shows the first 25 eigenfunctions, and
Figure \ref{energyKL} shows the running total of energy versus modes.
 The first mode to have $q \neq 1$ is the 41st mode (0,0,2).
The 90\% mark is reached at $D_{KL}=2,453$, where $D_{KL}$ can be considered as a measure of the intrinsic dimension of
the chaotic attractor describing the turbulence as discussed by Zhou and Sirovich
\cite{sirovich_chaos, zhou_sirovich}.  These eigenfunctions are the
preferred or natural basis function for turbulent pipe flow, and
insight is gained by observing their qualitative structure.

\begin{table}
\tbl{First 25 eigenvalues ranked in descending order of energy.}
{\begin{tabular}{lccccc}
\toprule
Index & m & n & q & Eigenvalue & Energy (\% Total) \\
\colrule
1 & 0 & 6 & 1 & 1.61 & 2.42 \% \\
2 & 0 & 5 & 1 & 1.48 & 2.22 \% \\
3 & 0 & 3 & 1 & 1.45 & 2.17 \% \\
4 & 0 & 4 & 1 & 1.29 & 1.93 \% \\
5 & 0 & 2 & 1 & 1.26 & 1.88 \% \\
6 & 1 & 5 & 1 & 0.936 & 1.40 \% \\
7 & 1 & 6 & 1 & 0.917 & 1.37 \% \\
8 & 1 & 3 & 1 & 0.902 & 1.35 \% \\
9 & 1 & 4 & 1 & 0.822 & 1.23 \% \\
10 & 0 & 1 & 1 & 0.805 & 1.20 \% \\
11 & 1 & 7 & 1 & 0.763 & 1.14 \% \\
12 & 1 & 2 & 1 & 0.683 & 1.02 \% \\
13 & 0 & 7 & 1 & 0.646 & 0.97 \% \\
14 & 2 & 4 & 1 & 0.618 & 0.92 \% \\
15 & 0 & 8 & 1 & 0.601 & 0.90 \% \\
16 & 2 & 5 & 1 & 0.580 & 0.87 \% \\
17 & 1 & 1 & 1 & 0.567 & 0.85 \% \\
18 & 2 & 7 & 1 & 0.524 & 0.78 \% \\
19 & 1 & 8 & 1 & 0.483 & 0.72 \% \\
20 & 2 & 6 & 1 & 0.476 & 0.71 \% \\
21 & 2 & 3 & 1 & 0.454 & 0.68 \% \\
22 & 2 & 2 & 1 & 0.421 & 0.63 \% \\
23 & 2 & 8 & 1 & 0.375 & 0.56 \% \\
24 & 1 & 9 & 1 & 0.358 & 0.54 \% \\
25 & 3 & 4 & 1 & 0.354 & 0.53 \% \\
\botrule
\end{tabular}}
\label{top25}
\end{table}

\subsection{Structure of the Eigenfunctions}
The eigenfunctions resulting from the KL decomposition can be
categorised into two distinct classes and six sub-classes based on their
characteristics, as listed in Table \ref{structureTBL} and depicted in
Figure \ref{subclassSpectra}.  It is
observed that certain characteristics of 
the eigenfunction are dependent on the wavenumber.  The modes with
higher azimuthal than axial wavenumber turn more than they lift, and
as such the near-wall stretching structures are found.  Likewise the modes with
higher axial than azimuthal wavenumber lift more than they turn, and
as such the lifting structures that extend from the near wall to the
outer region are found. 

\begin{table}
\tbl{Structures of turbulent pipe flow as classified by wavenumber.}
{\begin{tabular}{lllll}
\toprule
\bf Structure & \bf Definition & \bf Energy & \bf Description &
\bf Figure\\
\colrule
Propagating Modes & (m,n,q), $m \not= 0$ & 80.58\% & Constant phase
speed &\\
\hspace{2em}(a) Wall &  $n > m$ & 35.22\% & Structures that turn
azimuthally & Fig. \ref{eigen151}-f\\
& & & more than they lift axially &\\ 
& & & found near wall. &\\
\hspace{2em}(b) Lift &  $ m \geq n , n > 1$ & 29.68\% &   Structures that lift
axially more & Fig. \ref{eigen321}-f\\
& & & than they turn azimuthally, extend from & \\
& & & near wall to outer region &\\
\hspace{2em}(c) Asymmetry & $n=1$ & 9.09 \% & Structures that have
non-zero radial and & Fig. \ref{eigen111}-f \\
& & &  azimuthal velocity, typically with coherent & \\
& & & structures found in the outer
region & \\
\hspace{2em}(d) Ring & $n=0$ & 6.60\% & Ring-like structures in the
outer region& Fig. \ref{eigen101}-f\\
& & & & \\
Non-propagating Modes & $(0,n,q)$ & 19.42\% & Non-constant phase speed &\\
\hspace{2em}(a) Roll mode & $(0,n,q)$ $n \not= 0$ &  18.34\% & Near wall streamwise
vortices & Fig. \ref{eigen061}-f \\
\hspace{2em}(b) Shear mode & $(0,0,q)$ & 1.08\% & Non-zero centerline
streamwise velocity & Fig. \ref{eigen001real}-d\\
& & & No coherent vorticity & \\

\botrule
\end{tabular}}
\label{structureTBL}
\end{table}
This eigenfunction expansion, using both their structure and their time-dependent coefficients, provides a framework for
further analysis and comparison and has led to understanding
the mechanism of drag reduction by
spanwise wall oscillation \cite{duggleby_drPipe} and the mechanism of
relaminarization \cite{duggleby_relam}.
\begin{figure}
\epsfxsize=3 in
\centerline{\epsfbox{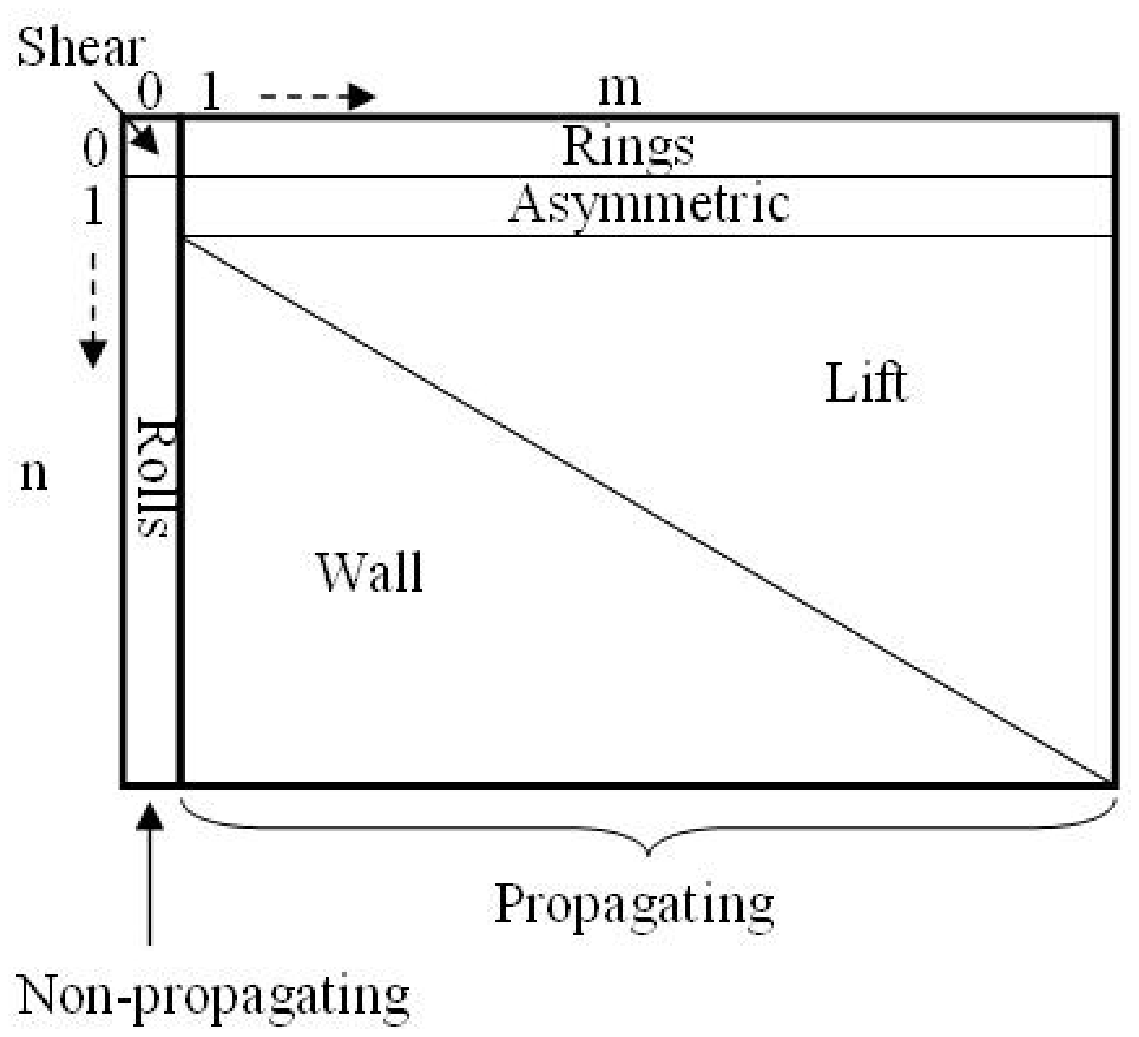}}
\caption{Subclass classification according to streamwise ($m$) and spanwise ($n$) wavenumbers.}
\label{subclassSpectra}
\end{figure}

In order to show the qualitative and quantitative classifications of
these eigenfunctions, the most energetic function of each class is presented in
Figures \ref{eigen151} through \ref{eigen001phase}.  For each class, subfigures (a) show iso-contours and slices of coherent
vorticity, and a slice displaying the Reynolds stress and coherent
vorticity together.   The coherent vorticity is defined as the
  rotating part of the fluid, defined as the
  imaginary component of the eigenvalues of the velocity strain rate
  tensor, following the work of Chong et al. \cite{chong}.  The next four subfigures for each class
are the plots of the eigenfunctions,
both the real (b) and imaginary component (d), and the time history
amplitude (c)  and phase (e).  To complete the class description, the
coherent vorticity of two more representative eigenfunctions are shown
in subfigures (f) and
(g).  For the shear modes (Figure \ref{eigen001real}), since they do not have coherent
vorticity, only the eigenfunction and time history are plotted.

The first subclass of interest is the wall eigenmodes.  The
coherent vorticity plots of three example wall modes are seen
in Figures \ref{eigen151}, \ref{eigen131}, and \ref{eigen241} for the
(1,5,1),(1,3,1), and (2,4,1) modes, respectively.  Each mode consists of a
travelling-wave coherent vortex near the wall, and all
eigenmodes with greater spanwise wavenumber than streamwise wavenumber
demonstrate this same structure.  A characteristic consistent of this
subclass is that the
Reynolds stress is generated near the wall.  The individual velocity
components of the (1,2,1) mode, the most energetic wall mode, is
shown in Figures \ref{eigen151real} and \ref{eigen151imag} for their
real and imaginary components, respectively, and in Figures
\ref{eigen151amp} and \ref{eigen151phase} for their amplitude squared and
phase time history.  The amplitude history shows the bursting nature
of these travelling-waves, and the near constant phase velocity
shows why these  modes are referred to as propagating or travelling-waves.  These wall functions constitute 35.22\% of the total
energy of the flow, and are the most energetic of the propagating modes.

The next mode of interest, still in the propagating mode class,
is the lift mode, found whenever the streamwise wavenumber is greater
than or equal to the spanwise wavenumber and the spanwise wavenumber is
greater than one.  Three examples of this subclass
are shown in Figures \ref{eigen221}, \ref{eigen321}, and \ref{eigen331}
for modes (2,2,1), (3,2,1), and (3,3,1), respectively.  In a lift
mode, the coherent vorticity and Reynolds stress extends from the wall
to near the centreline, and results in a lifting motion.  The real and imaginary velocity components are shown in Figures
\ref{eigen221real} and \ref{eigen221imag}, respectively, and the
amplitude and phase time history in Figures \ref{eigen221amp} and
\ref{eigen221phase}, respectively.  Again, the constant phase speed
classifies these structures as propagating modes, and the amplitude
bursts are also characteristic of a turbulent flow field.  The lift
structures constitute 29.68\% of the total energy of the flow.

The next  modes, the asymmetric modes, are responsible
for breaking the symmetry of the flow about the axis of the pipe.  These
 modes are found for a spanwise wavenumber of one, for any
streamwise wavenumber, and consist of a coherent vortex just outside
the log layer.  They are asymmetric because of their nonzero radial and azimuthal velocities at the origin, which is physical
only for azimuthal wavenumber $n=1$, since a positive radial velocity at
the origin, under a  rotation of $\pi$ around the
axis, results in a negative radial velocity.  The same is true for the
azimuthal velocity at the origin.  Three examples of
these modes are shown in Figures \ref{eigen111}, \ref{eigen211}
and \ref{eigen311} for modes (1,1,1), (2,1,1), and (3,1,1),
respectively, again having the Reynolds stress between the coherent
vorticity.  These modes are also propagating and turbulent, as seen
in the (1,1,1) mode's amplitude (Figure \ref{eigen111amp}) and phase time history
(Figure \ref{eigen111phase}), with its real and imaginary component
found in Figure \ref{eigen111real} and \ref{eigen111imag}.   This
 mode  constitutes
9.09\% of the total energy of the flow.

The last of the propagating modes is the ring mode, named for the ring
of coherent vorticity that is found for all modes with zero azimuthal
wavenumber.  Examples of this structure are shown in Figures
\ref{eigen101}, \ref{eigen102}, and \ref{eigen201} for modes (1,0,1),
(1,0,2), and (2,0,1), respectively.  Since the ring modes have no
azimuthal dependance, only a radial cross-section of coherent
vorticity is shown with Reynolds stress superimposed.  The real and imaginary velocities are shown in
Figures \ref{eigen101real} and \ref{eigen101imag}, and the amplitude
and phase in \ref{eigen101amp} and \ref{eigen101phase}. This mode
constitutes 6.60\% of the total energy of the flow.

The last two modes are the non propagating modes.  The first and more
energetic of the two is the roll mode, found for any mode with zero
streamwise wavenumber.  This consists of rolls of coherent vorticity,
as seen in the three example modes (0,6,1), (0,5,1), and (0,3,1) in
Figures \ref{eigen061}, \ref{eigen051}, and \ref{eigen031}, respectively.  The
Reynolds stress is strong between the coherent vortices but is an
order of magnitude less than those of the wall or lift structures. The
velocity components of the (0,6,1) mode are shown in Figure
\ref{eigen061real} and \ref{eigen061imag}.  Of note, these structures do not have a constant
phase velocity, seen in Figure \ref{eigen061phase}, and the decay rate
of the energy is slower than that of the propagating modes, seen if Figure
\ref{eigen061amp}.  This subclass constitutes 18.34\% of the total
energy of the flow.  

The other non propagating mode is the shear mode, found for zero
azimuthal and streamwise wavenumber, therefore corresponding to the
fluctuation of the mean flow rate.  Since these structures have no coherent vorticity nor imaginary components, only the real velocity
components (Figures \ref{eigen001real} and \ref{eigen002real}) and
their amplitude and phase time history (Figures \ref{eigen001amp} and
\ref{eigen001phase}) are plotted.  Like the roll modes, the shear
modes do not have a constant phase velocity, and since they do not
have an imaginary component, the phase oscillates between zero and
$\pi$.  The shear modes constitute 1.08\% of the total energy of the
flow.

\renewcommand{\thefigure}{\arabic{figure}\alph{subfigure}}

\begin{figure}[p]
\centering

\setcounter{subfigure}{1}
\includegraphics[width=5.5 in]{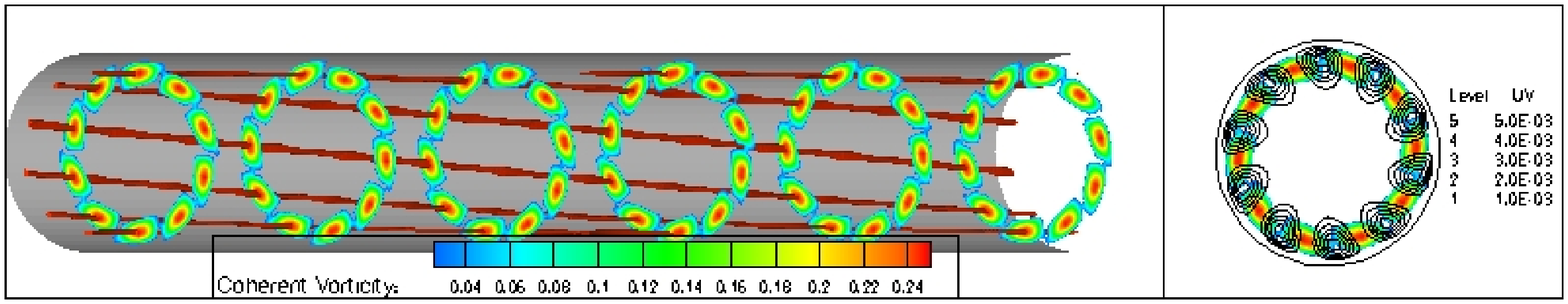}
\caption{Most energetic propagating wall mode (1,5,1). Coherent
  vorticity (left) and a cross-section of coherent vorticity with
  Reynolds stress superimposed (right).  }
\label{eigen151}

\begin{tabular}{cc}
\begin{minipage}{2.5 in}
\centering
\addtocounter{figure}{-1}
\addtocounter{subfigure}{1}
\includegraphics[width=2.5 in]{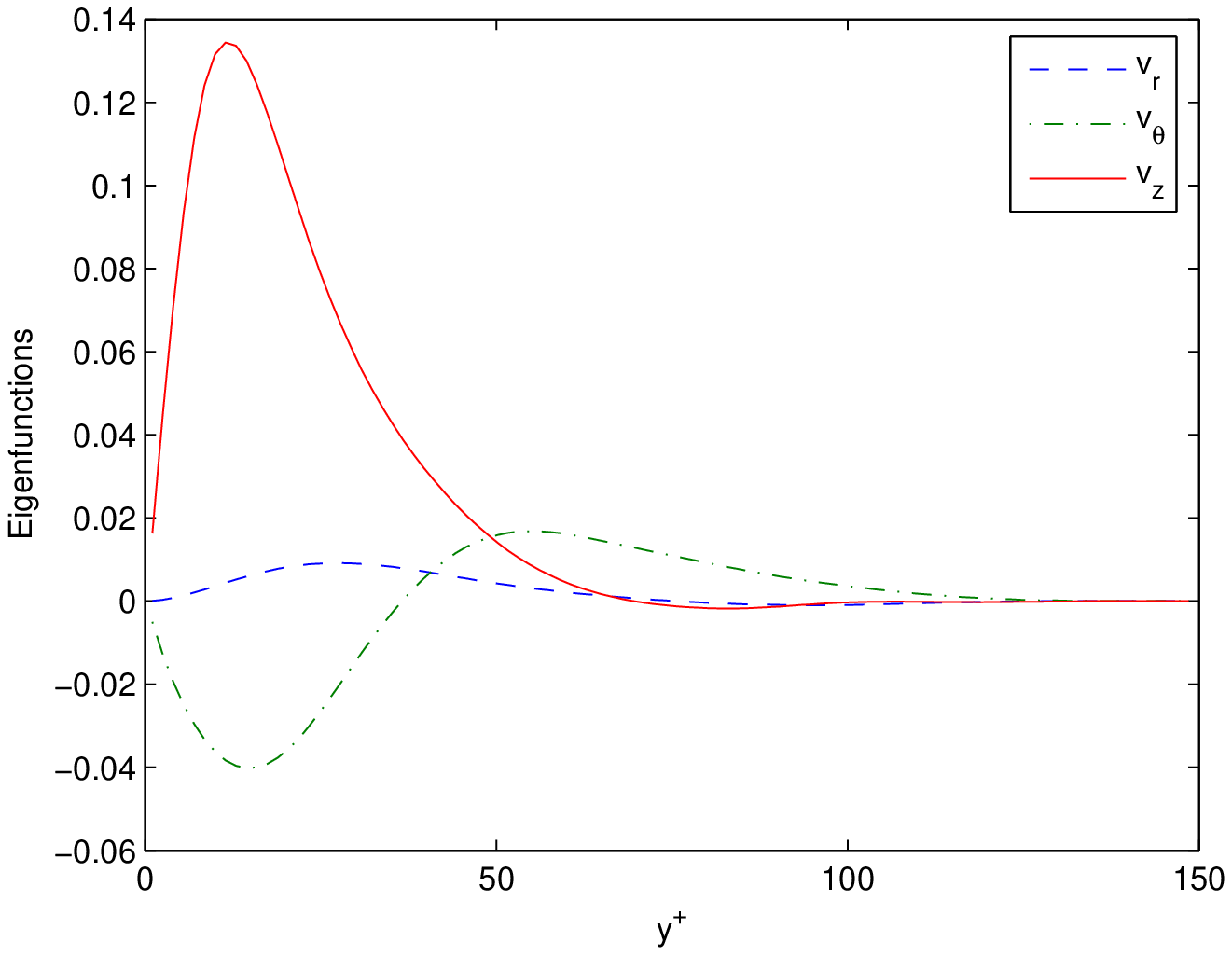}
\caption{Real component of (1,5,1).}
\label{eigen151real}
\end{minipage}
& 
\begin{minipage}{2.5 in}
\centering
\addtocounter{figure}{-1}
\addtocounter{subfigure}{1}
 \resizebox{2.5 in}{!}{\includegraphics{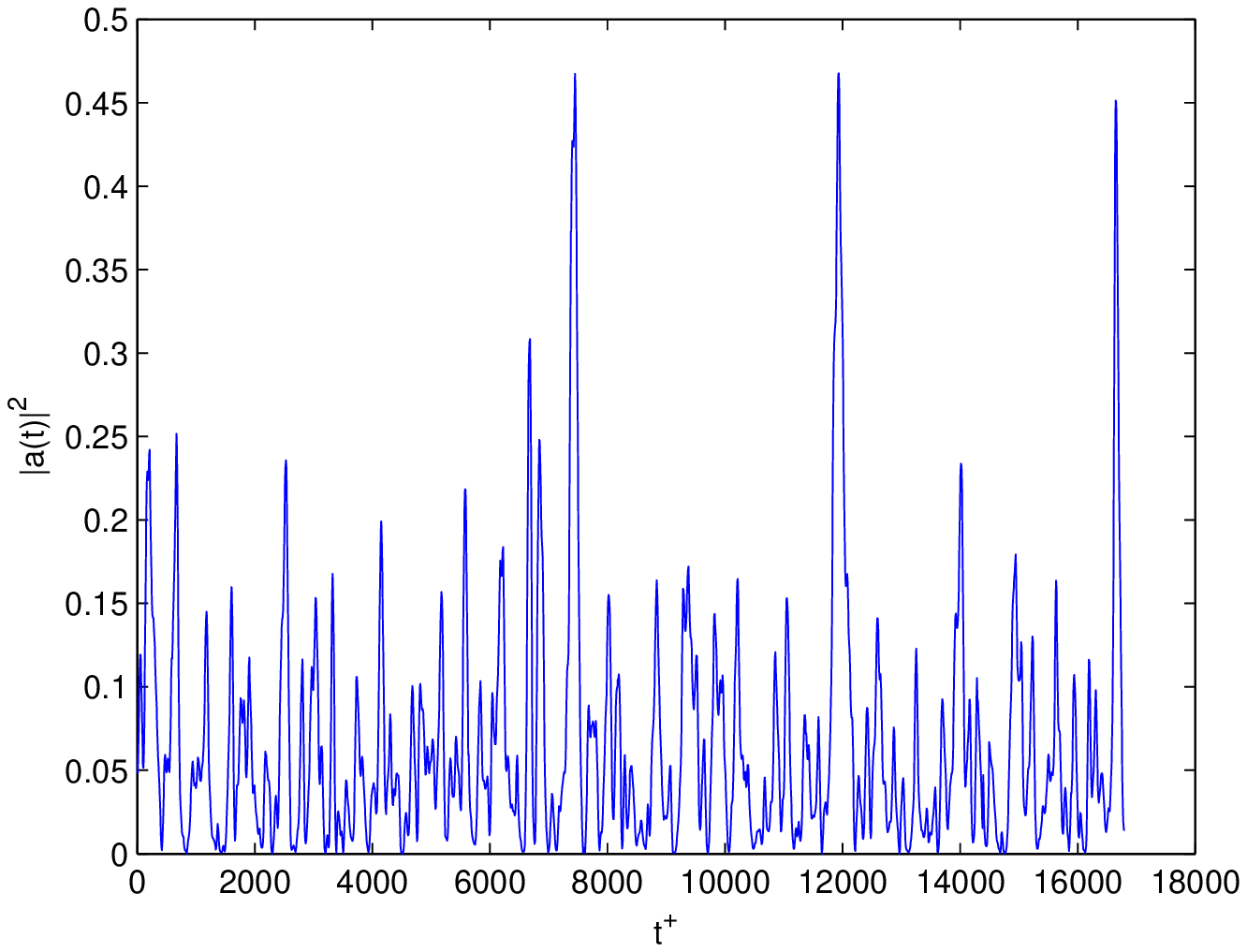}}
\caption{Time history amplitude of (1,5,1).}
\label{eigen151amp}
\end{minipage} \\
\begin{minipage}{2.5in}
\centering
\addtocounter{figure}{-1}
\addtocounter{subfigure}{1}
\includegraphics[width=2.5 in]{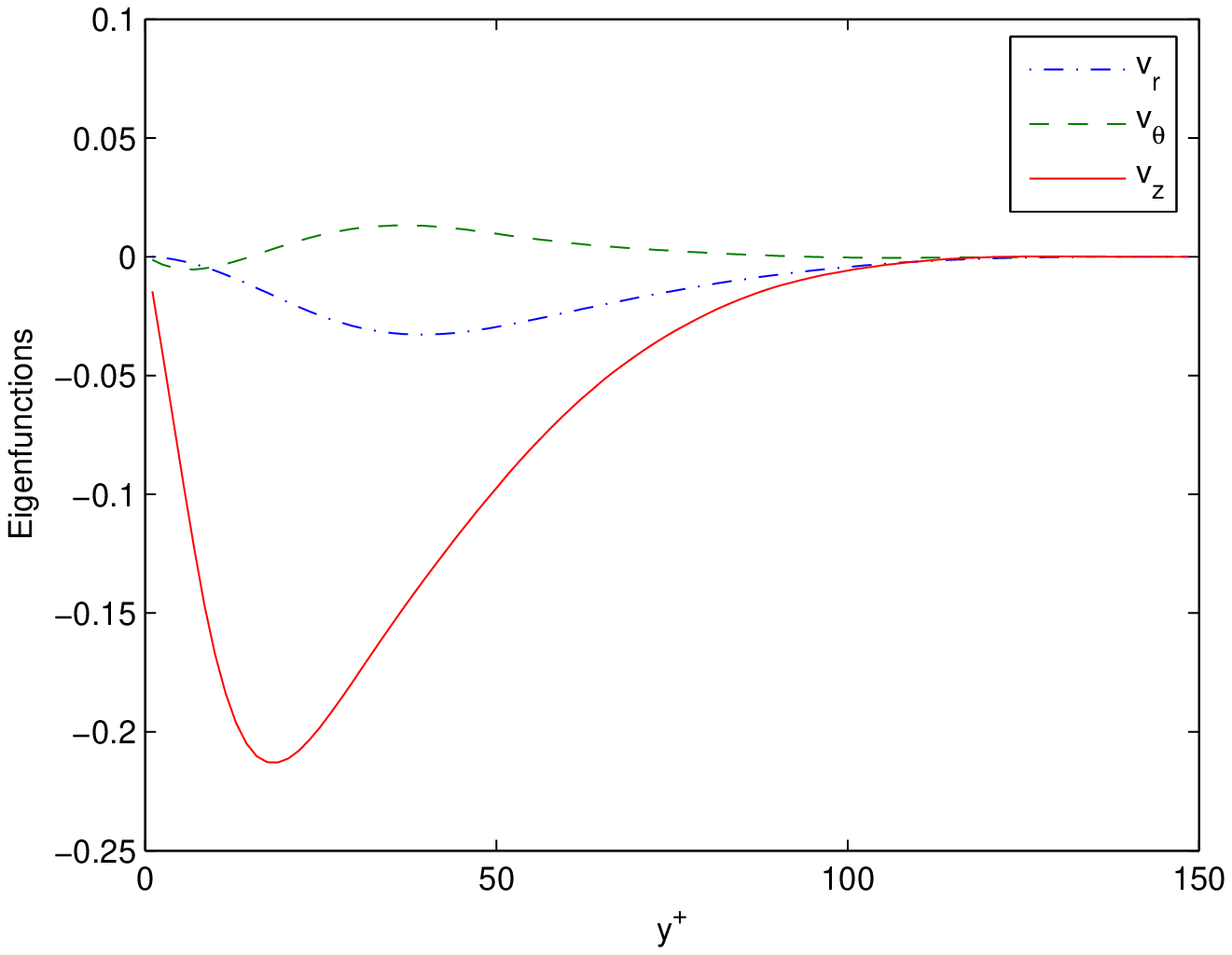}
\caption{Imaginary component of (1,5,1).}
\label{eigen151imag}
\end{minipage}
& 
\begin{minipage}{2.5 in}
\centering
\addtocounter{figure}{-1}
\addtocounter{subfigure}{1}
 \resizebox{2.5 in}{!}{\includegraphics{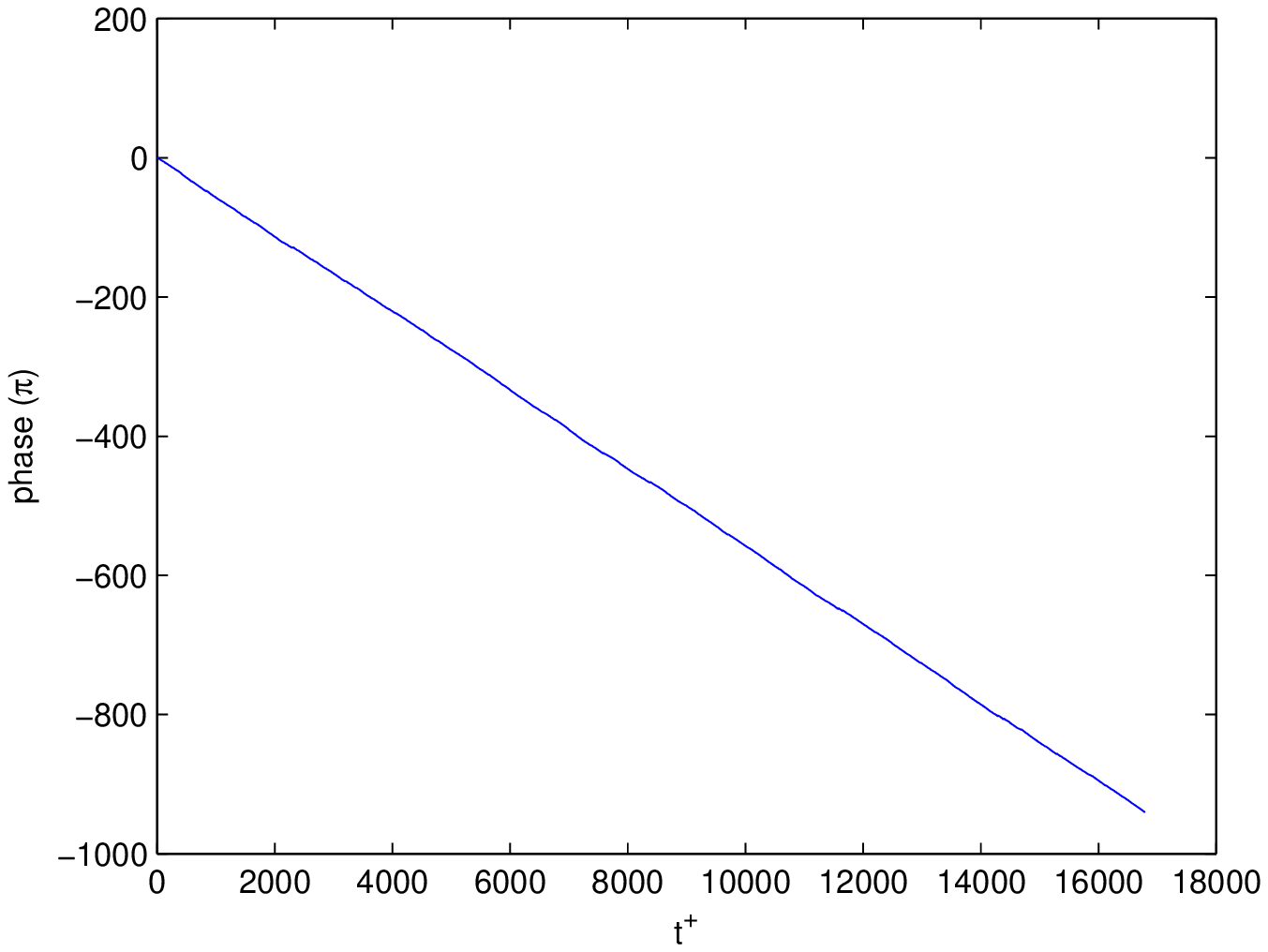}}
\label{eigen151phase}
\caption{Time history phase of (1,5,1).}
\end{minipage} \\
\end{tabular}

\addtocounter{figure}{-1}
\addtocounter{subfigure}{1}
\epsfxsize=5.5 in
\centerline{\epsfbox{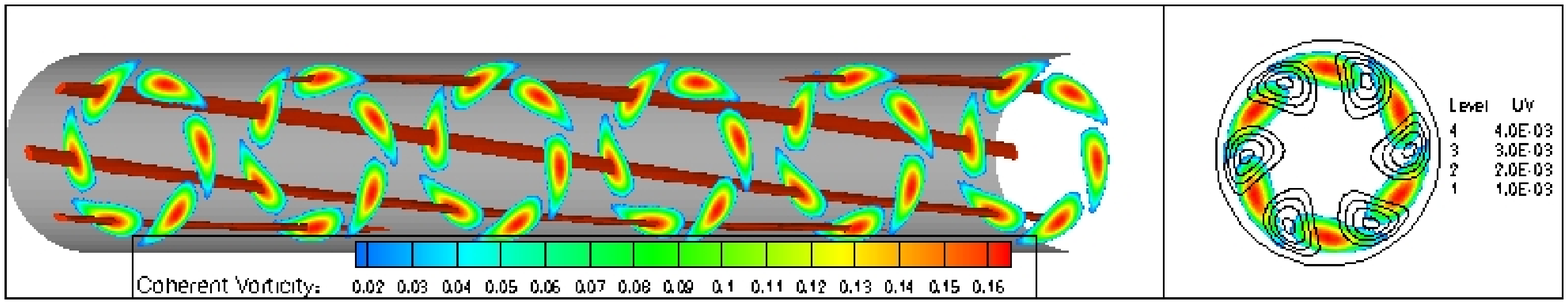}}
\caption{Propagating wall mode (1,3,1). }
\label{eigen131}

\addtocounter{figure}{-1}
\addtocounter{subfigure}{1}
\epsfxsize=5.5 in
\centerline{\epsfbox{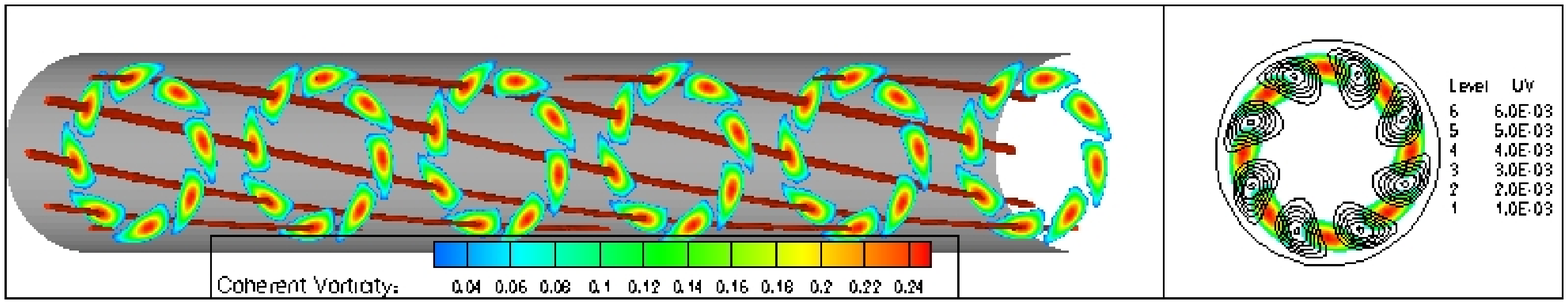}}
\caption{Propagating wall mode (2,4,1). }
\label{eigen241}
\end{figure}


\setcounter{subfigure}{1}
\begin{figure}[p]
\epsfxsize=5.5 in
\centerline{\epsfbox{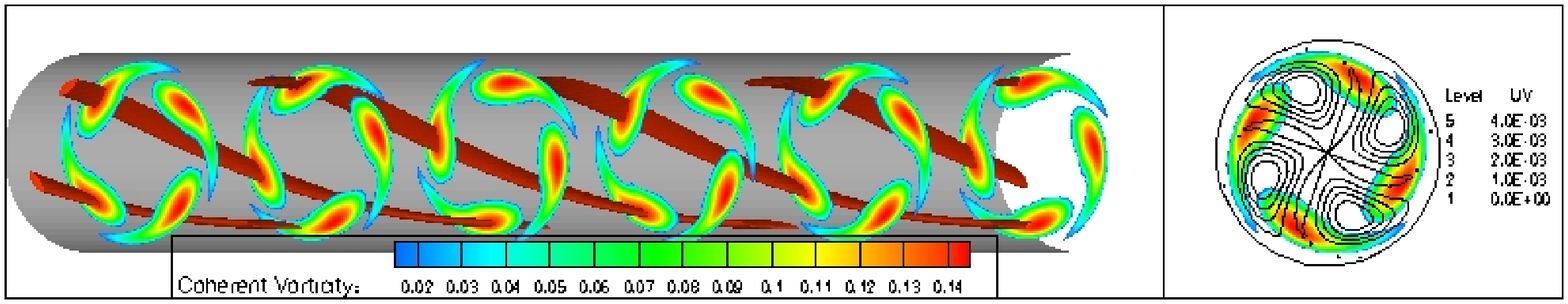}}
\caption{Most energetic propagating lift mode (2,2,1). Coherent
  vorticity (left) and a cross-section of coherent vorticity with
  Reynolds stress superimposed (right).  }
\label{eigen221}

\centering
\begin{tabular}{cc}
\begin{minipage}{2.5 in}
\centering
\addtocounter{figure}{-1}
\addtocounter{subfigure}{1}
\includegraphics[width=2.5 in]{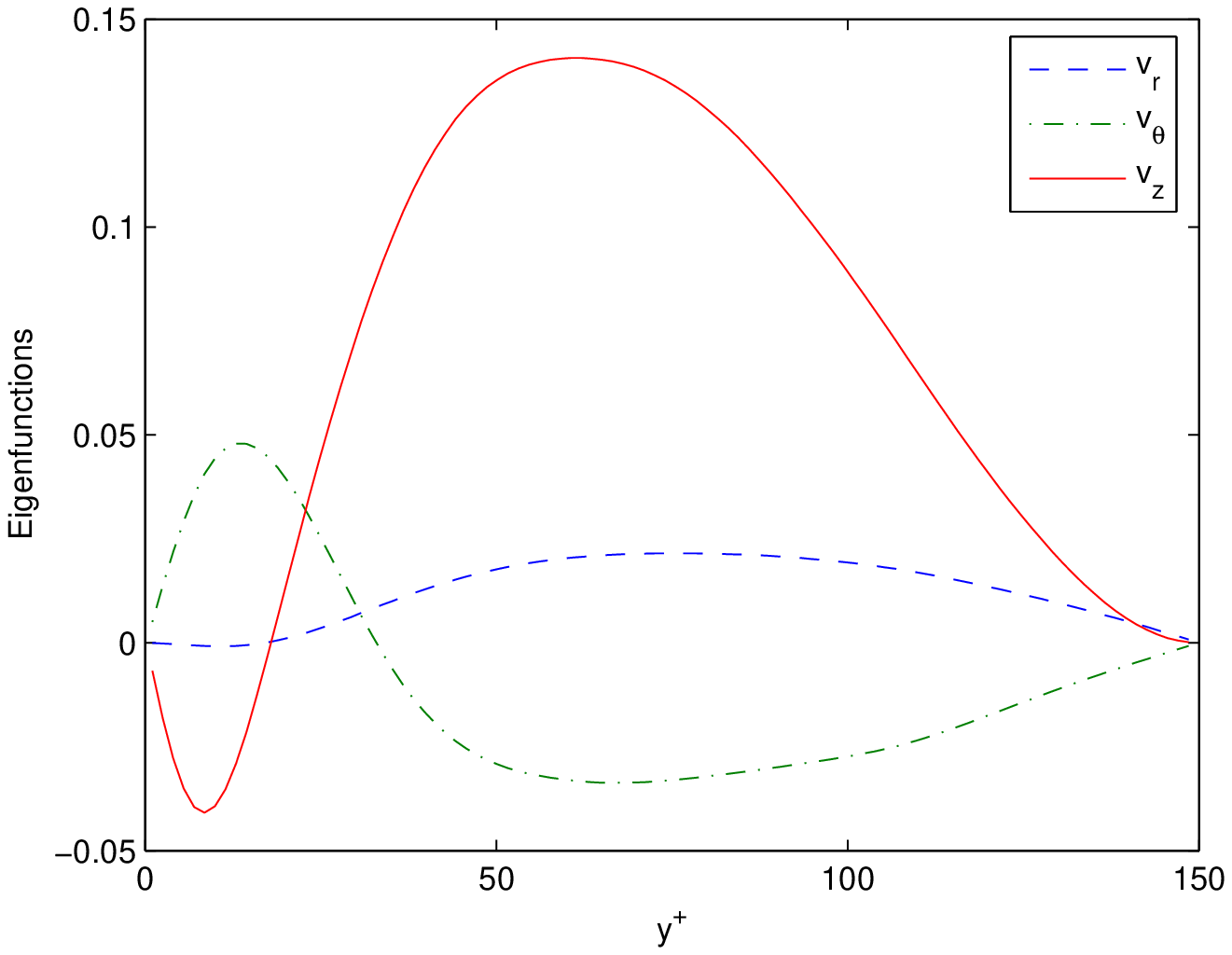}
\caption{Real component of (2,2,1)}
\label{eigen221real}
\end{minipage}
& 
\begin{minipage}{2.5 in}
\centering
\addtocounter{figure}{-1}
\addtocounter{subfigure}{1}
 \resizebox{2.5 in}{!}{\includegraphics{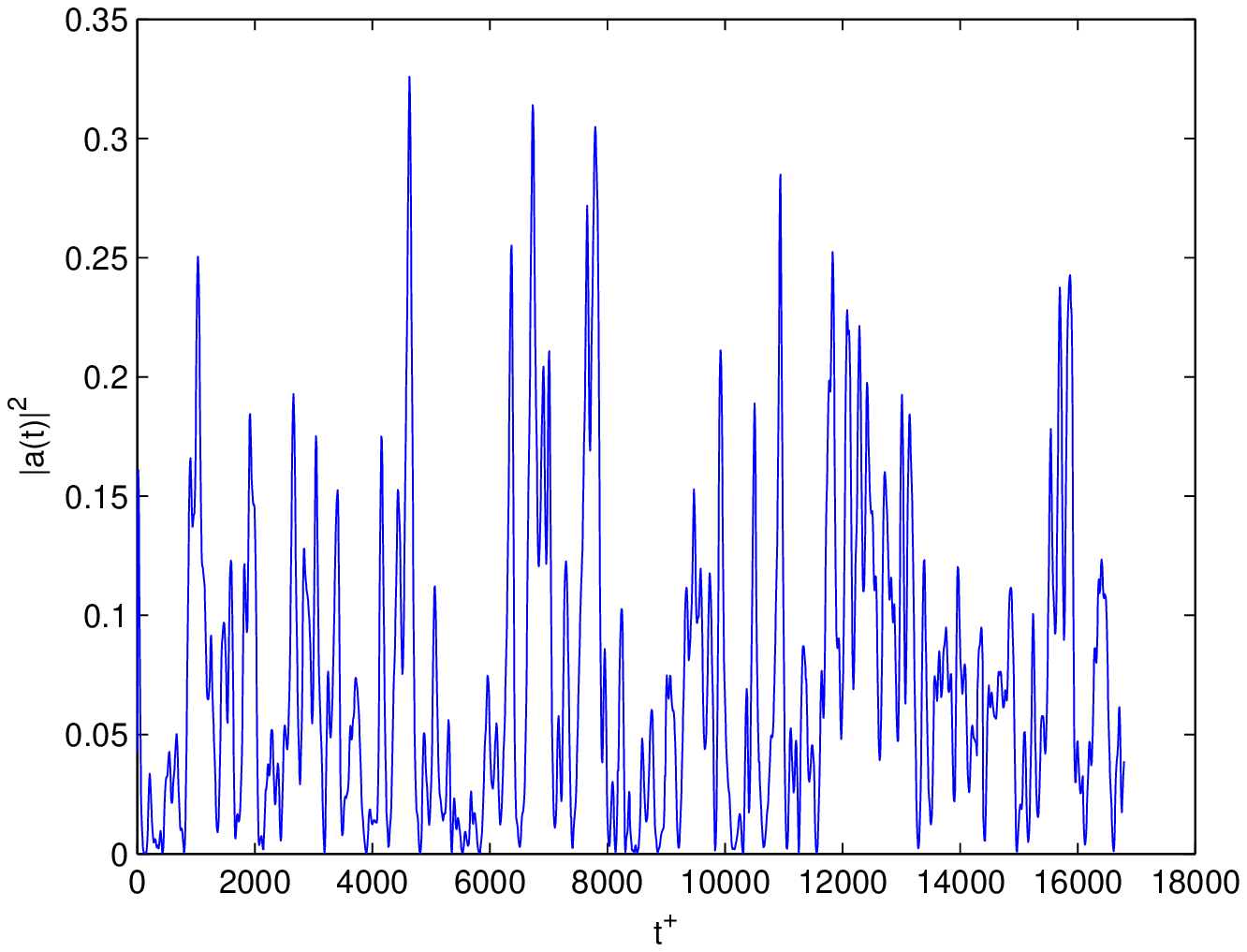}}
\caption{Time history amplitude of (2,2,1)}
\label{eigen221amp}
\end{minipage} \\
\begin{minipage}{2.5 in}
\centering
\addtocounter{figure}{-1}
\addtocounter{subfigure}{1}
\includegraphics[width=2.5 in]{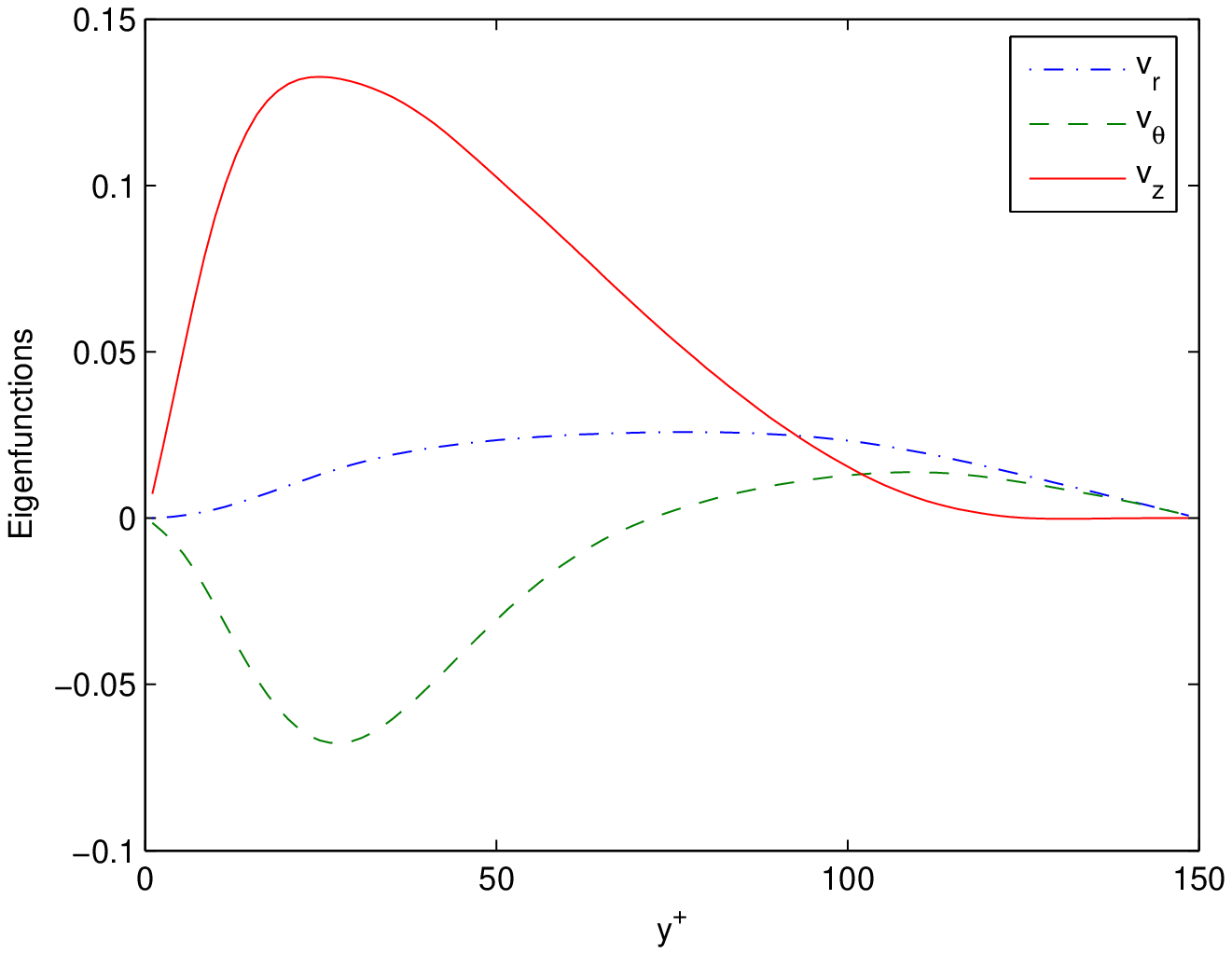}
\caption{Imaginary component of (2,2,1)}
\label{eigen221imag}
\end{minipage}
& 
\begin{minipage}{2.5 in}
\centering
\addtocounter{figure}{-1}
\addtocounter{subfigure}{1}
 \resizebox{2.5 in}{!}{\includegraphics{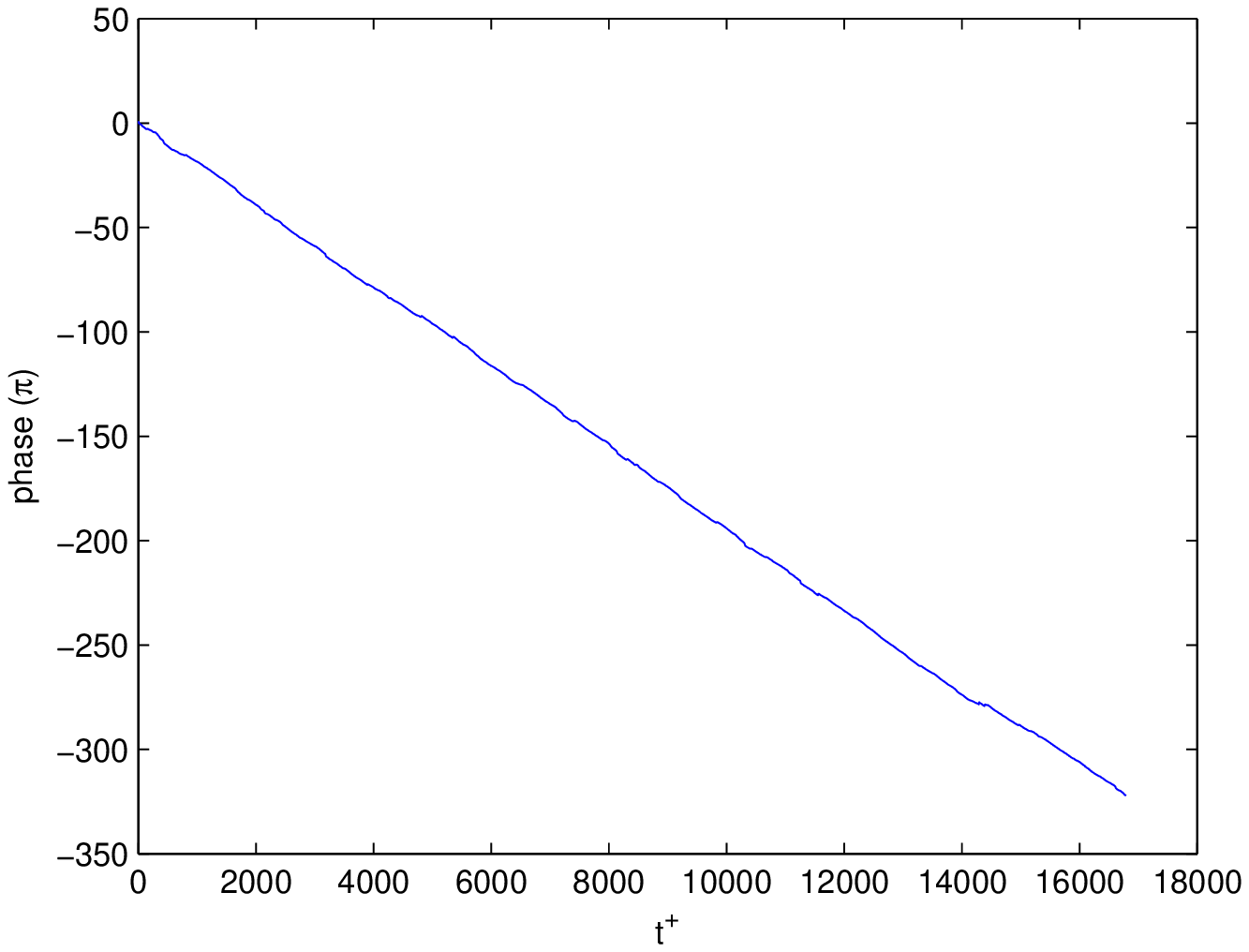}}
\caption{Time history phase of (2,2,1)}
\label{eigen221phase}
\end{minipage} \\
\end{tabular}
\addtocounter{figure}{-1}
\addtocounter{subfigure}{1}
\epsfxsize=5.5 in
\centerline{\epsfbox{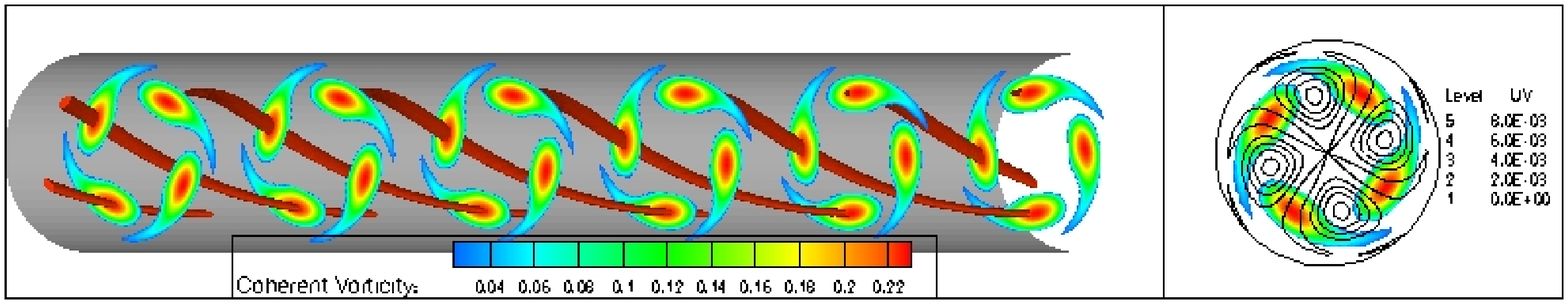}}
\caption{Propagating lift mode (3,2,1).}
\label{eigen321}
\addtocounter{figure}{-1}
\addtocounter{subfigure}{1}
\epsfxsize=5.5 in
\centerline{\epsfbox{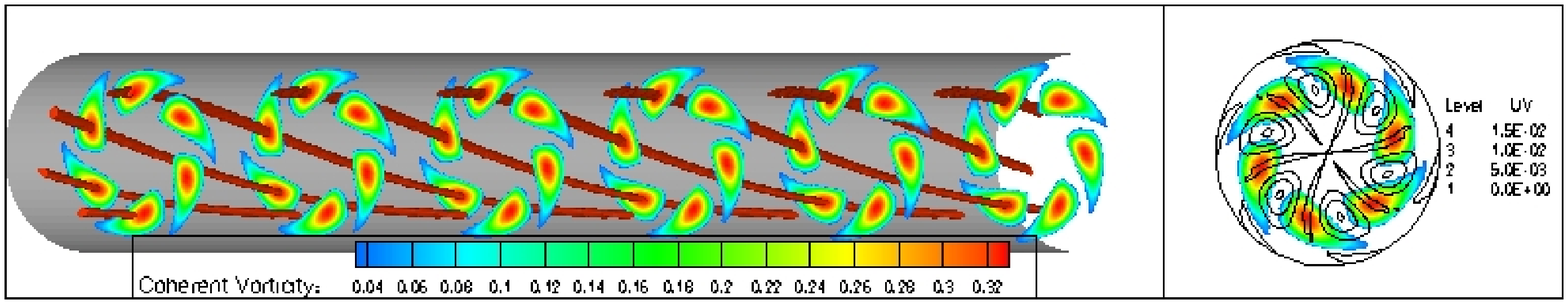}}
\caption{Propagating lift mode (3,3,1).}
\label{eigen331}
\end{figure}


\setcounter{subfigure}{1}
\begin{figure}[p]
\epsfxsize=5.5 in
\centerline{\epsfbox{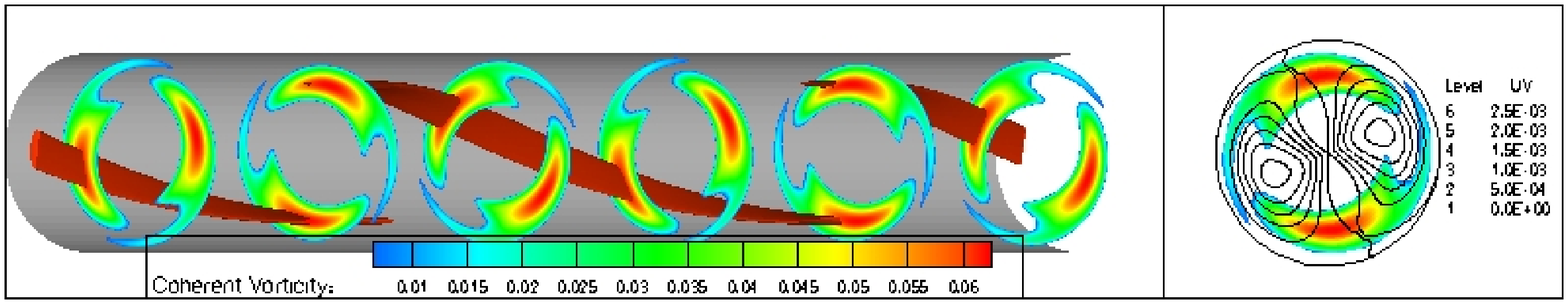}}
\caption{Most energetic propagating asymmetric mode (1,1,1). Coherent
  vorticity (left) and a cross-section of coherent vorticity with
  Reynolds stress superimposed (right).  }
\label{eigen111}

\centering
\begin{tabular}{cc}
\begin{minipage}{2.5 in}
\centering
\addtocounter{figure}{-1}
\addtocounter{subfigure}{1}
\includegraphics[width=2.5 in]{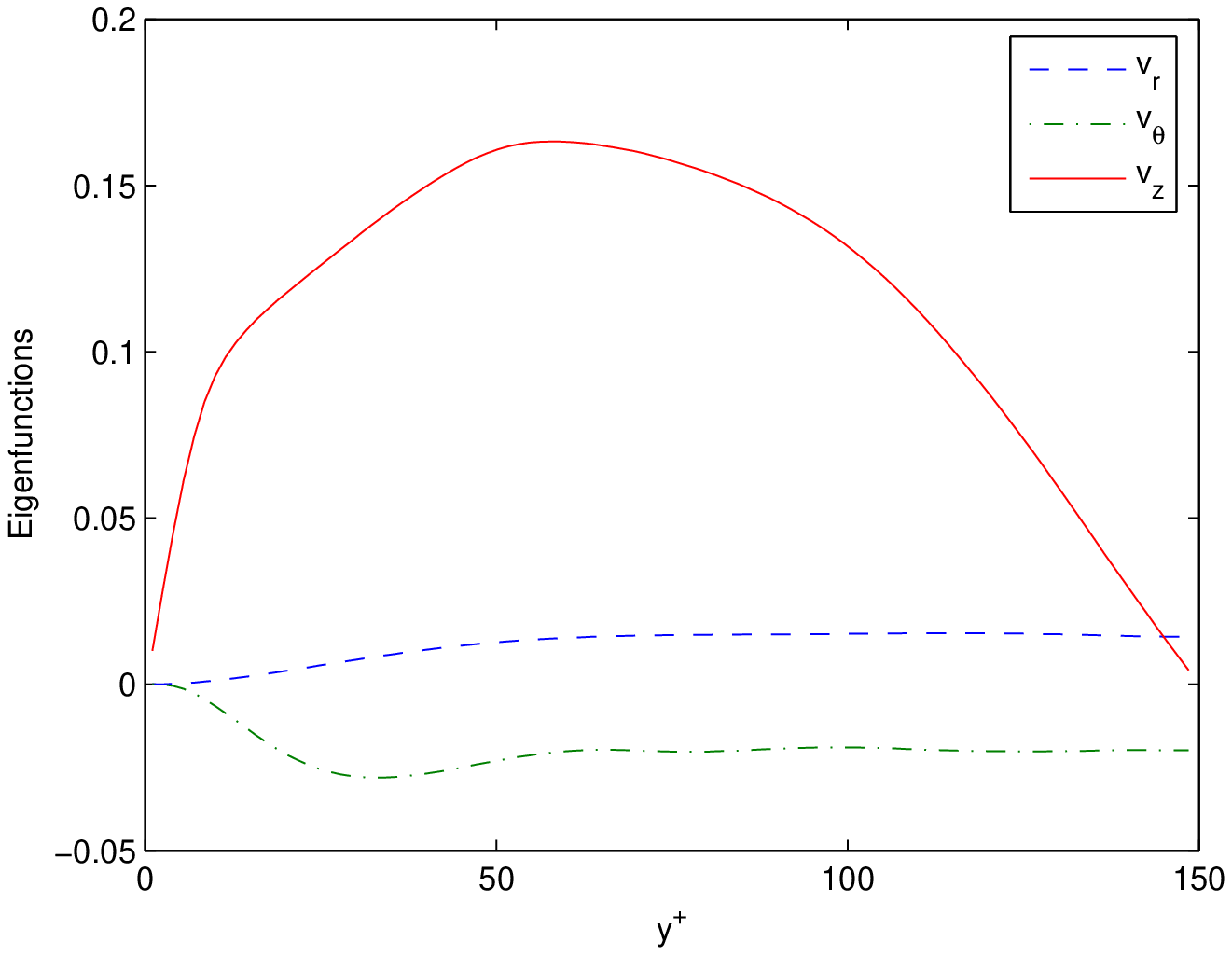}
\caption{Real component of (1,1,1).}
\end{minipage}
\label{eigen111real}
& 
\begin{minipage}{2.5 in}
\centering
\addtocounter{figure}{-1}
\addtocounter{subfigure}{1}
 \resizebox{2.5 in}{!}{\includegraphics{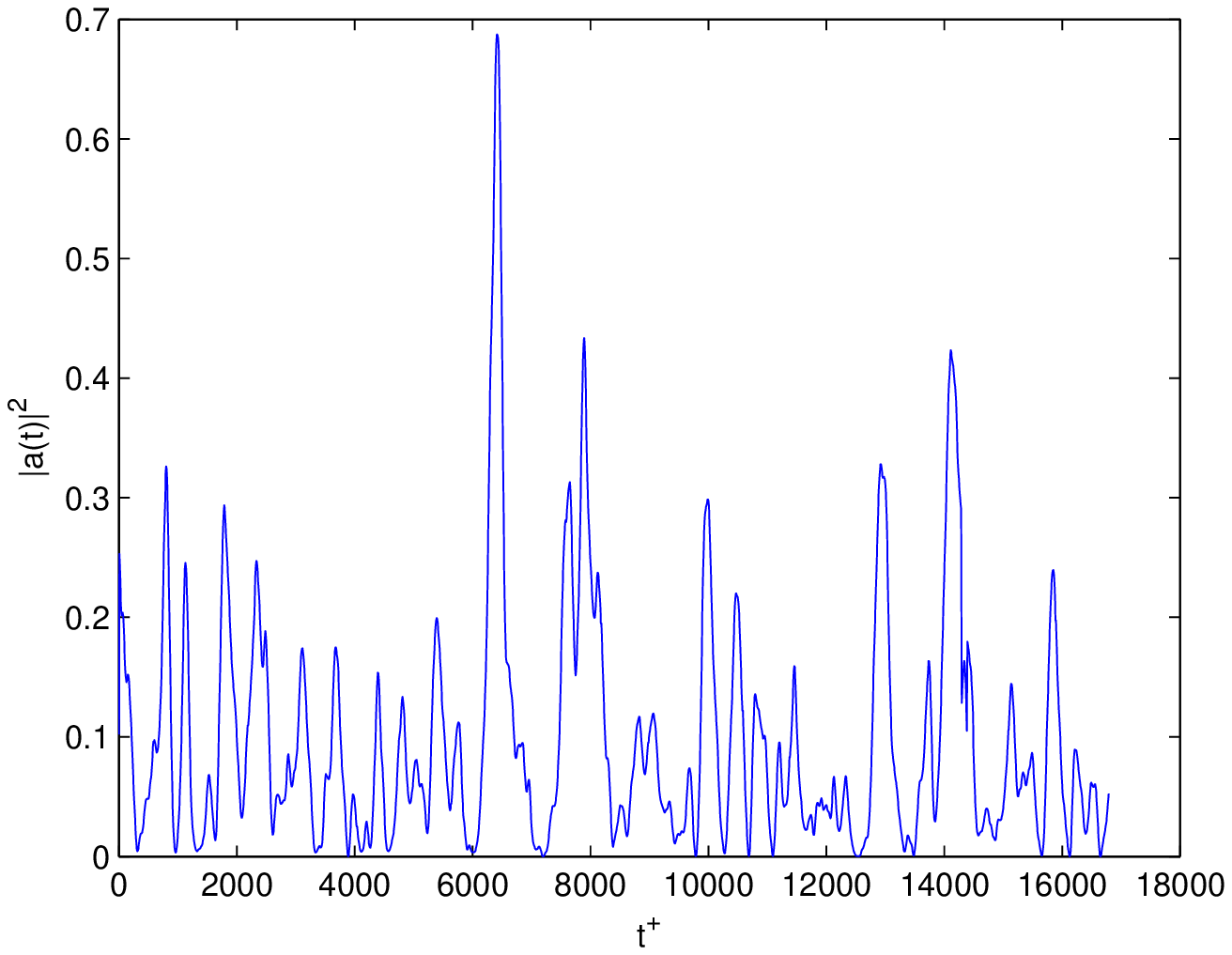}}
\caption{Time history amplitude of (1,1,1).}
\label{eigen111amp}
\end{minipage} \\
\begin{minipage}{2.5in}
\centering
\addtocounter{figure}{-1}
\addtocounter{subfigure}{1}
\includegraphics[width=2.5 in]{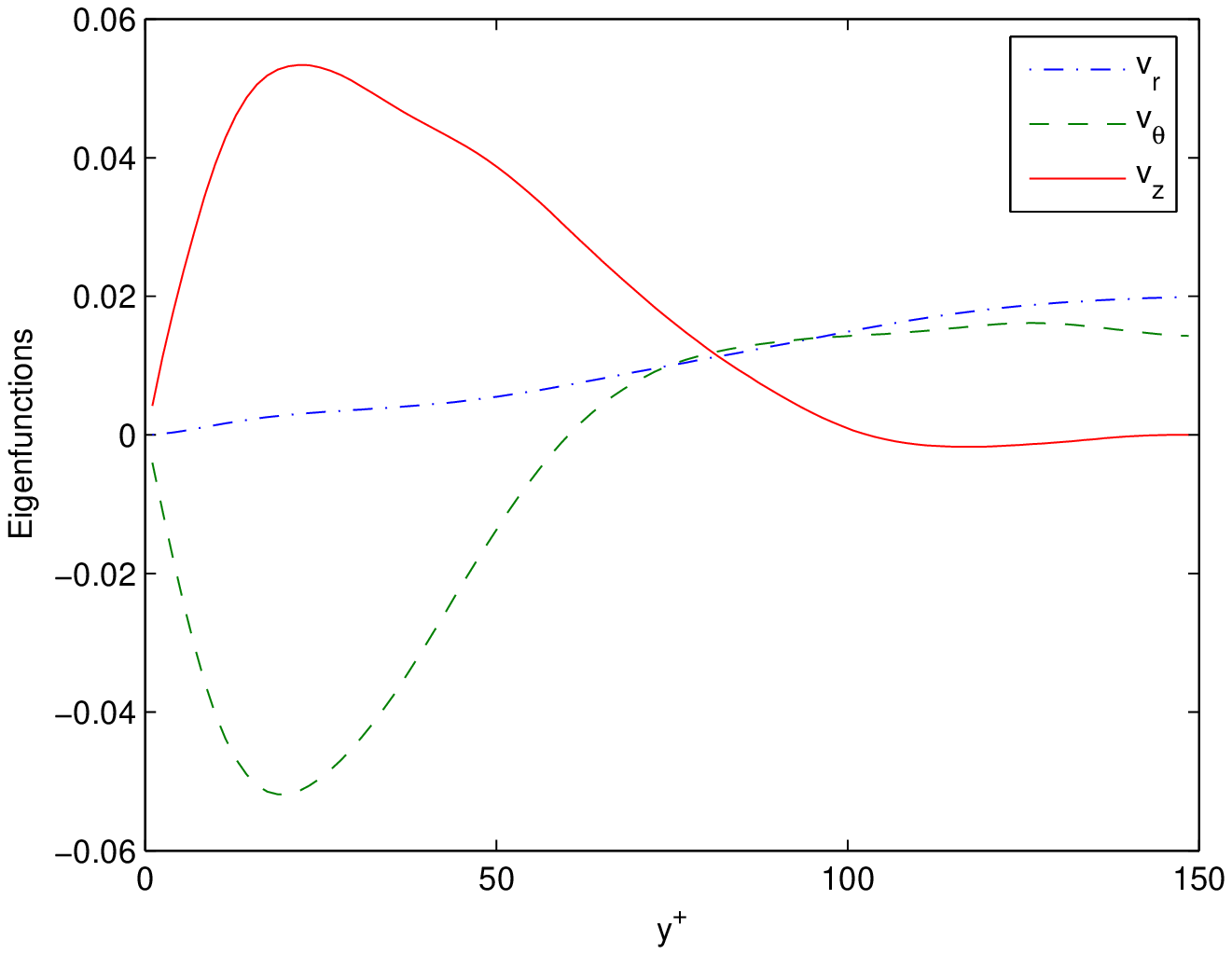}
\label{eigen111imag}
\caption{Imaginary component of (1,1,1).}
\end{minipage}
& 
\begin{minipage}{2.5 in}
\centering
\addtocounter{figure}{-1}
\addtocounter{subfigure}{1}
 \resizebox{2.5 in}{!}{\includegraphics{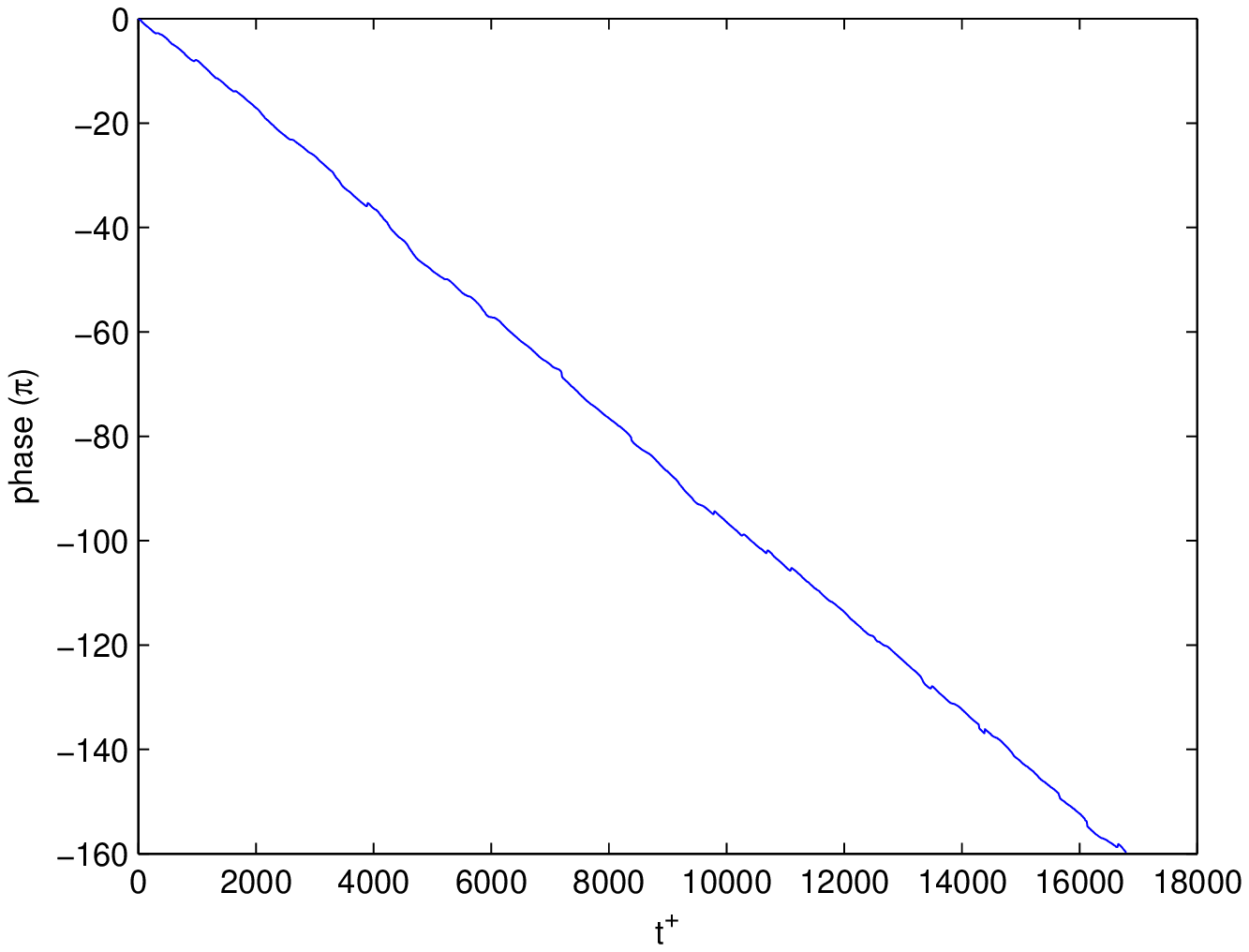}}
\caption{Time history phase of (1,1,1).}
\label{eigen111phase}
\end{minipage} \\
\end{tabular}

\addtocounter{figure}{-1}
\addtocounter{subfigure}{1}
\epsfxsize=5.5 in
\centerline{\epsfbox{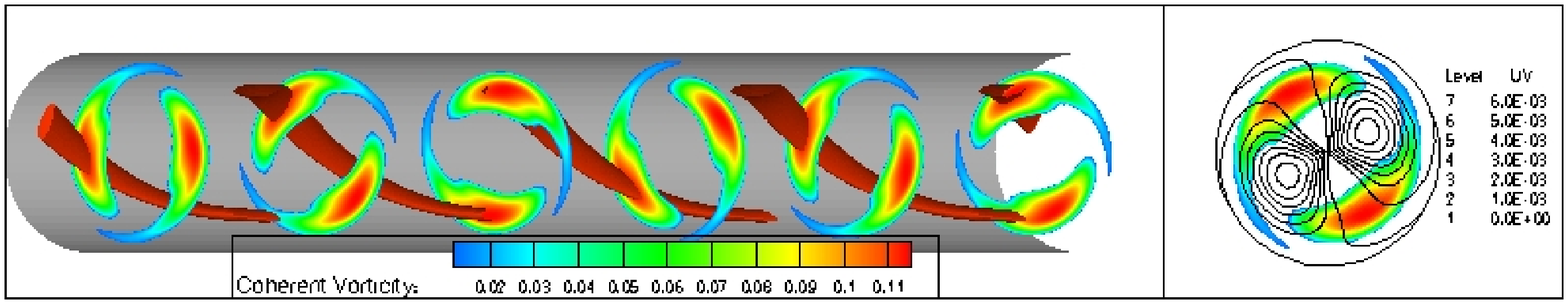}}
\caption{Propagating asymmetric mode (2,1,1). }
\label{eigen211}

\addtocounter{figure}{-1}
\addtocounter{subfigure}{1}
\epsfxsize=5.5 in
\centerline{\epsfbox{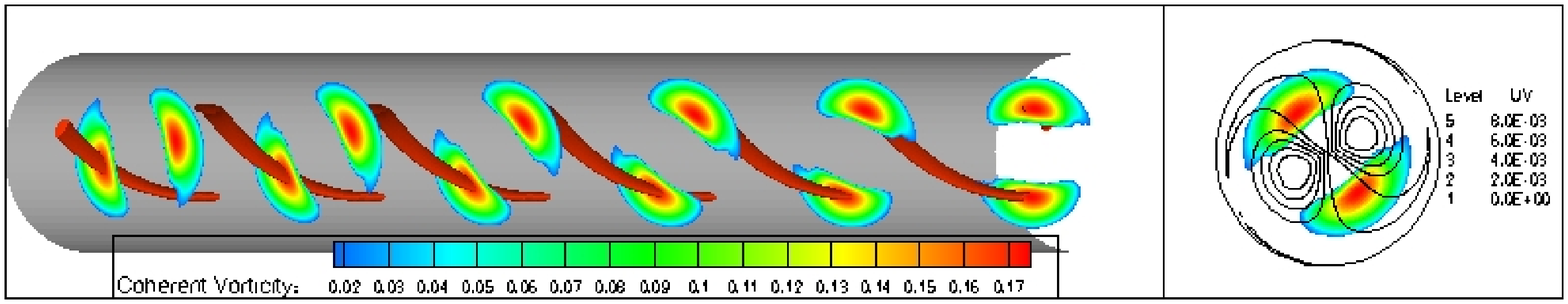}}
\caption{Propagating asymmetric mode (3,1,1). }
\label{eigen311}
\end{figure}


\setcounter{subfigure}{1}
\begin{figure}[p]
\epsfxsize=5.5 in
\centerline{\epsfbox{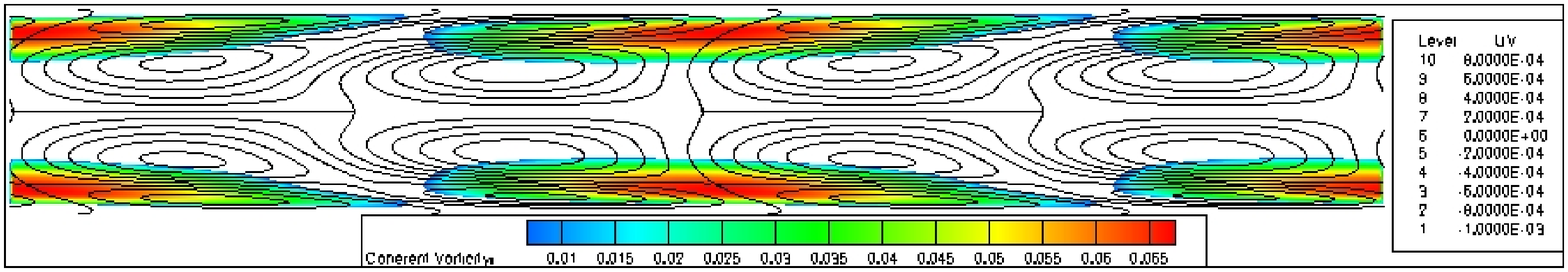}}
\caption{Most energetic propagating ring mode (1,0,1) cross-section. Coherent
  vorticity with
  Reynolds stress superimposed.  }
\label{eigen101}

\centering
\begin{tabular}{cc}
\begin{minipage}{2.5in}
\centering
\addtocounter{figure}{-1}
\addtocounter{subfigure}{1}
\includegraphics[width=2.5 in]{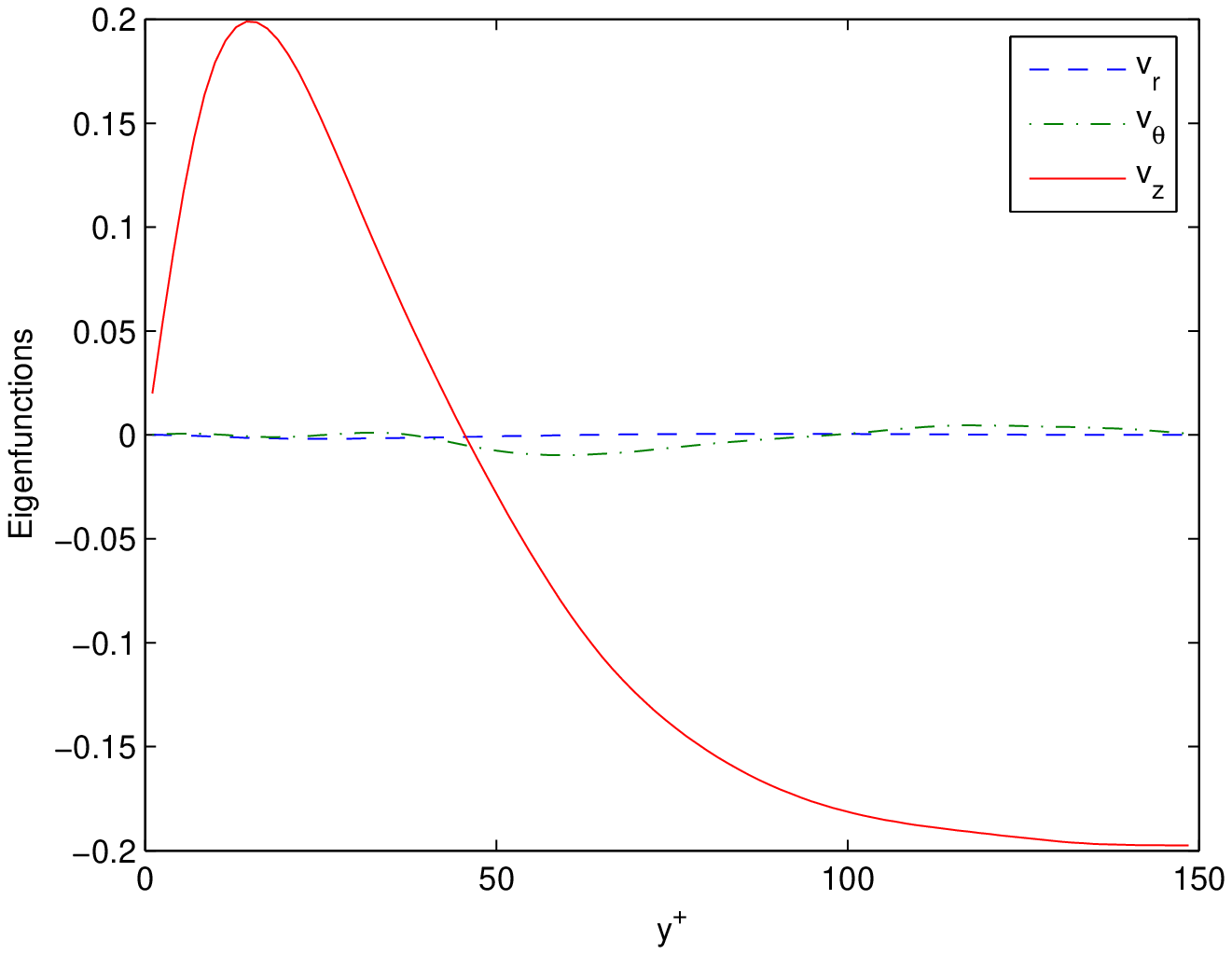}
\caption{Real component of (1,0,1).}
\label{eigen101real}
\end{minipage}
& 
\begin{minipage}{2.5 in}
\centering
\addtocounter{figure}{-1}
\addtocounter{subfigure}{1}
 \resizebox{2.5 in}{!}{\includegraphics{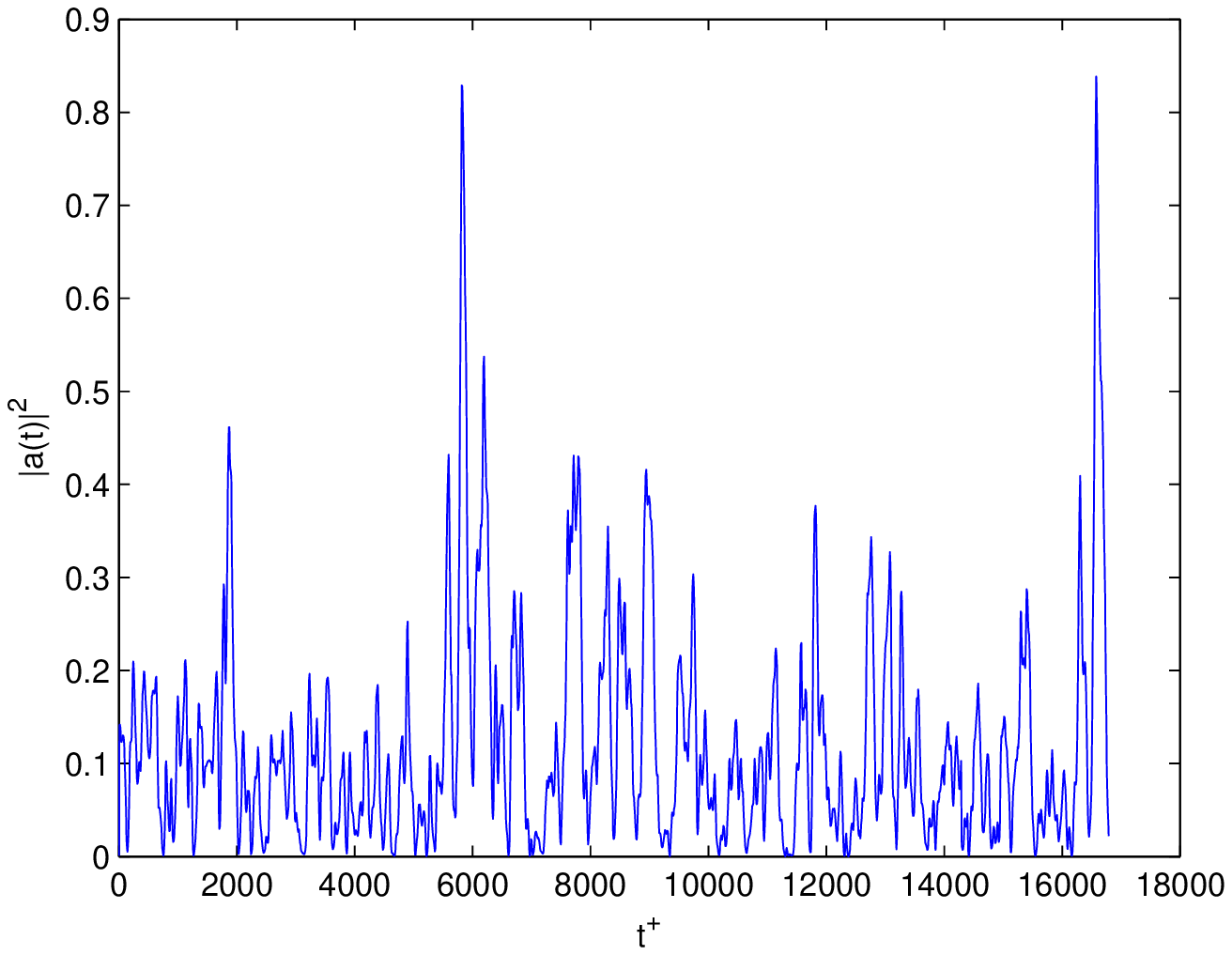}}
\label{eigen101amp}
\caption{Time history amplitude of (1,0,1).}
\end{minipage} \\
\begin{minipage}{2.5in}
\centering
\addtocounter{figure}{-1}
\addtocounter{subfigure}{1}
\includegraphics[width=2.5 in]{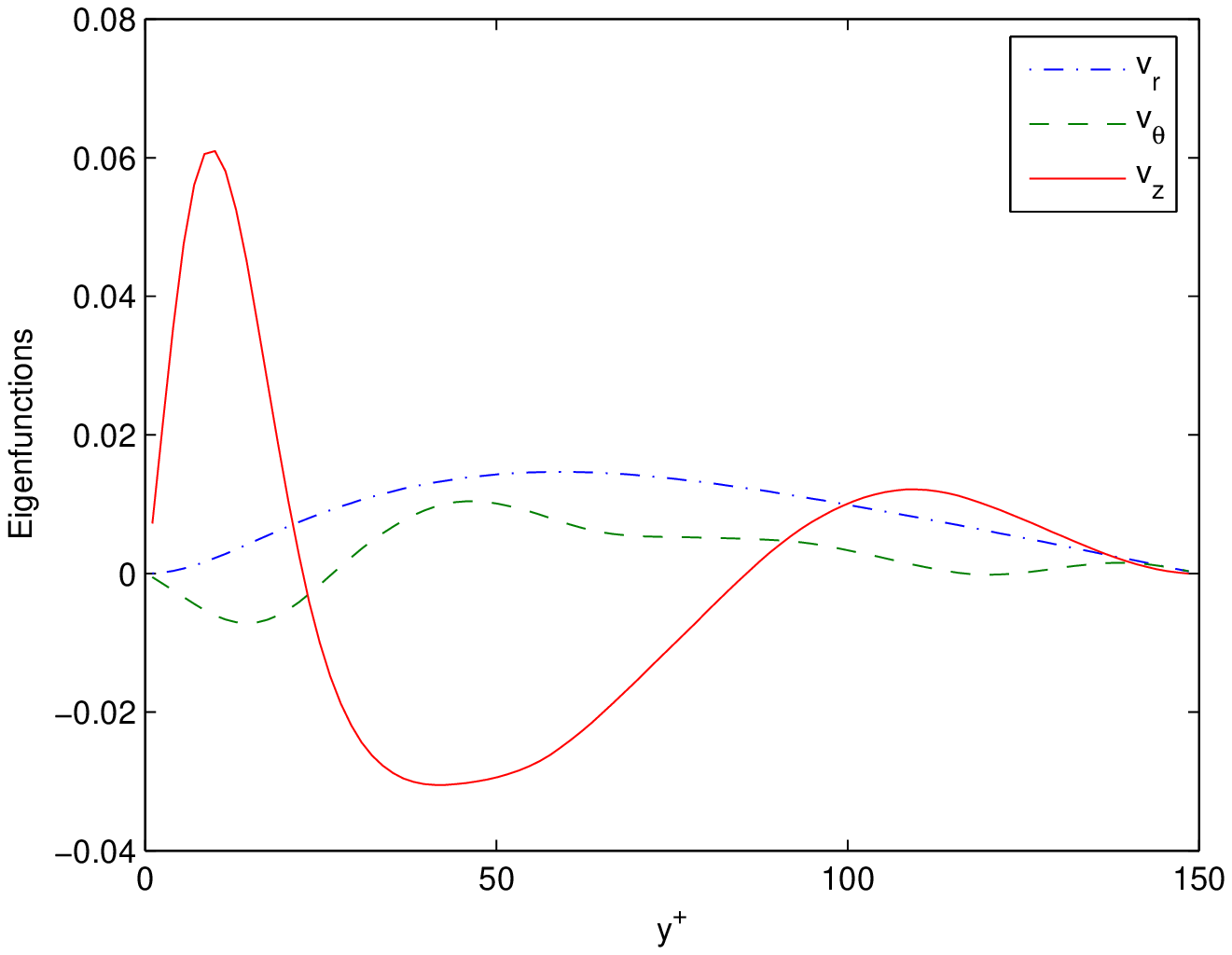}
\caption{Imaginary component of (1,0,1).}
\label{eigen101imag}
\end{minipage}
& 
\begin{minipage}{2.5 in}
\centering
\addtocounter{figure}{-1}
\addtocounter{subfigure}{1}
 \resizebox{2.5 in}{!}{\includegraphics{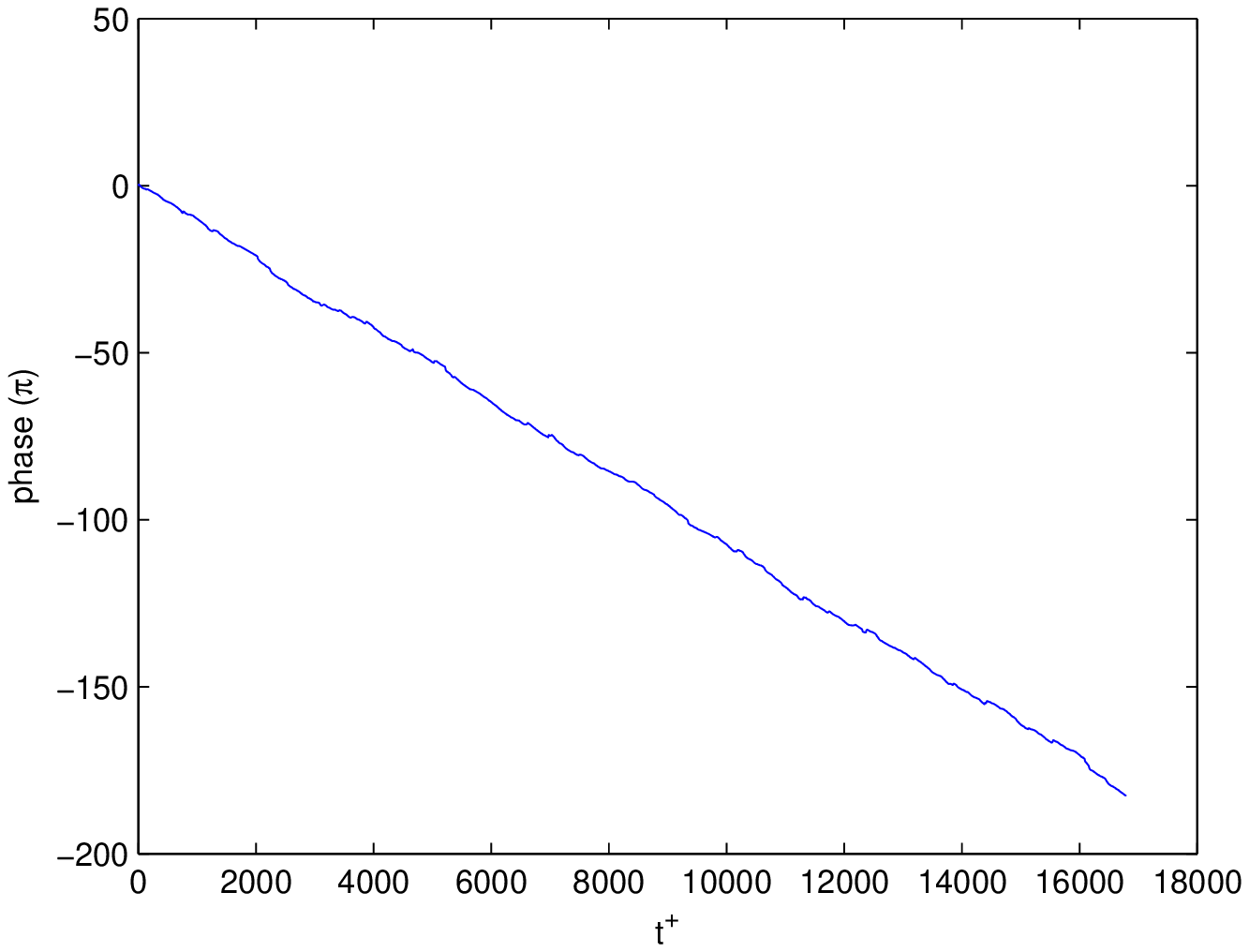}}
\caption{Time history phase of (1,0,1).}
\label{eigen101phase}
\end{minipage} \\
\end{tabular}

\addtocounter{figure}{-1}
\addtocounter{subfigure}{1}
\epsfxsize=5.5 in
\centerline{\epsfbox{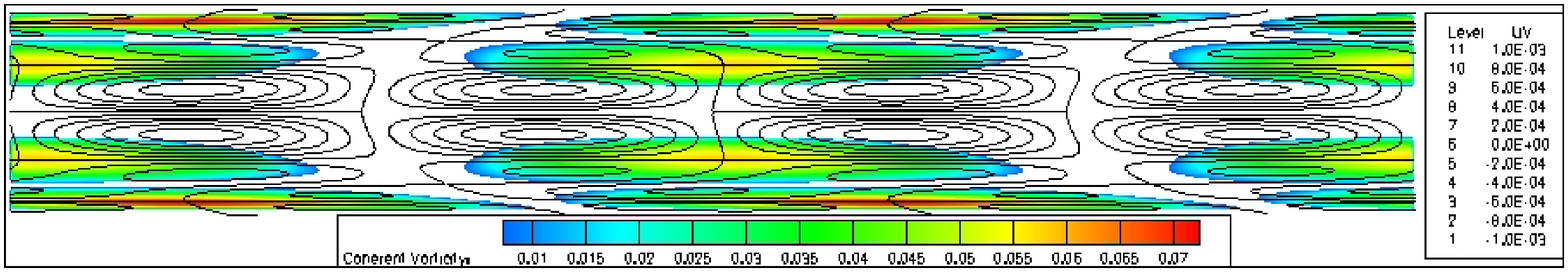}}
\caption{Propagating ring mode (1,0,2). }
\label{eigen102}

\addtocounter{figure}{-1}
\addtocounter{subfigure}{1}
\epsfxsize=5.5 in
\centerline{\epsfbox{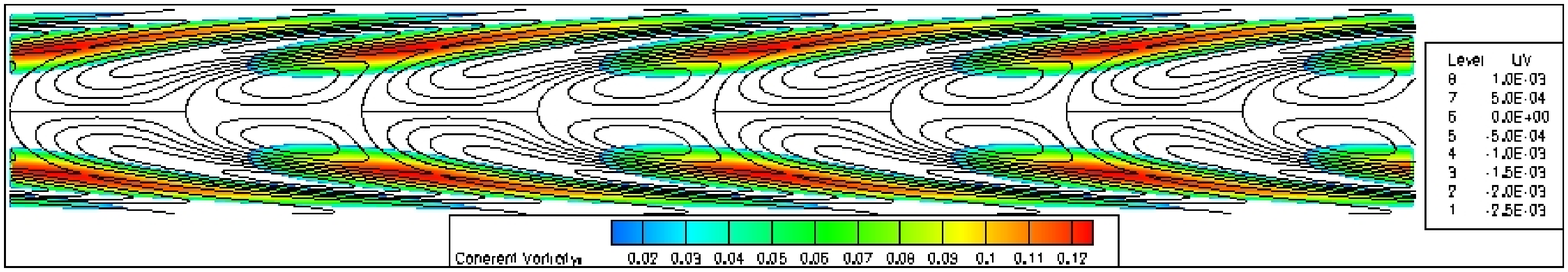}}
\caption{Propagating ring mode (2,0,1). }
\label{eigen201}
\end{figure}


\setcounter{subfigure}{1}
\begin{figure}[p]
\epsfxsize=5.5 in
\centerline{\epsfbox{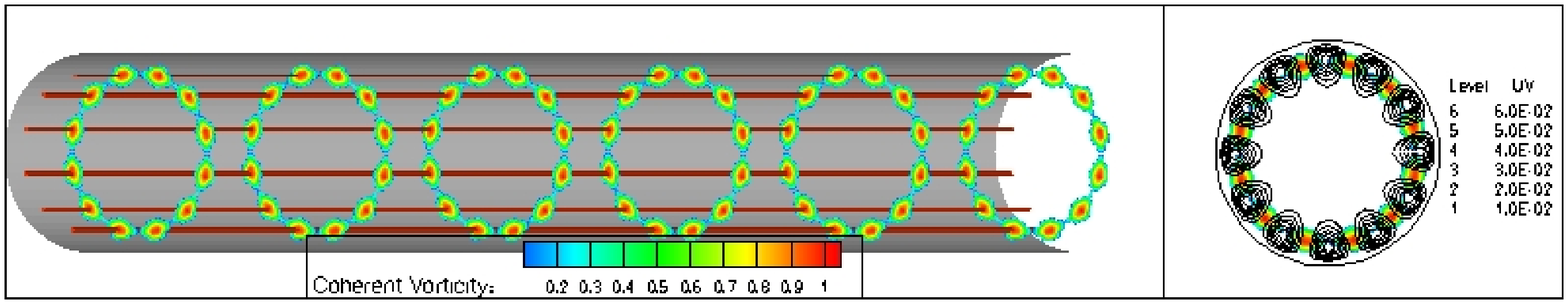}}
\caption{Most energetic non propagating roll mode (0,6,1). Coherent
  vorticity (left) and a cross-section of coherent vorticity with
  Reynolds stress superimposed (right).  }
\label{eigen061}
\centering
\begin{tabular}{cc}
\begin{minipage}{2.5in}
\centering
\addtocounter{figure}{-1}
\addtocounter{subfigure}{1}
\includegraphics[width=2.5 in]{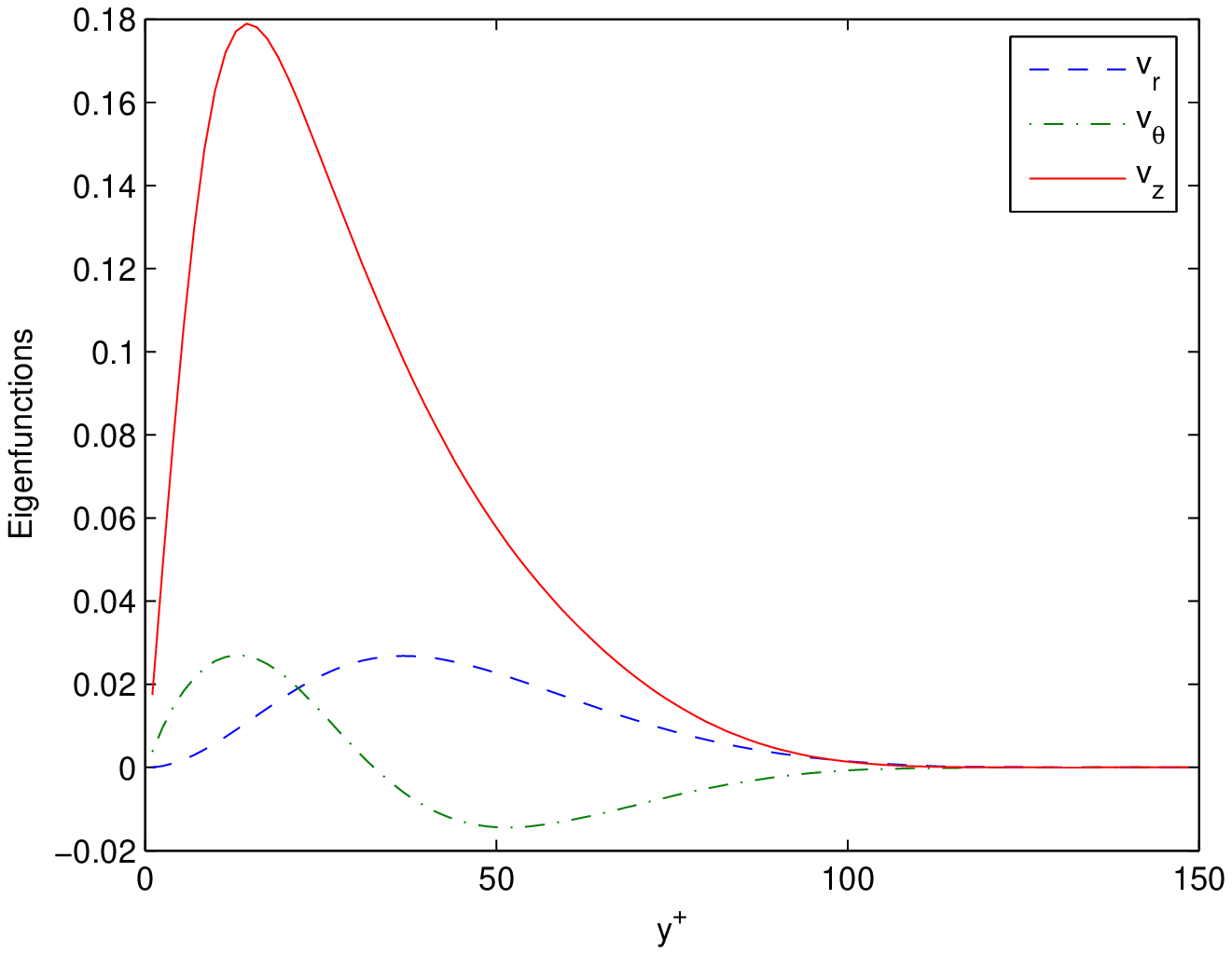}
\caption{Real component of (0,6,1).}
\label{eigen061real}
\end{minipage}
& 
\begin{minipage}{2.5 in}
\centering
\addtocounter{figure}{-1}
\addtocounter{subfigure}{1}
 \resizebox{2.5 in}{!}{\includegraphics{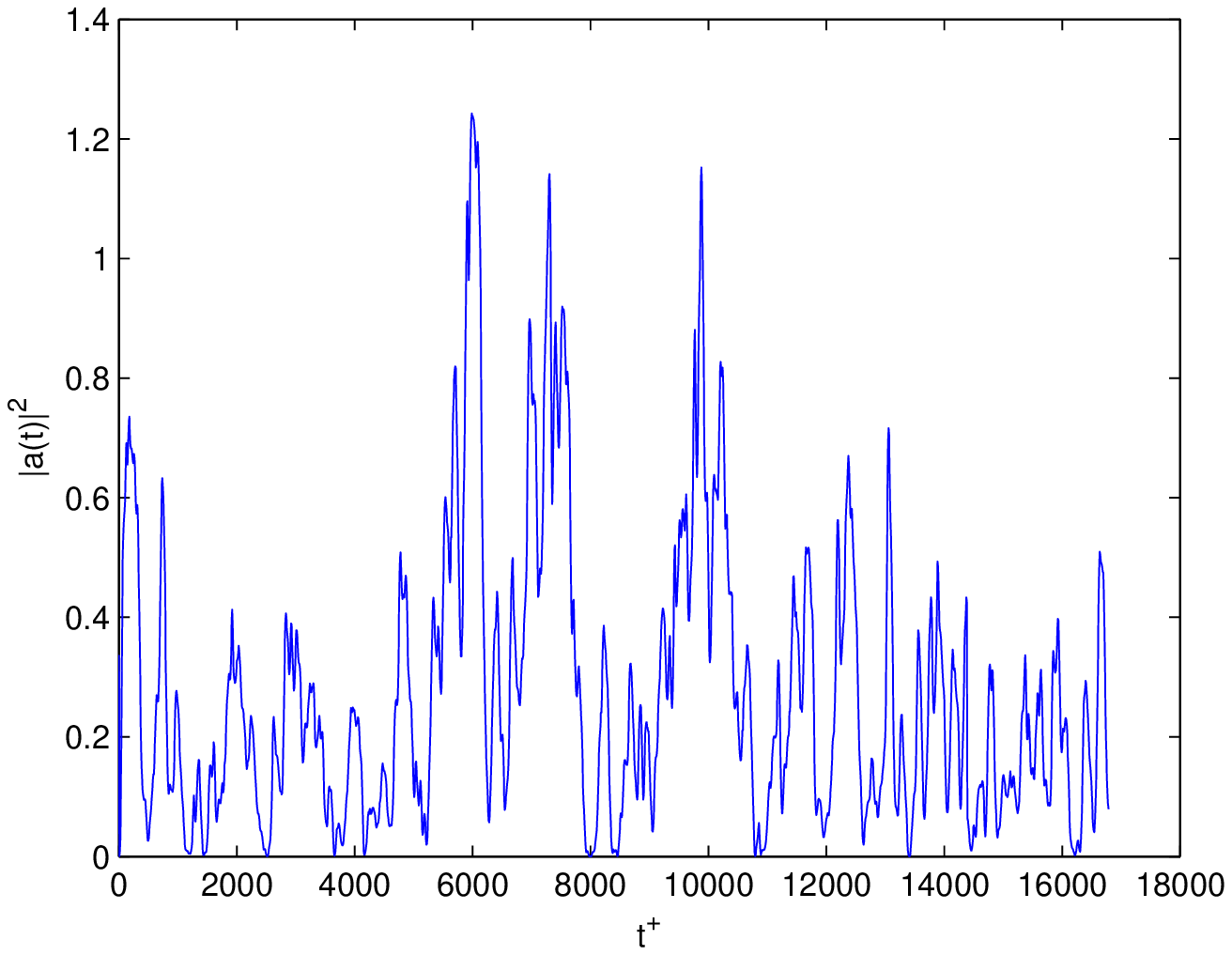}}
\caption{Time history amplitude of (0,6,1).}
\label{eigen061amp}
\end{minipage} \\
\begin{minipage}{2.5in}
\centering
\addtocounter{figure}{-1}
\addtocounter{subfigure}{1}
\includegraphics[width=2.5 in]{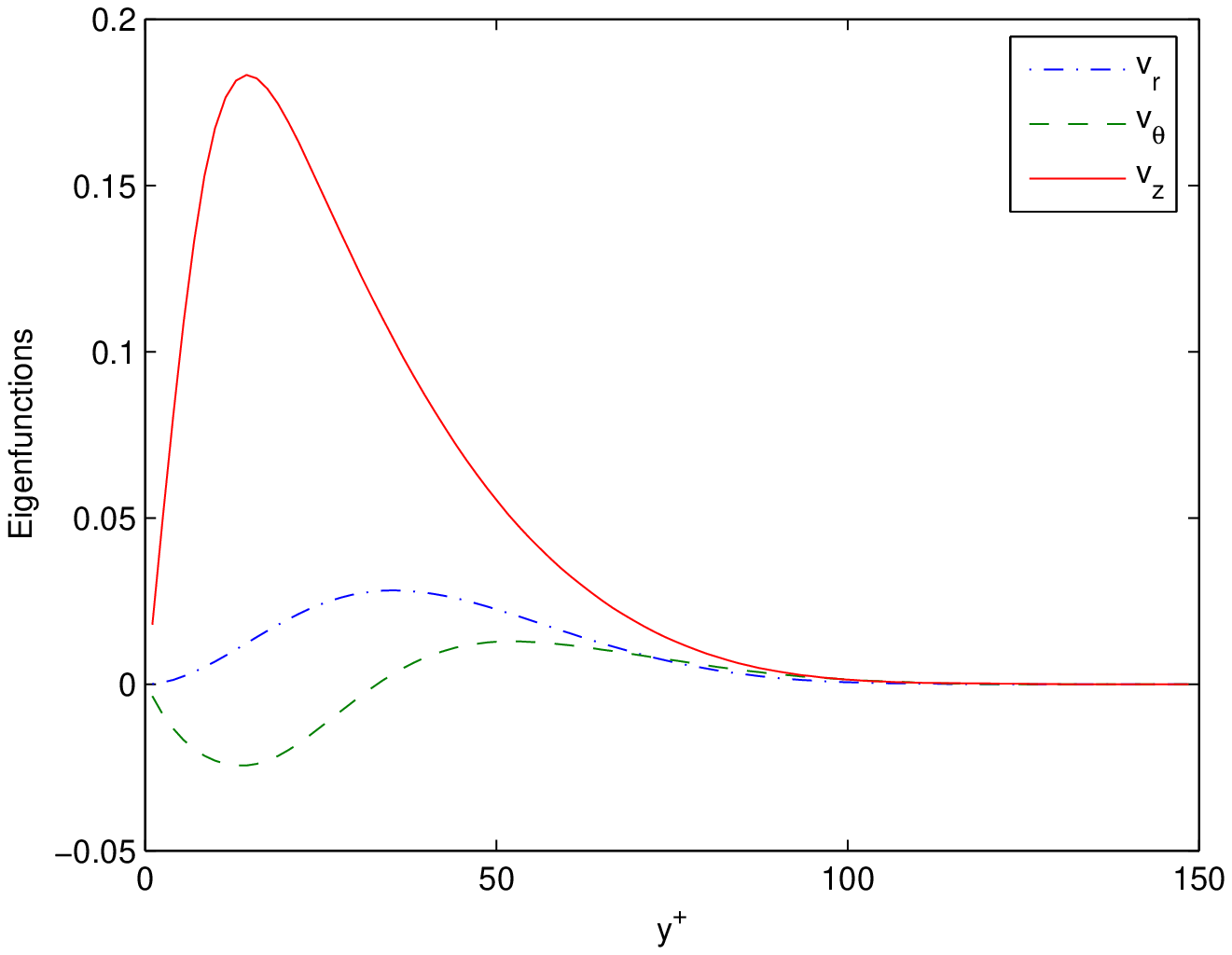}
\caption{Imaginary component of (0,6,1).}
\label{eigen061imag}
\end{minipage}
& 
\begin{minipage}{2.5 in}
\centering
\addtocounter{figure}{-1}
\addtocounter{subfigure}{1}
 \resizebox{2.5 in}{!}{\includegraphics{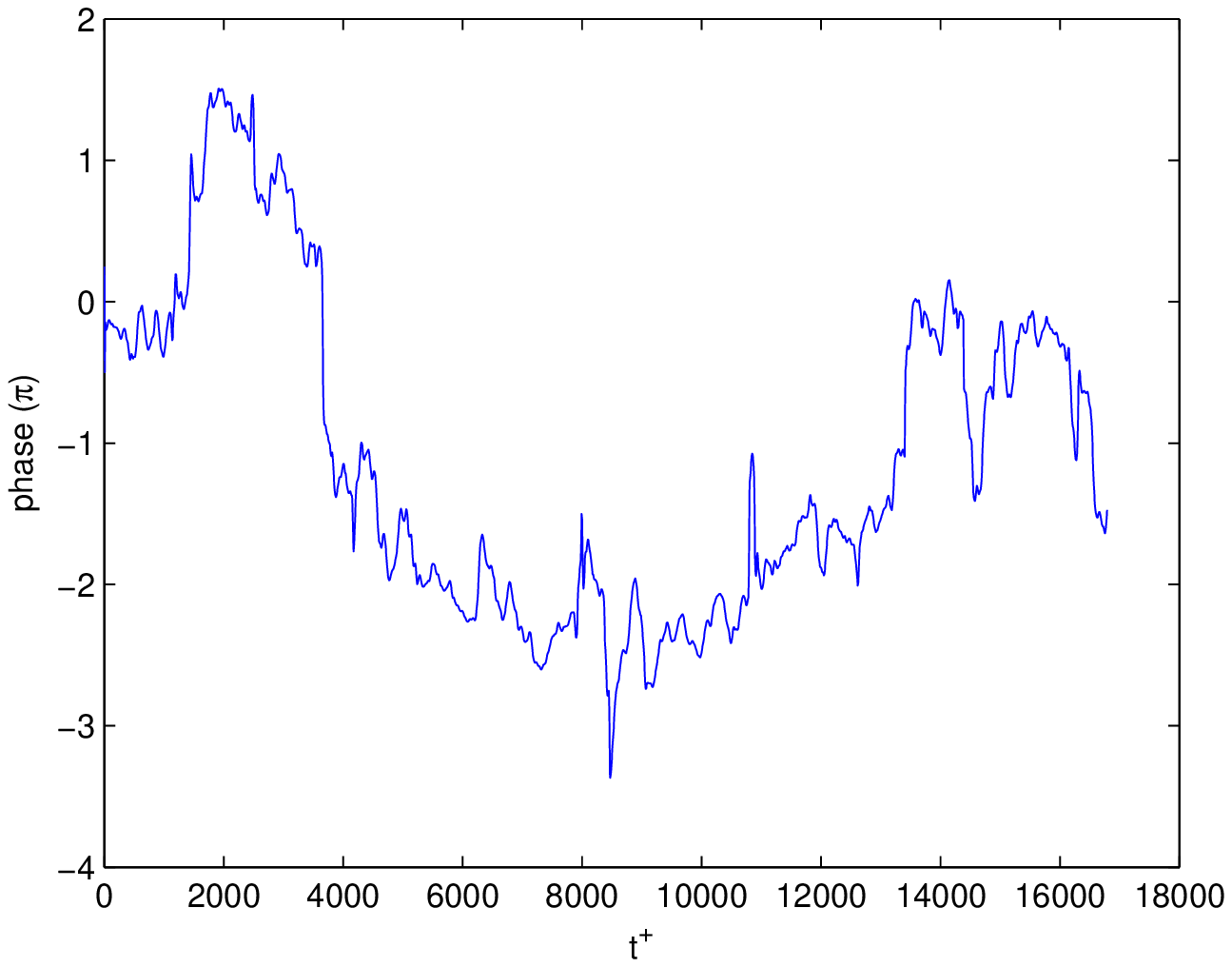}}
\caption{Time history phase of (0,6,1).}
\label{eigen061phase}
\end{minipage} \\
\end{tabular}

\addtocounter{figure}{-1}
\addtocounter{subfigure}{1}
\epsfxsize=5.5 in
\centerline{\epsfbox{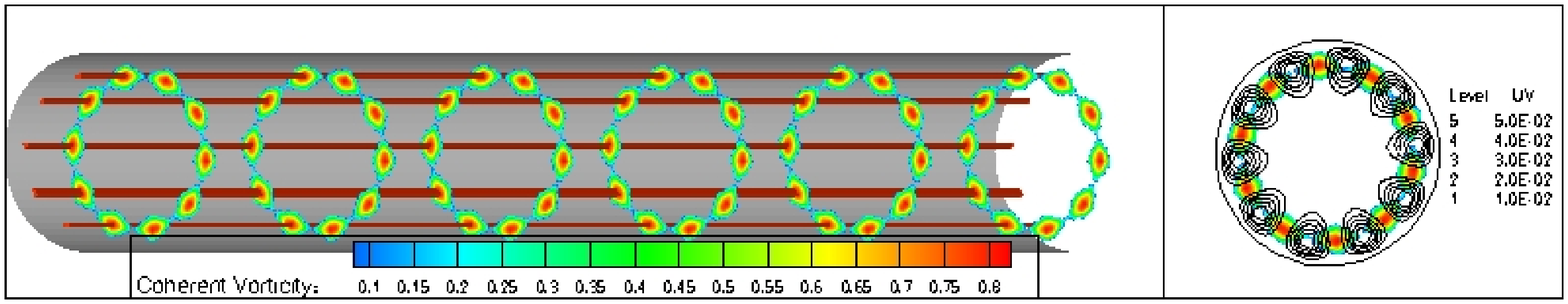}}
\caption{Non propagating roll mode (0,5,1). }
\label{eigen051}
\addtocounter{figure}{-1}
\addtocounter{subfigure}{1}
\epsfxsize=5.5 in
\centerline{\epsfbox{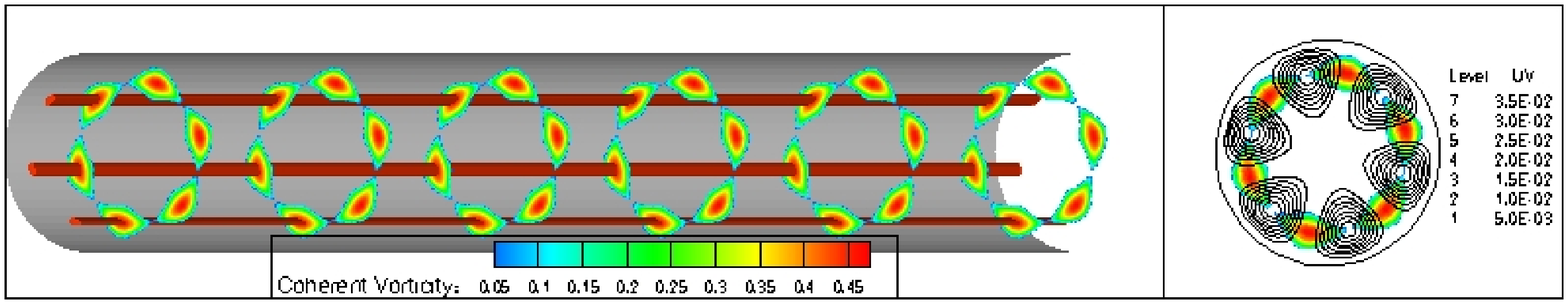}}
\caption{Non propagating roll mode (0,3,1). }
\label{eigen031}
\end{figure}


\begin{figure}[h]
\centering
\begin{tabular}{cc}
\begin{minipage}{2.5 in}
\centering
\setcounter{subfigure}{1}
\includegraphics[width=2.5 in]{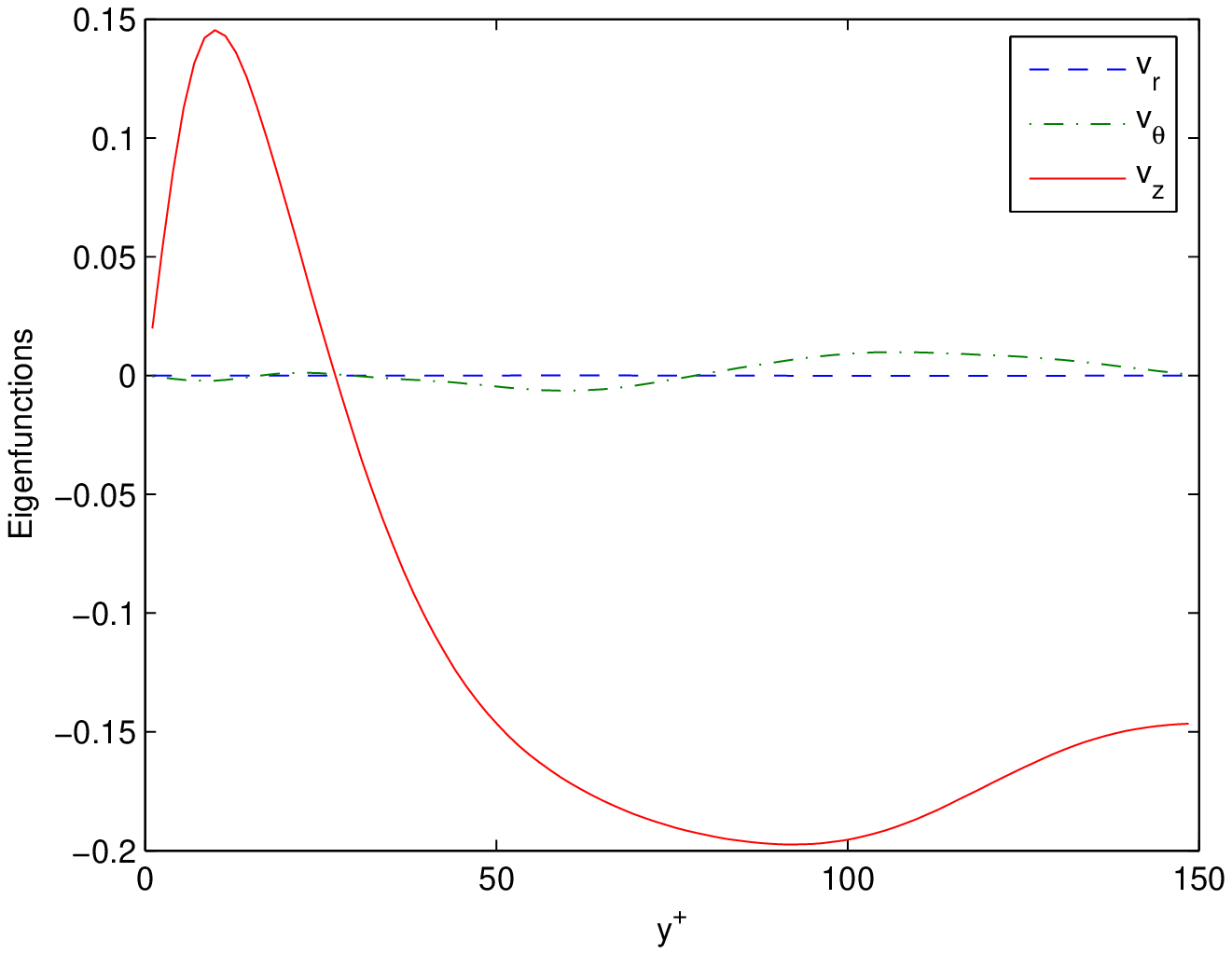}
\caption{Real component of (0,0,1).}
\label{eigen001real}
\end{minipage}
& 
\begin{minipage}{2.5 in}
\centering
\addtocounter{figure}{-1}
\addtocounter{subfigure}{1}
 \resizebox{2.5 in}{!}{\includegraphics{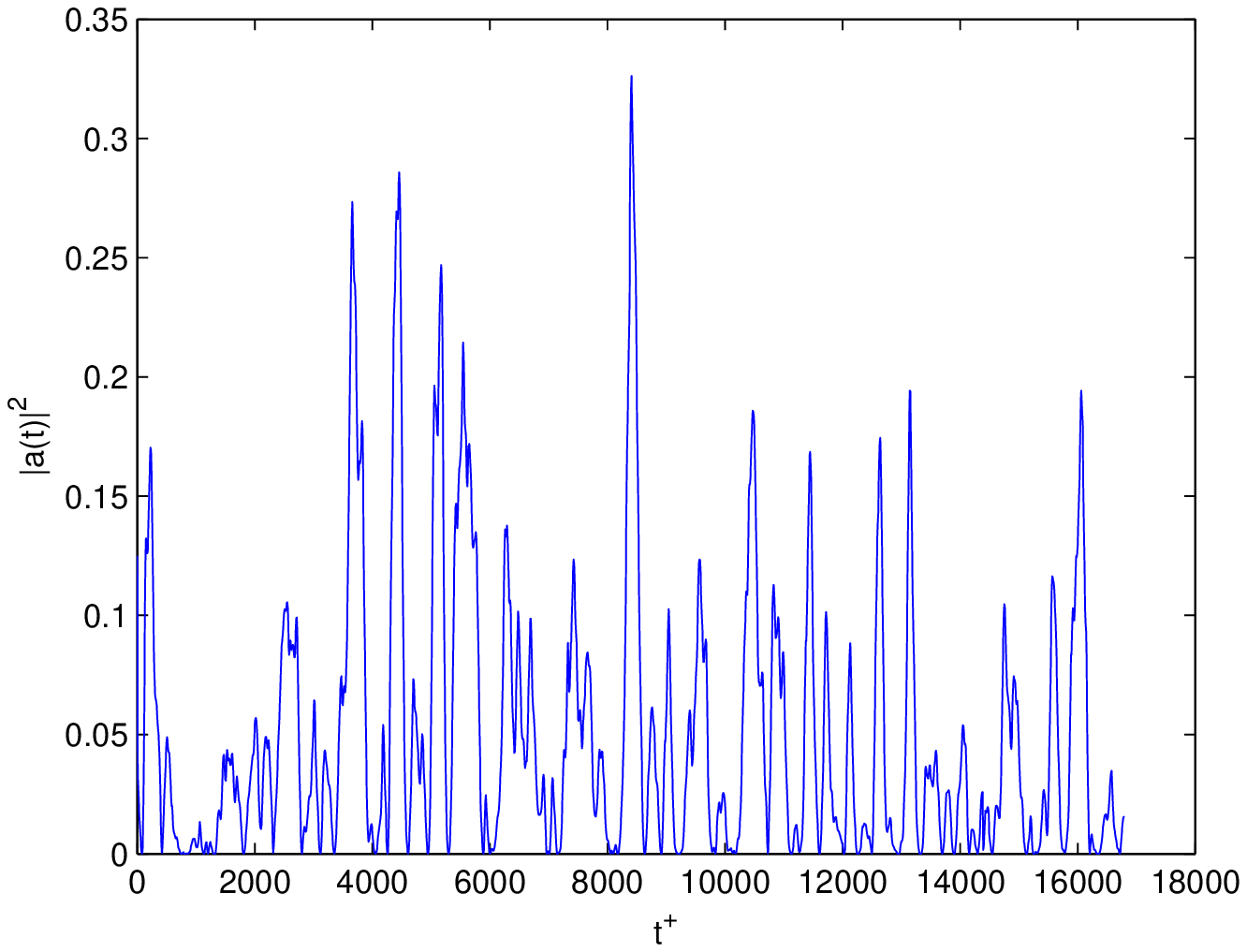}}
\caption{Time history amplitude of (0,0,1).}
\label{eigen001amp}
\end{minipage} \\
\begin{minipage}{2.5in}
\centering
\addtocounter{figure}{-1}
\addtocounter{subfigure}{1}
\includegraphics[width=2.5 in]{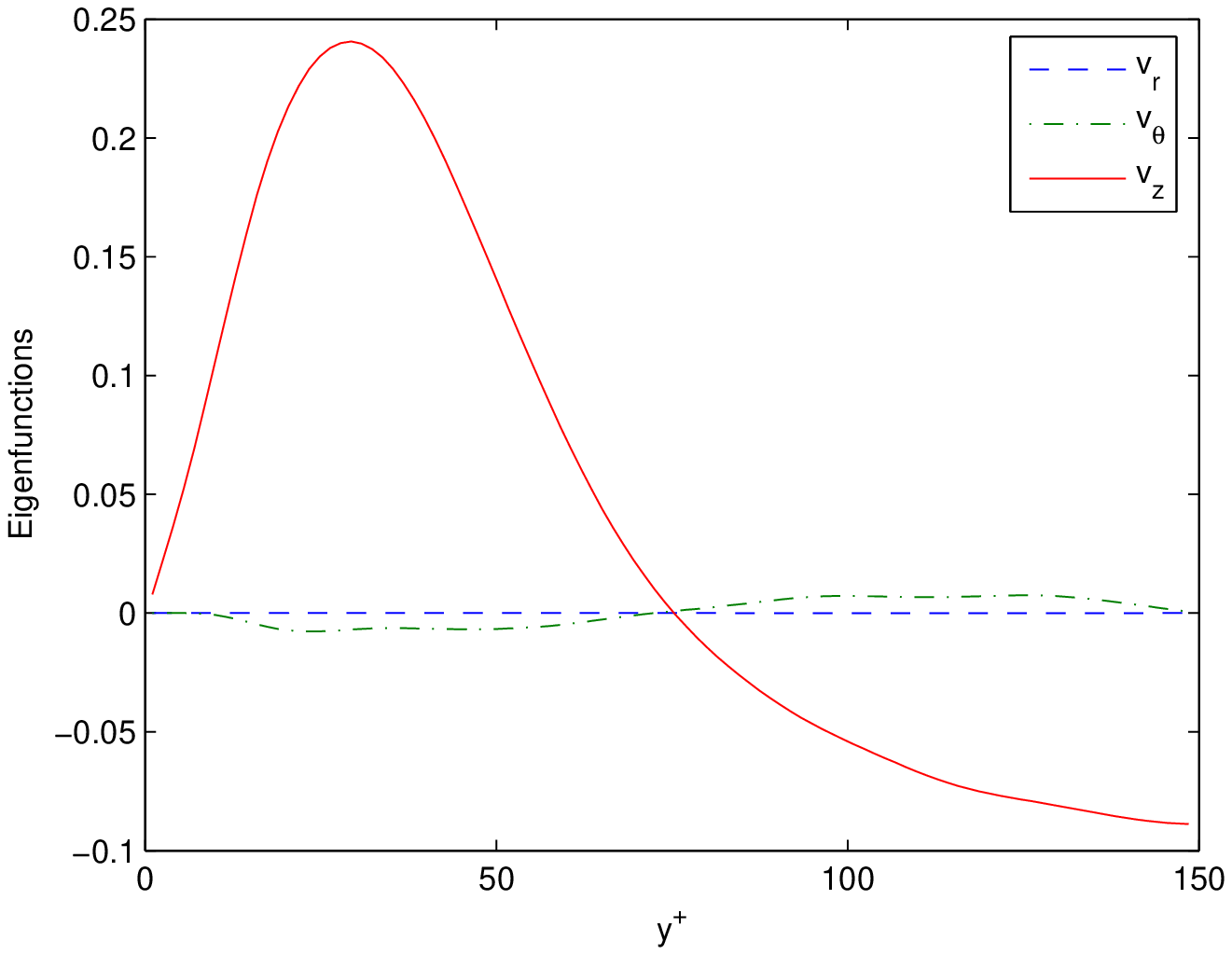}
\caption{Real component of (0,0,2).}
\label{eigen002real}
\end{minipage}
& 
\begin{minipage}{2.5 in}
\centering
\addtocounter{figure}{-1}
\addtocounter{subfigure}{1}
 \resizebox{2.5 in}{!}{\includegraphics{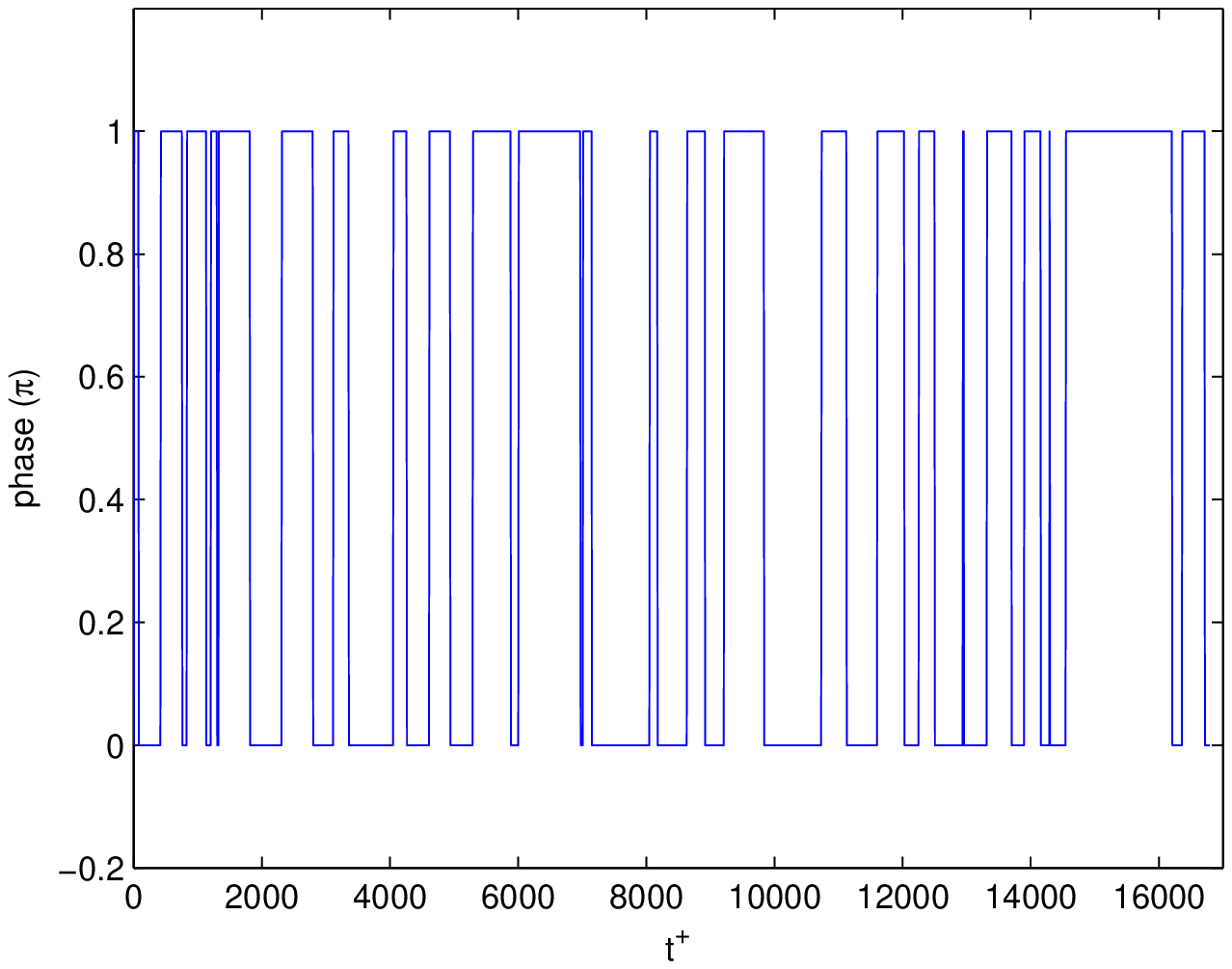}}
\caption{Time history phase of (0,0,1).}
\label{eigen001phase}
\end{minipage} \\
\end{tabular}
\end{figure}
\renewcommand{\thefigure}{\arabic{figure}}

The energy spectra of the propagating mode subclasses are shown in
Figure \ref{distribE}.  This shows that the tail wavenumber end of the inertial
range is populated primarily by the lift modes and the low end
of the spectra is populated predominantly by the wall modes. The
physical meaning of this is that the energy starts at the wall with
large scale structures and is
lifted away from the wall into the outer region in small scale structures, represented by the
lifting modes.

\begin{figure}[h]
\begin{center}
\includegraphics[width=4 in]{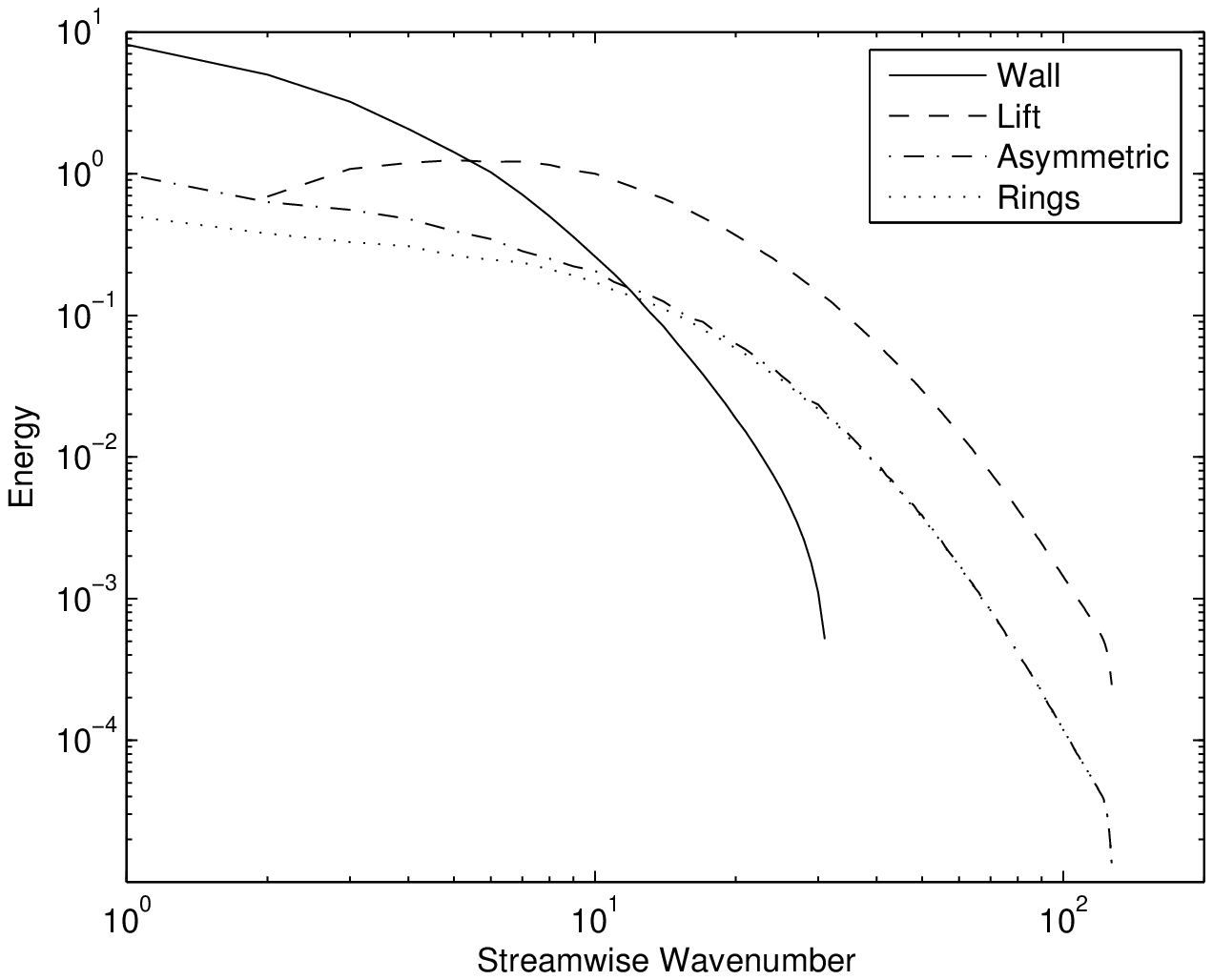}
\end{center}
\caption{Energy spectra for the propagating mode subclasses, wall
  (solid), lift (dashed), asymmetric (dash-dot), and rings (dots).
  The wall modes correspond to the low wavenumber energy and the lift modes represent the high wavenumber spectra commonly known
  as the inertial range.}
\label{distribE}
\end{figure}

The effect of higher radial quantum number $q>1$ is more zero
crossings of the velocities in the radial direction.  The effect on
the location of coherent vorticity and the number of vortex cores scales with the
radial quantum number $q$, making the subclasses invariant with $q$.  The wall and lift modes retain their
characteristics, in that even with more vortices, they remain close
to the wall for the wall mode and close to the centreline for the
lift modes, shown in Figures \ref{623non} through \ref{265non}.

\renewcommand{\thefigure}{\arabic{figure}\alph{subfigure}}

\setcounter{subfigure}{1}
\begin{figure}[h]
\centering
\begin{tabular}{cc}
\begin{minipage}{2.5 in}
\includegraphics[width=2.5 in]{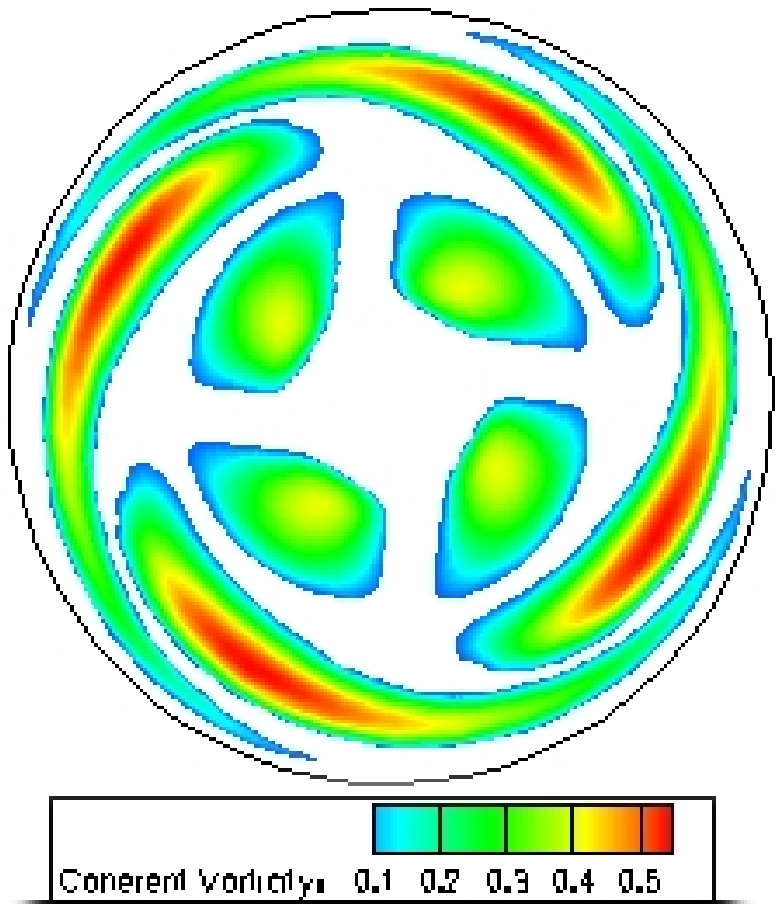}
\caption{(6,2,3) lift mode, contours of coherent vorticity.}
\label{623non}
\end{minipage}
&
\begin{minipage}{2.5 in}
\addtocounter{figure}{-1}
\addtocounter{subfigure}{1}
\includegraphics[width=2.5 in]{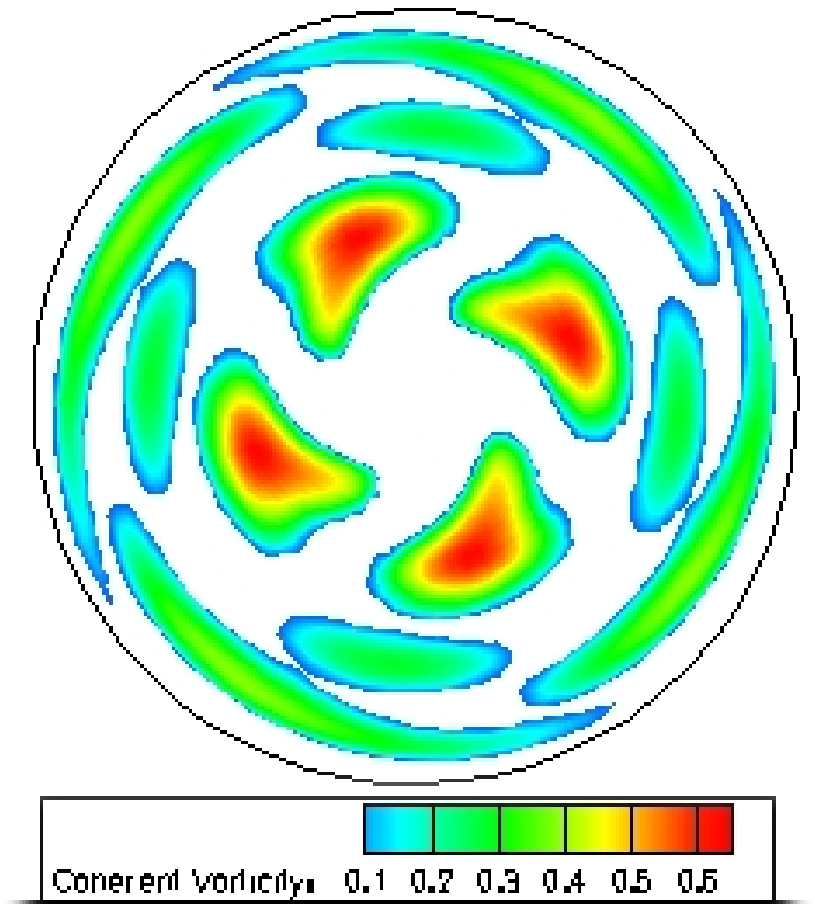}
\caption{(6,2,5) lift mode, contours of coherent vorticity. }
\label{625non}
\end{minipage}\\

\setcounter{subfigure}{1}
\begin{minipage}{2.5 in}
\includegraphics[width=2.5 in]{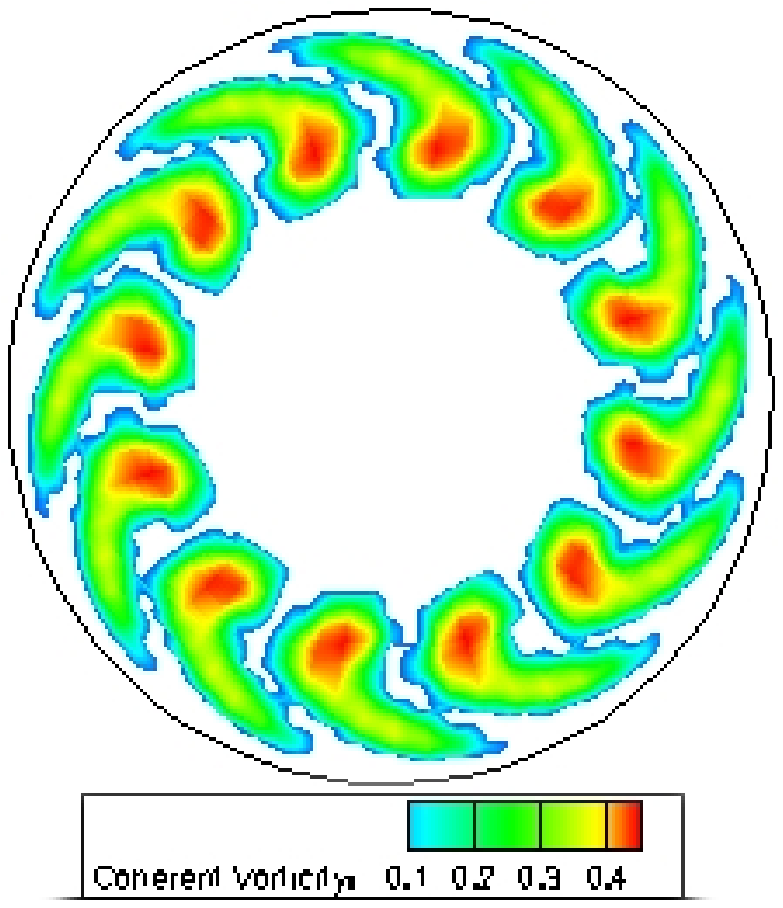}
\caption{(2,6,3) wall mode, contours of coherent vorticity. }
\label{263non}
\end{minipage}
&
\begin{minipage}{2.5 in}
\addtocounter{figure}{-1}
\addtocounter{subfigure}{1}
\includegraphics[width=2.5 in]{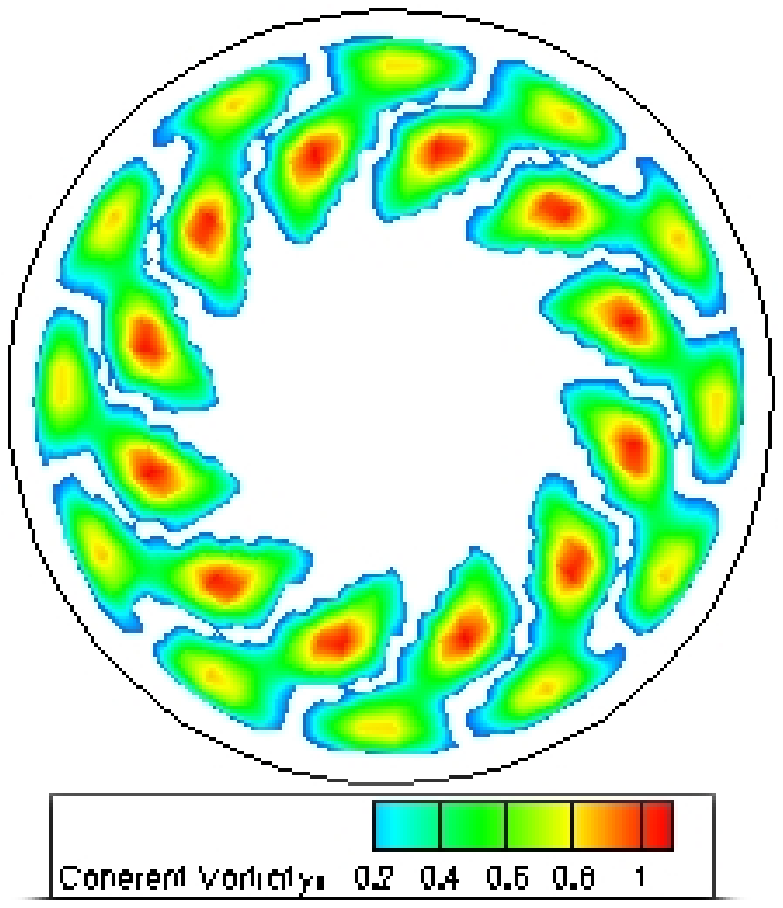}
\caption{(2,6,5) wall mode, contours of coherent vorticity. }
\label{265non}
\end{minipage}\\
\end{tabular}
\end{figure}

\renewcommand{\thefigure}{\arabic{figure}}

\section{Discussion}
Although the Karhunen-Lo\`{e}ve method yields a preferred or natural basis
for turbulence, one must be careful in the conclusions drawn from
the results, as any structure or feature can be reconstructed from any given
orthogonal basis.  The strength of this method is that it is the most
optimal basis, and by focusing on the most energetic modes a
low order dynamical representation of the flow is realised.  This low
order dynamical system reveals three important results, and paints a
picture of the energy dynamics in turbulent pipe flow.

\begin{figure}
\begin{center}
\includegraphics[width=4 in]{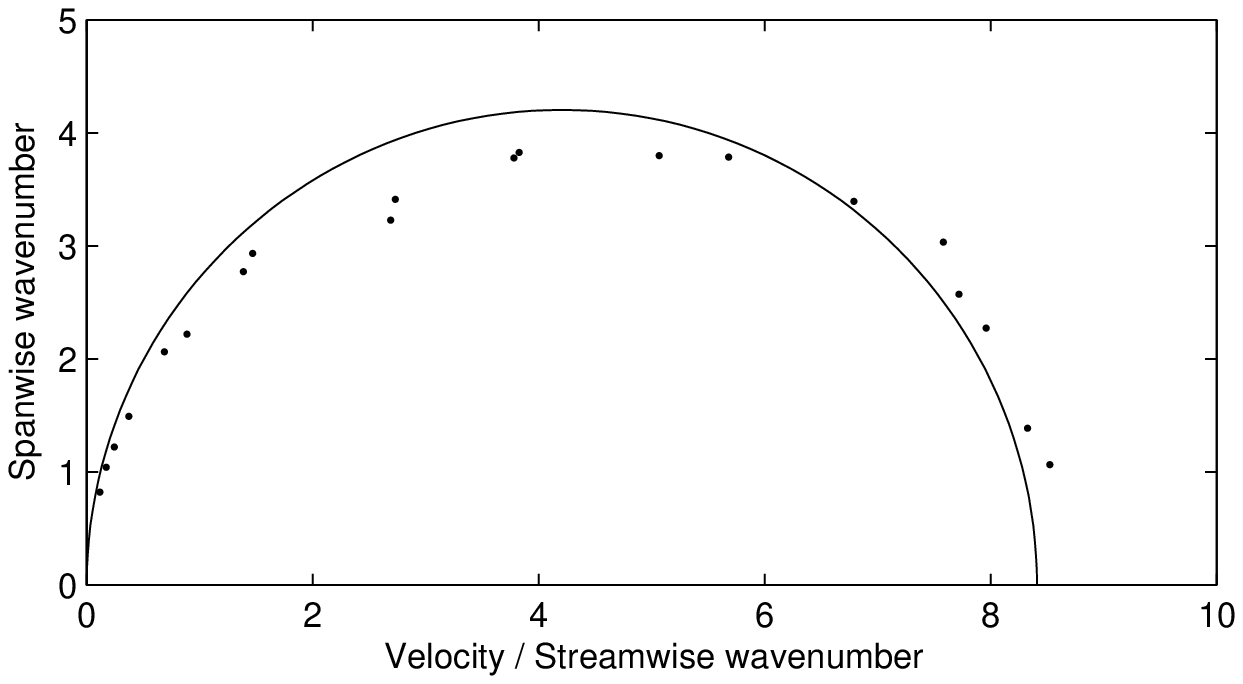}
\end{center}
\caption{Normal speed locus of top 25 most energetic propagating
  eigenfunctions, normalised with shear velocity $u_\tau$.  The circle
  is a least-squares fit to the data and represents that the wave
  packets are acting together as a group with speed of  8.41 $U_\tau$, the point
  where the circle intersects the y-axis.}
\label{locus}
\end{figure}

The first result is the constant phase speed of the propagating
modes.  This was also found in studies of turbulent channel flow, as the structures
represented by the basis advect with a constant group velocity, the
same average velocity of burst events \cite{ball, sirovich1, sirovich2}.  The
normal speed locus of the propagating waves is shown in Figure
\ref{locus} for the propagating modes found in the top 50 most energetic modes.  For this, the phase speed $\omega / \|\mathbf{ k} \|$ is plotted
in the direction $\mathbf{k} / \|\mathbf{k}\|$.  The locus is nearly circular, which is evidence that these structures
propagate as a wave packet or envelope that travels with speed of
{\bf $8.41 U_\tau$}, the point at which the circle intersects the
x-axis in Figure \ref{locus}, which corresponds to the mean velocity
in the buffer region at $y^+=9.6$.  

The second interesting result is that none of the modes with azimuthal wave
number $n=1$ exhibit any travelling waves near the wall, and are
without exception a streamwise or inclined streamwise vortex in the
outer region.  The reason is that the $n=1$ mode does not allow for a
near-wall travelling wave as found by Kerswell \cite{kerswell}, and as such, the basis expansions
for these modes have only near-centreline structures. 

The third result of note is that the different KL structures, as observed
from our results, can be seen qualitatively as an expansion that represents
the horseshoe (hairpin) vortex representation of turbulent
structures.  This horseshoe structure is supported by a large number
of researchers in the turbulence
community as the self-sustaining mechanism for turbulence
\cite{panton_overview}.  In the representative horseshoe structure (see, for
example, the figure by Theodorsen \cite{theodorsen}), the
structures found in the KL can be seen.  The wall modes represent the
leg structure and its perturbation near the wall.  The lift modes
represent the structure lifting off the wall.  The asymmetric modes
represent the secondary and tertiary horseshoes that are formed. The
turn modes represent the spanwise head of the horseshoe.  As mentioned
before, since the KL decomposition forms a basis, any flow can be
recreated from these eigenmodes, so this result is reported as an
interesting qualitative result, and as an understanding that the
propagating modes are the structures of interest in turbulence.

 The final dynamical picture of the KL modes in summarising the work done
  by Kerswell \cite{kerswell}, Webber et al. \cite{webber1, webber2},
  Sirovich et al. \cite{sirovich1}, and this current work is as follows.  Webber et
  al. found that the energy
  enters the flow from the pressure gradient to the shear modes.  The
  shear modes, interacting with the roll modes present in
  fully developed turbulence transfers the energy from the shear modes
 to the roll modes.  Shown in this study, and
  theoretically by Kerswell, the roll modes decay much more slowly than the
  propagating
modes.  Because of this slow decay, these streamwise rolls provide an energy
storage role in the turbulence, similar to a capacitance.  This allows
  enough time for the roll to interact with the propagating wave
  packet shown by Sirovich et al.  Because this interaction requires a finite
time to occur, the energy storage and slow decaying nature of the
streamwise rolls play an integral role in the self-sustaining
nature of turbulence.   It is also surmised
that the travelling
modes found by Kerswell and other researchers are represented in the
KL formulation as the most
unstable wave packet of KL modes.  Following again the work of Webber et al. and
  applying the classifications found in this study, the majority of the energy from the roll modes
  are transferred to the wall modes.  The wall modes then interact
  with themselves through an asymmetric mode catalyst, and with the
  lift modes through a ring mode catalyst.  Physically, this gives a
  picture of wall turbulence energy being transferred near the wall and
  then lifted up to the outer layer.  This interaction between the
  wall and the lift modes is what
  populates the inertial range shown in figure \ref{distribE}.  The
  energy is ultimately dissipated by viscosity, faster for the higher
  wavenumber modes.

\section{Conclusions}
We have presented the use of the Karhunen-Lo\`{e}ve expansion method with the results of a globally high-order
direct numerical simulation of turbulent pipe flow.  The
results reveal the structure of the
turbulent pipe flow as propagating (80.58\% total energy) and non propagating modes (19.42\%
total energy).  The
propagating modes are characterised by a constant phase speed and have four distinct classes:  wall, 
lift, asymmetric, and ring modes.  These propagating modes form
a travelling wave envelope, forming a circular, normal-speed locus,
with advection speed of $8.41 U_\tau$ corresponding to the mean
velocity in the buffer region near $y^+=9.6$.  The non propagating modes
have two subclasses: streamwise roll and shear modes.  These
represent the energy storage and mean fluctuations of the turbulent
flow, respectively.   The energy is transferred from the streamwise
rolls to the wall modes first, and then later to the lift modes,
physically representing the energy transfer from the wall to the outer
region.  This eigenfunction expansion, using both their structure and
their time-dependent coefficients, provides a framework for
understanding the dynamics of turbulent pipe flow, and will provide a
basis for
further analysis and comparison leading to understanding
the mechanism of drag reduction by
spanwise wall oscillation \cite{duggleby_drPipe} and the mechanism of
relaminarization \cite{duggleby_relam}.

\section*{Acknowledgements}
We gratefully acknowledge the use of the Virginia Tech Terascale Computing Facility
and the Teragrid San Diego Supercomputing facility.




\begin{thebibliography}{26}

\bibitem{panton_overview}
R. Panton, 2001.  Overview of the Self-sustaining Mechanisms of Wall
Turbulence.  {\itshape Prog. Aero. Sci.,} {\bfseries 37}, 343--383.


\bibitem{patel_head}
Patel, V.C., and Head, M.S., 1969.  Some observations on skin friction
and velocity profiles in fully developed pipe and channel
flows. {\itshape J. Fluid Mech.,}, {\bfseries 38}, 181--201.

\bibitem{durst}
Durst, F., Jovanovic, J., and Sender, J., 1995.  LDA measurements in
the near-wall region of a turbulent pipe flow. {\itshape J. Fluid
  Mech.,} {\bfseries 295}, 305--335.

\bibitem{orszag_patera}
Orszag, S.A., and Patera, A.T., 1983. Secondary instability in wall-bounded
shear flows. {\itshape J. Fluid  Mech.,} {\bfseries 128}, 347--385.

\bibitem{oSullivan_breuer}
O'Sullivan, P.L, and Breuer, K.S., 1994.  Transient growth in
circular pipe flow, 1: Linear disturbances.   {\itshape
  Phys. Fluids,} {\bfseries 6}, 11, 3643--3651.

\bibitem{ma}
Shan, H., Ma, B., Zhang, Z., and Nieuwstadt, F.T.M., 1999.  Direct numerical
simulation of a puff and a slug in transitional cylindrical pipe
flow. {\itshape J. Fluid Mech.,} {\bfseries 387}, 39--60.

\bibitem{loulou}
Loulou, P., Moser, R.D., Mansour, N.N., and Cantwell, B.J., 1997.  Direct
Numerical Simulation of Incompressible Pipe Flow Using a B-Spline
Spectral Method.  {\itshape NASA Technical Memorandum-110436}.

\bibitem{eggels}
Eggels, J.G.M., Unger, F., Weiss, W.H., Westerweel, J., Adrian, R.J., Friedrich, R., and
Nieuwstadt, F.T.M., 1994.  Fully developed turbulent pipe flow: A comparison
between direct numerical simulation and experiment.  {\itshape J. Fluid
  Mech.,} {\bfseries 268}, 175--209.

\bibitem{verzicco}
Verzicco, R., and Orlandi, P., 1996.  A finite-difference scheme for
three-dimensional incompressible flows in cylindrical coordinates {\itshape
  J. Comp. Phys,} {\bfseries 123}, 402--414.

\bibitem{fukagata}
Fukagata, K., and Kasagi, N., 2002.  Highly energy-conservative finite
difference method for the cylindrical coordinate system. {\itshape J. Comp
  Phys.,} {\bfseries 181}, 478--498.


\bibitem{lumley1}
Lumley, J.L., 1967.  The structure of inhomogeneous turbulent
flows. {\itshape Atmospheric Turbulence and Radio Wave Propagation,}
 Yaglom, A.M. and Tatarski, V.I. (eds), (Nauka, Moscow)  166--178.

\bibitem{lumley2}
Lumley, J.L. 1970.  {\it Stochastic Tools in Turbulence.}  Academic
  Press, (New York).

\bibitem{ball}
Ball, K.S., Sirovich, L., and Keefe, L.R., 1991.  Dynamical eigenfunction
decomposition of turbulent channel flow.  {\itshape
  Int. J. Numer. Meth. Fluids,} {\bfseries 12}, 585--604.

\bibitem{sirovich1}
Sirovich, L., Ball, K.S., and  Keefe, L.R., 1990.  Plane waves and structures
in turbulent channel flow.  {\itshape Phys. Fluids A,} {\bfseries 12},
2217--2226.

\bibitem{sirovich2}
Sirovich, L., Ball, K.S., and Handler, R.A,,1991.  Propagating structures
in wall-bounded turbulent flows.  {\itshape Theoret. Comput. Fluid
  Dynamics,} {\bfseries 2}, 307--317.

\bibitem{aubry}
Aubry, N, Holmes, P., Lumley, J.L., Stone, E., 1988.  The dynamics of
coherent structures in the wall region of a turbulent boundary layer. {\itshape
  J. Fluid Mech.,} {\bfseries 192}, 115--173. 

\bibitem{zhou_sirovich}
Zhou,X., and Sirovich, L., 1992. Coherence and chaos in a model of the turbulent
boundary layer. {\itshape Phys. Fluids A,} {\bfseries 4},
2855--2874. 

\bibitem{sirovich_xhou}
Sirovich L., and Zhou, X., 1994.  Dynamical model of wall bounded turbulence.
 {\itshape Phys. Rev. Lett.,} {\bfseries 72}, 340--343. 

\bibitem{smith2005}
Smith, T.R., J. Moehlis, and P. Holmes, 2005.  Low-dimensional models for
turbulent pipe Coutette flow in a minimal flow unit.
 {\itshape J. Fluid. Mech.,} {\bfseries 538}, 71--110. 

\bibitem{webber1}
Webber, G.A., Handler, R.A., and Sirovich L., 1997.  The Karhunen-Lo\`{e}ve
decomposition of minimal channel flow.  {\itshape Phys. Fluids,}
{\bfseries 9}, 1054--1066.

\bibitem{webber2}
Webber, G.A., Handler, R.A, and Sirovich, L., 2002.  Energy dynamics in a
turbulent channel flow using the Karhunen-Lo\`{e}ve approach. {\itshape
  Int. J. Numer. Meth. Fluids,} {\bfseries 40}, 1381--1400.

\bibitem{boyd}
Boyd, J.P., 2000.  {\it Chebyshev and Fourier Spectral Methods.}  DOVER
  Publications, Mineola, New York.

\bibitem{fischer_patera}
Fischer, P.F., Ho, L.W., Karniadakis, G.E, Ronouist, E.M., and
 Patera, A.T., 1988.  Recent advances in parallel spectral element simulation of unsteady incompressible flows.
{\itshape Comput. \& Struct.,} {\bfseries 30}, 217--231.

\bibitem{tufo}
Tufo, H.M., and Fischer, P.F., 1999.  Terascale spectral element
algorithms and implementations, {\itshape Proceedings of the ACM/IEEE SC99
  Conference on High Performance Networking and Computing,}  IEEE
  Computer Soc., Gordon Bell Prize paper.

\bibitem{lottes}
Lottes, J.W. and Fischer, P.F., 2004. Hybrid multigrid/Schwarz algorithms for
the spectral element method.  {\itshape J. Sci. Comp.,} {\bfseries 24}, 45--78.

\bibitem{paul}
Paul, M.R., Chiam, K.H., and Cross, M.C., 2004.  Rayleigh-Benard
convection in large-aspect-ratio domains.  {\itshape
  Phys. Rev. Lett.,} {\bfseries 93}, 064503.

\bibitem{gullbrand}
Gullbrand, J., 2000.  An evaluation of a conservative fourth order DNS code in
turbulent channel flow.  {\itshape Center for Turbulence Research, Anual
  Research Briefs}.

\bibitem{jimenez}
Jim\'{e}nez, J. 1998.  The largest scales of turbulent wall flows  {\itshape Center for Turbulence Research, Annual
  Research Briefs}.

\bibitem{sirovichKL1}
Sirovich, L., 1987.  Turbulence and the dynamics of coherent
structures, Part I: Coherent Structures. {\itshape Q. Appl. Math.,}
{\bfseries XLV 3}, 561--571.

\bibitem{sirovichKL2}
Sirovich, L., 1987.  Turbulence and the dynamics of coherent
structures, Part II:  Symmetries and transformations. {\itshape Q. Appl. Math.,}
{\bfseries XLV 3}, 573--582.

\bibitem{sirovichKL3}
Sirovich, L., 1987.  Turbulence and the dynamics of coherent
structures, Part III:  Dynamics and scaling. {\itshape Q. Appl. Math.,}
{\bfseries XLV 3}, 583--590.

\bibitem{sirovich_chaos}
Sirovich, L., 1989.  Chaotic dynamics of coherent structures.  {\itshape
  Phys. D,} {\bfseries 37}, 126--143.

\bibitem{chong}
Chong, M.S., Perry, A. E., and Cantwell, B. J., 1990.  A general
classification of three-dimensional flow fields {\itshape Phys. Fluids
  A,} {\bfseries 2}, 765--777.

\bibitem{kerswell}
Kerswell, R.R., 2005.  Recent progress in understanding the transition
to turbulence in a pipe.  {\itshape Nonlinearity,} {\bfseries 18},
R17--R44.

\bibitem{theodorsen}
Theodorsen, T. 1952.  Mechanism of turbulence.  {\itshape Proceedings
  of the Second Midwestern Conference on Fluid Mechanics, Ohio State
  University,} 1--18.

\bibitem{duggleby_drPipe}
Duggleby, A., Ball, K.S., and Paul, M.R., 2006. Dynamics of propagating turbulent pipe flow
    structures. Part I: Effect of drag reduction by spanwise wall
    oscillation. {\itshape Phys. Fluids,} in review.

\bibitem{duggleby_relam}
Duggleby, A., Ball, K.S., and Paul, M.R., 2006. Dynamics of propagating turbulent pipe flow
    structures. Part II: Relaminarization {\itshape Phys. Fluids,} in review.

\end{thebibliography}
\end{document}